\documentclass{pas}

\usepackage{graphicx}	
\usepackage{amsmath}	
\usepackage{multirow}
\usepackage{soul}
\usepackage{pdflscape}
\usepackage[shortcuts]{extdash}
\usepackage{orcidlink}
\usepackage[normalem]{ulem}
\usepackage{fixfoot}
\usepackage{booktabs} 
\usepackage{eso-pic}
\usepackage{rotating}
\usepackage{textcase}

\begin{document}

\lefttitle{Combining Pantheon$+$ and DES-SN5YR}
\righttitle{R.~Camilleri~\textit{et al.}}

\jnlPage{1}{4}
\jnlDoiYr{2021}
\doival{10.1017/pasa.xxxx.xx}

\articletitt{Research Paper}

\title{Supernovae Unite: Combining Pantheon$+$ and DES-SN5YR}

\author{
\gn{Ryan} \sn{Camilleri}$^{\orcidlink{0000-0002-7436-3950}}$,$^{1}$
\gn{Jason} \sn{Lee}$^{\orcidlink{0000-0001-6633-9793}}$,$^{2}$
\gn{Tamara M.} \sn{Davis}$^{\orcidlink{0000-0002-4213-8783}}$,$^{1}$
\gn{David} \sn{Rubin}$^{\orcidlink{0000-0001-5402-4647}}$,$^{3}$
\gn{Paul} \sn{Shah}$^{\orcidlink{0000-0002-8000-6642}}$,$^{4}$
\gn{Dan} \sn{Scolnic}$^{\orcidlink{0000-0002-4934-5849}}$,$^{5}$
\gn{Chris} \sn{Lidman}$^{\orcidlink{0000-0003-1731-0497}}$,$^{6}$
\gn{Brodie} \sn{Popovic}$^{\orcidlink{0000-0002-8012-6978}}$,$^{7}$
\gn{Koby} \sn{Grech}$^{\orcidlink{0009-0002-2654-843X}}$,$^{1}$
\gn{Maria} \sn{Vincenzi}$^{\orcidlink{0000-0001-8788-1688}}$,$^{8}$
\gn{Dillon} \sn{Brout}$^{\orcidlink{0000-0001-5201-8374}}$,$^{9}$
\gn{Maria} \sn{Acevedo}$^{\orcidlink{0000-0002-5389-7961}}$,$^{5}$
\gn{Patrick} \sn{Armstrong}$^{\orcidlink{0000-0003-1997-3649}}$,$^{6,10,11}$
\gn{Bruce A.} \sn{Bassett}$^{\orcidlink{0000-0001-7700-1069}}$,$^{12,13}$
\gn{Keith} \sn{Bechtol}$^{\orcidlink{0000-0001-8156-0429}}$,$^{2}$
\gn{Rebecca C.} \sn{Chen}$^{\orcidlink{0000-0003-3917-0966}}$,$^{14,15,16}$
\gn{Herman T.} \sn{Diehl}$^{\orcidlink{0000-0002-8357-7467}}$,$^{17,18,19}$
\gn{Josh} \sn{Frieman}$^{\orcidlink{0000-0003-4079-3263}}$,$^{17,18,19}$
\gn{Llu\'is} \sn{Galbany}$^{\orcidlink{0000-0002-1296-6887}}$,$^{20,21}$
\gn{Anais} \sn{M\"oller}$^{\orcidlink{0000-0001-8211-8608}}$,$^{22}$
\gn{Masao} \sn{Sako}$^{\orcidlink{0000-0003-2764-7093}}$,$^{23}$
\gn{Bruno O.} \sn{S\'anchez}$^{\orcidlink{0000-0002-8687-0669}}$,$^{24}$
\gn{Mark} \sn{Sullivan}$^{\orcidlink{0000-0001-9053-4820}}$$^{7}$
and
\gn{Brad E.} \sn{Tucker}$^{\orcidlink{0000-0002-4283-5159}}$$^{6}$
}

\affil{
$^{1}$School of Mathematics and Physics, University of Queensland, Brisbane, QLD 4072, Australia, 
$^{2}$Department of Physics, University of Wisconsin Madison, WI 53706-1390, USA, 
$^{3}$Department of Physics and Astronomy, University of Hawai‘i at Mānoa, Honolulu, Hawai‘i 96822, 
$^{4}$Department of Physics and Astronomy, University College London, Gower Street, London, UK, 
$^{5}$Department of Physics, Duke University, Durham, NC 27708, USA, 
$^{6}$Research School of Astronomy and Astrophysics \& Centre for Gravitational Astrophysics, The Australian National University, Canberra, ACT, Australia, 
$^{7}$School of Physics and Astronomy, University of Southampton, Southampton, SO17 1BJ, UK, 
$^{8}$Department of Physics, University of Oxford, Denys Wilkinson Building, Keble Road, Oxford OX1 3RH, United Kingdom, 
$^{9}$Departments of Astronomy and Physics, Boston University, Boston, MA 02215, 
$^{10}$Department of Physics, University of California Berkeley, Berkeley, CA 94720, USA, 
$^{11}$E.O. Lawrence Berkeley National Laboratory, 1 Cyclotron Rd., Berkeley, CA, 94720, USA, 
$^{12}$Wits MIND Institute and School of Computer Science and Applied Mathematics, University of the Witwatersrand, Johannesburg, South Africa, 
$^{13}$School for Data Science and Computational Thinking, Stellenbosch University, South Africa, 
$^{14}$Kavli Institute for Particle Astrophysics \& Cosmology, P. O. Box 2450, Stanford University, Stanford, CA 94035, USA, 
$^{15}$SLAC National Accelerator Laboratory, Menlo Park, CA 94025, USA, 
$^{16}$Department of Physics, Stanford University, 382 Via Pueblo Mall, Stanford, CA 94305, USA, 
$^{17}$Fermi National Accelerator Laboratory, P. O. Box 500, Batavia, IL 60510, USA, 
$^{18}$Kavli Institute for Cosmological Physics, University of Chicago, Chicago, IL 60637, USA, 
$^{19}$Department of Astronomy and Astrophysics, University of Chicago, Chicago, IL 60637, USA, 
$^{20}$Institute of Space Sciences (ICE-CSIC), Campus UAB, Carrer de Can Magrans, s/n, E-08193 Barcelona, Spain, 
$^{21}$Institut d'Estudis Espacials de Catalunya (IEEC), 08860 Castelldefels (Barcelona), Spain, 
$^{22}$Centre for Astrophysics and Supercomputing, Swinburne University of Technology, John St, Hawthorn, VIC 3122, Australia, 
$^{23}$Department of Physics and Astronomy, University of Pennsylvania, Philadelphia, PA 19104, USA and 
$^{24}$Aix Marseille Univ, CNRS/IN2P3, CPPM, Marseille, France
}

\corresp{R.~Camilleri, Email: uqrcamil@uq.edu.au}

\citeauth{R.~Camilleri, J.~Lee, T.~M.~Davis, D.~Rubin, P.~Shah, D.~Scolnic, C.~Lidman, B.~Popovic, K.~Grech, M.~Vincenzi, D.~Brout, M.~Acevedo, P.~Armstrong, B.~A.~Bassett, K.~Bechtol, R.~C.~Chen, H.~T.~Diehl, J.~Frieman, L.~Galbany, A.~Möller, M.~Sako, B.~O.~Sánchez, M.~Sullivan, B.~E.~Tucker, Supernovae Unite: Combining Pantheon$+$ and DES-SN5YR. {\it Publications of the Astronomical Society of Australia} {\bf 00}, 1--12. https://doi.org/10.1017/pasa.xxxx.xx}

\history{(Received xx xx xxxx; revised xx xx xxxx; accepted xx xx xxxx)}

\defcitealias{Jones17}{J17}
\defcitealias{Vincenzi19}{V19}
\defcitealias{Vincenzi_2021_}{V21}
\defcitealias{Fitzpatrick99}{F99}
\defcitealias{Cardelli89}{CCM}


\newcommand\aap{A\&A}                
\let\astap=\aap                          
\newcommand\aapr{A\&ARv}             
\newcommand\aaps{A\&AS}              
\newcommand\actaa{Acta Astron.}      
\newcommand\afz{Afz}                 
\newcommand\aj{AJ}                   
\newcommand\ao{Appl. Opt.}           
\let\applopt=\ao                         
\newcommand\aplett{Astrophys.~Lett.} 
\newcommand\apj{ApJ}                 
\newcommand\apjl{ApJ}                
\let\apjlett=\apjl                       
\newcommand\apjs{ApJS}               
\let\apjsupp=\apjs                       
\newcommand\apss{Ap\&SS}             
\newcommand\araa{ARA\&A}             
\newcommand\arep{Astron. Rep.}       
\newcommand\aspc{ASP Conf. Ser.}     
\newcommand\azh{Azh}                 
\newcommand\baas{BAAS}               
\newcommand\bac{Bull. Astron. Inst. Czechoslovakia} 
\newcommand\bain{Bull. Astron. Inst. Netherlands} 
\newcommand\caa{Chinese Astron. Astrophys.} 
\newcommand\cjaa{Chinese J.~Astron. Astrophys.} 
\newcommand\fcp{Fundamentals Cosmic Phys.}  
\newcommand\gca{Geochimica Cosmochimica Acta}   
\newcommand\grl{Geophys. Res. Lett.} 
\newcommand\iaucirc{IAU~Circ.}       
\newcommand\icarus{Icarus}           
\newcommand\japa{J.~Astrophys. Astron.} 
\newcommand\jcap{J.~Cosmology Astropart. Phys.} 
\newcommand\jcp{J.~Chem.~Phys.}      
\newcommand\jgr{J.~Geophys.~Res.}    
\newcommand\jqsrt{J.~Quant. Spectrosc. Radiative Transfer} 
\newcommand\jrasc{J.~R.~Astron. Soc. Canada} 
\newcommand\memras{Mem.~RAS}         
\newcommand\memsai{Mem. Soc. Astron. Italiana} 
\newcommand\mnassa{MNASSA}           
\newcommand\mnras{MNRAS}             
\newcommand\na{New~Astron.}          
\newcommand\nar{New~Astron.~Rev.}    
\newcommand\nat{Nature}              
\newcommand\nphysa{Nuclear Phys.~A}  
\newcommand\pra{Phys. Rev.~A}        
\newcommand\prb{Phys. Rev.~B}        
\newcommand\prc{Phys. Rev.~C}        
\newcommand\prd{Phys. Rev.~D}        
\newcommand\pre{Phys. Rev.~E}        
\newcommand\prl{Phys. Rev.~Lett.}    
\newcommand\pasa{Publ. Astron. Soc. Australia}  
\newcommand\pasp{PASP}               
\newcommand\pasj{PASJ}               
\newcommand\physrep{Phys.~Rep.}      
\newcommand\physscr{Phys.~Scr.}      
\newcommand\planss{Planet. Space~Sci.} 
\newcommand\procspie{Proc.~SPIE}     
\newcommand\rmxaa{Rev. Mex. Astron. Astrofis.} 
\newcommand\qjras{QJRAS}             
\newcommand\sci{Science}             
\newcommand\skytel{Sky \& Telesc.}   
\newcommand\solphys{Sol.~Phys.}      
\newcommand\sovast{Soviet~Ast.}      
\newcommand\ssr{Space Sci. Rev.}     
\newcommand\zap{Z.~Astrophys.}       
\newcommand{\pplus}{Pantheon$+$}
\newcommand{\des}{DES-SN5YR}
\newcommand{\saltunite}{SALT3.UNITE} 
\newcommand{\dest}{DES-SN3YR}
\newcommand{\ppdes}{Pantheon$+$ and DES-SN5YR}
\newcommand{\sntot}{2884}
\newcommand{\kmsMpc}{km\,s$^{-1}$\,Mpc$^{-1}$}
\newcommand{\om}{\Omega_{\rm m}}
\newcommand{\oll}{\Omega_{\rm \Lambda}}
\newcommand{\ok}{\Omega_{\rm K}}
\newcommand{\mycite}[2]{(\citeauthor{#1} \citeyear{#1}, hereafter #2)}
\newcommand{\mycitee}[2]{(#2;~\citeauthor{#1} \citeyear{#1})}
\newcommand{\red}{\color{red}}



\def\mflcdmS{0.310^{+0.012}_{-0.011}} 
\def\mflcdmSB{0.302\pm0.007} 
\def\mflcdmSC{0.314\pm0.005} 
\def\mflcdmSBC{0.303\pm0.003} 

\def\mlcdmS{0.212\pm0.044} 
\def\mlcdmSB{0.288\pm 0.011} 
\def\mlcdmSC{0.321\pm0.010} 
\def\mlcdmSBC{0.304\pm0.003} 

\def\klcdmS{0.26\pm0.11} 
\def\klcdmSB{0.054^{+0.031}_{-0.033}} 
\def\klcdmSC{-0.003\pm0.003} 
\def\klcdmSBC{0.003\pm0.001} 

\def\mfwcdmS{0.197^{+0.056}_{-0.054}} 
\def\mfwcdmSB{0.293\pm0.008} 
\def\mfwcdmSC{0.313\pm0.006} 
\def\mfwcdmSBC{\textcolor{lightgray}{0.303\pm 0.004}} 

\def\wfwcdmS{-0.764^{+0.078}_{-0.096}} 
\def\wfwcdmSB{-0.939\pm0.031} 
\def\wfwcdmSC{-1.002\pm0.019} 
\def\wfwcdmSBC{\textcolor{lightgray}{-1.002\pm0.017}} 

\def\fwasig{3.3}
\def\mfwacdmS{0.297^{+0.086}_{-0.141}} 
\def\mfwacdmSB{0.307^{+0.012}_{-0.014}} 
\def\mfwacdmSC{0.298\pm0.007} 
\def\mfwacdmSBC{0.305\pm0.004} 

\def\wfwacdmS{-0.779^{+0.099}_{-0.098}}  
\def\wfwacdmSB{-0.890^{+0.057}_{-0.051}} 
\def\wfwacdmSC{-0.796^{+0.073}_{-0.075}} 
\def\wfwacdmSBC{-0.861^{+0.044}_{-0.042}} 

\def\afwacdmS{-0.86^{+1.01}_{-1.78}} 
\def\afwacdmSB{-0.47^{+0.41}_{-0.40}} 
\def\afwacdmSC{-1.00^{+0.37}_{-0.36}} 
\def\afwacdmSBC{-0.60^{+0.17}_{-0.19}} 

\begin{abstract}
We present the Hubble diagram and cosmological constraints resulting from the combination of the \ppdes~supernova samples (including the DES-Dovekie updates), which we refer to as the `Unite' sample. Unite updates Pantheon+, including methodology improvements that the DES sample enabled, and adds the new supernovae from the DES survey. This represents the most comprehensive and internally consistent type Ia supernovae (SNe~Ia) dataset currently available, consisting of \sntot~likely SNe~Ia. This work features consistent SN Ia modelling, sample selection, and bias corrections. We also redetermine host-galaxy stellar masses for over 98\% of the sample using a consistent framework, as the host stellar mass is known to correlate with SN Ia luminosity after light curve width and colour corrections. Using Unite alone, for a spatially flat universe with a cosmological constant (Flat-$\Lambda$CDM), we find $\om = \mflcdmS$.  We further present constraints using the Unite sample combined with Cosmic Microwave Background (CMB) and both DES and DESI Baryon Acoustic Oscillation data. We find strong evidence of tension between CMB measurements and the combined BAO and supernova datasets when fitting a model with constant dark energy equation of state ($w$), which is alleviated when allowing $w$ to vary with time.

For a flat universe with dark energy that has a time-evolving equation of state parametrised by $w_0$ and $w_a$ we find $(\om, w_0, w_a) = (\mfwacdmSBC, \wfwacdmSBC, \afwacdmSBC)$ with a dark energy Figure of Merit of 315. This corresponds to a $\sim$$30\%$ reduction in the area of the $w_0-w_a$ confidence region relative to previous constraints from the combination of DES Y6, DESI-DR2, and CMB data \citep{DESCOMBINED}. The Bayesian evidence indicates only weak preference for time evolving dark energy, whereas frequentist analyses yield a preference over Flat-$\Lambda$CDM at a significance of $\fwasig\sigma$ ($3.1\sigma$) when using the maximum \textit{a posteriori} probability (maximum likelihood).
\end{abstract}


\maketitle

\section{Introduction}\label{sec:intro}
Type Ia supernovae (SNe~Ia) have been instrumental in shaping our understanding of cosmic expansion.  Early observations led to the discovery of the accelerating universe, which implies the existence of a dark energy component driving this acceleration \citep{Riess_1998, Perlmutter_1999}. Over the past few decades, increasingly large and more precise SN Ia samples have made significant improvements in both understanding systematics and cosmological parameter constraints. Three of the most comprehensive supernova datasets to date are \pplus~\citep{scolnic2021pantheon, 2022_pantheon_analysis}, Union3.1 \citep[][]{Rubin_2025, Rubin_2026, Hoyt2026} and the Dark Energy Survey 5-Year Supernova sample \citep[DES-SN5YR;][]{vincenzi24, DES-SN5YR}, which was recently reanalysed in \citet[][hereafter DES-Dovekie]{desdov}. \pplus~and \des~include a significant overlap in their analysis methodology but differ in the SN samples used.  \pplus~combines data from multiple low- and high-redshift surveys, while \des~provides a homogeneous, high-redshift sample with well-controlled systematics. Contrastingly, \pplus~and Union3.1 have significant overlap in the SN samples that make up their respective compilations; however, they differ in their methodology, offering a perspective on analysis choices and systematic uncertainties. 

The constraining power on cosmological models by SNe~Ia are significantly enhanced by combining constraints from multiple probes \citep{DESCOMBINED}. Cosmological results from the measurement of Baryon Acoustic Oscillations (BAO) were presented using the first data release (DR1) from the Dark Energy Spectroscopic Instrument (DESI) collaboration \citep{desicollaboration2024desi}. Within that analysis, the DESI-DR1 measurements were combined with Planck \citep{2020_planck} Cosmic Microwave Background (CMB) observations along with each of the \pplus, Union3 and \des~datasets individually. Notably, the combined BAO, CMB and SN constraints show a preference for time-varying dark energy over the cosmological constant, independent of which SN sample was used. However, the strength of the evidence varies depending on the SN sample included, with preferences for time-varying dark energy over the cosmological constant at the $\sim$$2.5\sigma$, $\sim$$3.5\sigma$ and $\sim$$3.9\sigma$ significance levels for \pplus, Union3 and \des, respectively. When combined with the second data release (DR2) from DESI \citep{desidr2}, these preferences further increase to $\sim$$2.8\sigma$, $\sim$$3.8\sigma$ and $\sim$$4.2\sigma$ for \pplus, Union3 and \des, respectively, with all significances based on the maximum \textit{a posteriori} (MAP) probability.

More recently, \citet{desdov} performed a re-analysis of \des, improving the photometric cross-calibration and replacing a numerical approximation of extinction due to dust in the host galaxy with the exact relation. \citet{desdov} find, using the
maximum likelihood, that the preference for time-varying dark energy reduces to $\sim$$3.2 \sigma$ (with a weak Bayesian model preference), when combined with CMB data from Planck, the Atacama Cosmology Telescope (ACT) and the South Pole Telescope (SPT) as well as DESI-DR2 BAO measurements. This $\sim$$1\sigma$ shift arises even though the supernova only constraints on time-varying dark energy shift by less than $0.2 \sigma$. This occurs because of the different degeneracy directions of the various probes, and demonstrates the sensitivity of the result to having complementary data sets. 

Host-galaxy stellar mass is typically used in the standardisation of SN Ia luminosities to account for observed correlations with host properties. Work by \citet{Hoyt2026} redetermined the host masses for both the Union3 and \pplus~compilations and find that the host masses of $z < 0.10$ SNe in Union3 were, on average, overestimated relative to the rest of the sample, while the opposite was true for $z < 0.15$ SNe in \pplus. Applying the new self consistent host masses to Union3.1 and \pplus, \citet[][using MAP statistics]{Hoyt2026} finds that the preference for time-varying dark energy increases from $2.8\sigma$ to $3.2\sigma$ for \pplus, and decreases from $3.8\sigma$ to $3.4\sigma$ for Union3.1.


The preference for dynamical dark energy, along with the apparent SN dependence on the strength of that evidence, has prompted scrutiny, in particular with regards to the \ppdes~analyses \citep{Notari24, efstathiou25, huang25}. A common point that arises is an offset in the difference between the magnitudes of low-$z$ and high-$z$ SNe when comparing common events between the two samples. However, as shown in \cite{vincenzi25} these differences are well justified and largely due to (1) well-motivated updates to the modelling of SN intrinsic scatter and host galaxy properties, and (2) object-to-object comparisons that failed to account for the bias correction framework, which naturally shifts individual SN distance measurements to account for the supernovae we do not see (selection effects). In particular, these critiques did not account for DES-SN5YR's higher sample completeness, which meant it required smaller bias corrections than the DES 3-year sample that appeared in Pantheon+.

\begin{figure*}
    \centering \includegraphics[width=\linewidth]{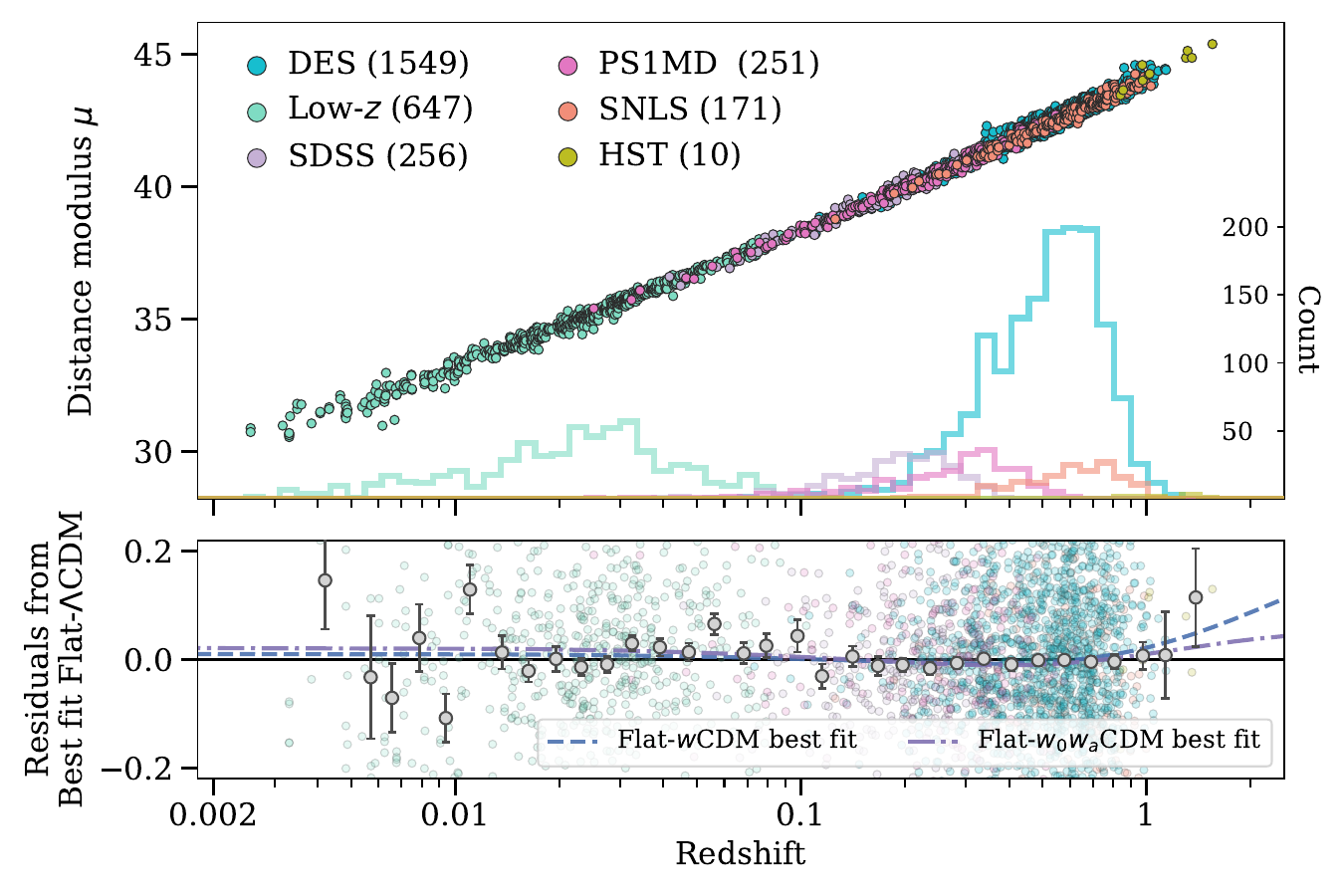}
    \caption{Hubble diagram of the Unite sample containing \sntot~likely SN Ia. The number of SNe as a function of redshift is overlaid on the upper panel. The difference between the data (single events and redshift-binned SN) and best fit Flat-$\Lambda$CDM model from Unite alone is shown in the lower panel. For comparison, we also overplot the best fit Flat-$w$CDM and Flat-$w_0 w_{\rm a}$CDM models from Unite alone. Note that, in this analysis, we apply a cut on the SN Ia probability of $P_{\rm Ia}>0.8$ to the DES-SN (see Section~\ref{sec:samplecuts}).}
    \label{fig:hubble_diagram}
\end{figure*}

Here, we unify the \ppdes~samples and present Unite,\footnote{When using the Unite Hubble diagram, please cite this work together with the major source analyses from which the sample is derived, including \citet{DES-SN5YR}, \citet{desdov}, and \citet{2022_pantheon_analysis}.} the largest Hubble diagram assembled to date consisting of \sntot~likely SNe~Ia (see Figure~\ref{fig:hubble_diagram}). In this analysis, we use consistent SN Ia modelling, sample selection, and bias corrections. We improve the \pplus~data with all the methodology improvements that the DES sample enabled. We further redetermine the host stellar masses for over 98\% of the sample using a consistent framework \citet{lee25mass} and cross-calibrate our surveys using the open-source code\footnote{\url{https://github.com/bap37/Dovekie/}} released for the DES-Dovekie analysis in \cite{p25a}. As such, we advocate its adoption for future cosmological analyses.

With the Unite Hubble diagram, we then derive cosmological constraints on the standard cosmological model and some basic extensions.
The Unite Hubble diagram can be used in both anchored and unanchored forms,\footnote{Anchoring refers to calibrating the absolute magnitude of the supernovae.} with both implementations provided in the public likelihood code. In this work, we present cosmological constraints from the unanchored Hubble diagram, which does not constrain the Hubble constant, $H_0$. In a companion paper, \citet[][in prep.]{lee25h0}, uses the anchored Hubble diagram, which incorporates Cepheid-based distances to SN host galaxies to infer the sound horizon scale and $H_0$.

This paper is structured as follows. We begin in Section~\ref{sec:Data} by introducing both the \ppdes~SN samples, along with the external probes that we combine with Unite when constraining cosmological models. In Section~\ref{sec:cosmology}, we summarise the models we test and describe our cosmology inference procedure. To produce the Unite Hubble diagram, we mostly follow the \des~analysis detailed in \cite{vincenzi24} and the subsequent re-analysis in \cite{desdov} with some minor changes, as detailed in Section~\ref{sec:analysis}. In Section~\ref{sec:hubble_diagram} we present the Unite Hubble diagram, while our systematic error budget is discussed in Section~\ref{sec:systbudget}. We present our cosmological results in Section~\ref{sec:results}, assess model consistency and tension in Section~\ref{sec:modelcomp_tensions}, and compare the Unite Hubble diagram to its constituent counterparts in Section~\ref{sec:comparisons}. We conclude in Section~\ref{sec:conclusion}.

\section{Data}\label{sec:Data}
\subsection{\des}
The \des~sample is the largest and deepest single sample survey to date, consisting of 1635 SN Ia-like candidates (DES-SN) ranging in redshift from $0.05$ to $1.13$ with 1499 photometrically classified as SNe~Ia (with $>0.5$ probability) using the open source algorithm \texttt{SuperNNova} \citep{2020MNRAS.491.4277M, maria_2021, 10.1093/mnras/stac1691}. The majority of host-galaxy redshifts were obtained through the OzDES follow-up program using the AAOmega spectrograph \citep{Smith04,Lidman20}.

The DES sample is complemented by 194 spectroscopically confirmed low-$z$ SN Ia from CfA3 \citep{2009ApJ...700..331H}, CfA4 \citep{2012ApJS..200...12H}, CSP \citep{2017AJ....154..211K} (DR3) and the Foundation SN sample \citep{2018_Foley_foundation}. These external surveys span a redshift range of $0.025 <z< 0.1$. In the \des~analysis, only low-$z$ SNe above redshift $0.025$ were included in the \des~analysis to mitigate the effects of peculiar velocities. For more details, see \citet{vincenzi24} and \citet{DESDATA24}.

\subsection{\texorpdfstring{\pplus}{Pantheon+}}
The \pplus~sample consists of 1701 light-curves of 1550 unique, spectroscopically confirmed SNe~Ia ranging in redshift from $z = 0.001$ to $2.26$. The SN photometry used within \pplus~are from seventeen different sources, namely \dest~\citep[a small spectroscopically classified DES sample from the first 3 years of observations;][]{Brout_2019, Smith_2020}, Foundation \citep{2018_Foley_foundation}, PS1 \citep{Scolnic_2018}, SNLS \citep{2006A&A...447...31A, 2011ApJ...737..102S, 2014JLA}, SDSS \citep{Kessler2009SDSS,Sako_2011}, HST \citep{Gilliland_1999, Riess_2001, Riess_2004, Riess_2007, Suzuki_2012, Riess_2018}, LOSS1 \citep{Ganeshalingam_2010}, LOSS2 \citep{stahl_2019}, SOUSA \citep{Brown2014}, CNIa0.02 \citep{Chen_2022}, CSP \citep{2017AJ....154..211K}, CfA1 \citep{Riess_1999}, CfA2 \citep{Jha_2006}, CfA3 \citep{2009ApJ...700..331H}, CfA4 \citep{2012ApJS..200...12H}, and other low-$z$ samples \citep{Zhang_2010, Milne_2010, Stritzinger_2010, tsvetkov2010sn2008fvtypeia, 2017AJ....154..211K, Gall_2018, Burns_2018, Burns_2020, Kawabata_2020}. For more details, see \cite{scolnic2021pantheon} and \cite{2022_pantheon_analysis}.

\subsection{Combining \texorpdfstring{\ppdes}{Pantheon+ and DES}}
In this section we describe how SNe in common to \ppdes~have been handled in the context of this analysis.

All of the low-$z$ samples in \des\ are also included within the \pplus~compilation; however, \des\ applied a $z>0.025$ redshift cut. In the Unite Hubble diagram, we do not include this cut. This decision was driven by our goal to present a Hubble diagram combining both \ppdes\ that is able to be anchored to low-$z$ calibrators to derive $H_0$.

The \dest~sample (consisting of 203 SNe, included in \pplus) is also contained within the \des~sample with 145 common DES SN events. In our combined Hubble diagram, we only include the 145 common SN events used in \des~due to different sample selection functions between \dest~and \des~\citep[see][for more details]{Vincenzi_2021_}. More specifically, spectroscopically followed-up DES SNe were selected in \dest, regardless of host spectroscopic redshift information; however, \des~(and this analysis) only consider DES SNe with host galaxy spectroscopic redshifts. 

\subsection{External probes}
In this section, we detail the complementary external probes that we combine with the Unite sample to obtain more precise constraints, all of which are publicly available and have been implemented in \texttt{CosmoSIS} \citep{ZUNTZ201545}.

\subsubsection{BAO}
Following six years of observations, \citet{desbao24} released the final dataset from the BAO analysis. The galaxy sample was optimised for BAO science in the redshift range $0.6 < z < 1.2$, with an effective redshift at $z_{\rm eff}= 0.85$, and constrained the ratio of the angular distance to the scale of the sound horizon to $D_M(0.85)/r_d = 19.51\pm0.41$. Furthermore, the second data release (DR2) of DESI BAO measurements \citep[DESI-BAO;][]{desidr2} provide twelve measurements in seven redshift bins in the redshift range $0.1 < z < 4.2$ and improves on the statistical precision from DESI DR1 by 30-50\%. 

The DES- and DESI-BAO measurements share a small overlap in the area and redshift range that could give rise to a covariance among these datasets. Although the covariance is expected to be small, recent work by \citet{baocombo} analysed the DES-BAO in the area that does not overlap with DESI and finds $D_M(0.85)/r_d = 19.74\pm0.60$. We use this measurement and provide constraints simultaneously using both DES-BAO and DESI-BAO, which we refer to as `BAO'.

\subsubsection{CMB}
In this work, we use CMB measurements from Planck \citep{2020_planck}, the Atacama Cosmology Telescope (ACT) and the South
Pole Telescope (SPT) and include lensing reconstructions.

For measurements of temperature and polarisation power spectra (TTTEEE) we use the following: For Planck, we use the Python implementation \texttt{Planck-py}\footnote{\url{https://github.com/heatherprince/planck-lite-py}} \citep{PrinceDunkley19} of both the \texttt{Plik-lite} likelihood using $\ell < 1000, 600, 600$ for TT, and TE, EE respectively and the combined \texttt{simall} and \texttt{Commander} likelihoods for $\ell < 30$; the \texttt{ACT-DR6-Lite}\footnote{\url{https://github.com/ACTCollaboration/DR6-ACT-lite}} likelihood \citep[with $\ell> 1000, 600, 600$ for TT, and TE, EE respectively;][]{ActDr6maps, ActDr6like} for ACT; and SPT-3G data \citep{Camphuis2025} as wrapped in the \texttt{candl} likelihood \citep{Balkenhol2024}.\footnote{\url{https://github.com/Lbalkenhol/candl}}

For lensing reconstructions, we use data based on the combined ACT and Planck maps \citep{Madhavacheril2024, Qu2024, Carron2022} incorporated into the \texttt{ACT-DR6-Lens} likelihood,\footnote{\url{https://github.com/ACTCollaboration/act_dr6_lenslike}} specifically the \texttt{actplanck-baseline} option. This is combined with SPT-3G lensing \citep{Pan2023} that is also implemented in the \texttt{candl} likelihood. 

The priors on the CMB normalisation nuisance parameters for the likelihoods discussed above are given in Table~\ref{tab:priors}.
\begin{table*}
    \centering
    \caption{Cosmological model variations tested in this paper, their Friedmann Equations, and the free parameters in the fit.}
    \renewcommand{\arraystretch}{1.5}
    \begin{tabular}{lll}
    \hline\hline
        Cosmological Model & Friedmann Equation: $E(z,\Theta)\equiv H(z, \Theta)/H_0=$ & Fit Parameters, $\Theta$ \\ \hline
        Flat-$\Lambda$CDM & $\left[\om (1+z)^3 + (1-\om)\right]^{1/2}$ & $\om$ \\
        $\Lambda$CDM & $\left[\om (1+z)^3 + \oll + (1-\om-\oll)(1+z)^{2}\right]^{1/2}$ & $\om, \oll$ \\
        Flat-$w$CDM & $\left[\om (1+z)^3 + (1-\om)(1+z)^{3(1+w)}\right]^{1/2}$ & $\om, w$ \\
        Flat-$w_0$$w_a$CDM & $\left[\om (1+z)^3 + (1-\om) (1+z)^{3(1+w_0+w_a)}e^{-3w_a z/(1+z)}\right]^{1/2}$ & $\om, w_0, w_a$ \\[1mm]
        \hline
    \end{tabular}
    \label{tab:models}
\end{table*}

\section{Cosmology}\label{sec:cosmology}
In this work, we use the Unite sample to constrain the standard cosmological model, a spatially flat universe with cold dark matter and a cosmological constant (Flat-$\Lambda$CDM), and three basic extensions. The first relaxes the assumption of spatial flatness ($\Lambda$CDM). The final two models (Flat-$w$CDM \& Flat-$w_0 w_a$CDM) retain spatial flatness while allowing for a constant ($w$) or linear parameterisation \citep[where $w=w_0 + w_a(1-a)$;][]{chevallier01, Linder_2003} of the dark energy equation of state parameter. Here, we describe the relevant theory used to constrain the cosmological parameters of these models. 

\subsection{Theory}\label{sec:theory}
We constrain cosmological parameters by finding the minimum of the $\chi^2$ likelihood given by,
\begin{equation}\label{eq:chi2_sn}
-2\mathrm{ln}(\mathcal{L}) = \chi^{2}_{\mathrm{SN}}( \Theta)=\vec{D}^{T}\mathcal{C}^{-1}_{\mathrm{SN}} \vec{D},
\end{equation}
where $\vec{D}$ is the difference in the distance moduli ($\mu_i$) between data and theory for every $i$th SN and is therefore computed as,
\begin{equation}
    D_i(\Theta) = \mu_{{\rm obs},i} - \mu_{\mathrm{theory}}(z, \Theta)
\end{equation}
where $z$ is the cosmological redshift, $\Theta$ is the set of cosmological parameters and $\mathcal{C}^{-1}_{\mathrm{SN}}$ is the inverse covariance matrix including both statistical and systematic errors. The theoretical distance moduli, $\mu_{\mathrm{theory}}$ are calculated as,
\begin{equation}
    \mu_{\mathrm{theory}}(z, \Theta) = 5 \log_{10}[D_L(z, \Theta)/1~\mathrm{Mpc}] + 25,
\end{equation}
where $D_L$ is the luminosity distance defined as, 
\begin{equation}\label{eq:dl}
D_L(z_{\rm obs}, z, \Theta)=(1+z_{\rm obs})R_0S_k\left(\chi(z,\Theta)\right).
\end{equation}
%
$z_{\rm obs}$ is the observed heliocentric redshift, $R_0 = c/(H_0  \sqrt{|\ok|})$ is the present day scale factor with dimensions of distance and the comoving distance is given by
\begin{equation}
    R_0\chi(z,\Theta) = \frac{c}{H_0}\int_0^{z} \frac{d z'}{E(z', \Theta)},
\end{equation}
where $E(z, \Theta)$ describes the expansion history for each model. Finally, the curvature is captured by $S_k(\chi)=\sin\chi$, $\chi$, and $\sinh\chi$ for closed ($\ok<0$), flat ($\ok=0$), and open ($\ok>0$) universes respectively. Both $E(z, \Theta)$ and $\Theta$ for each model constrained in this work are given in Table~\ref{tab:models}.

We note that the cosmological constraints presented in this work use the unanchored Hubble diagram and therefore the SN alone offer no information regarding $H_0$ or the SN Ia peak absolute magnitude due to the degeneracy between these two parameters. Therefore, we follow the treatment outlined in Appendix A.1 of \citet{Goliath_2001} and combine them into a single parameter that we analytically marginalise over. In a companion paper, \citet[][in prep.]{lee25h0} anchors the Unite Hubble diagram to low-$z$ calibrators and measures $H_0$ and the sound horizon scale.

\section{Analysis}\label{sec:analysis}

In this work, we follow the \des~analysis detailed in \cite{vincenzi24} and \cite{DES-SN5YR} along with updates from the DES-Dovekie re-analysis \citep{desdov} unless explicitly stated. As part of the \des~analysis, several improvements were implemented\footnote{Further details on the improvements made in the \des~analysis and how they impacted cosmology is examined in \cite{vincenzi25}.} along with stricter sample cuts which we apply to the remaining \pplus~samples. Additionally, we adopt two corrections implemented in \cite{desdov}. In this section, we detail all relevant analysis changes, which are summarised in Table~\ref{tab:analysis_changes}. 

\begin{table*}
\centering
\caption{Summary of the analysis changes between \pplus, \des, DES-Dovekie, and Unite.}
\label{tab:analysis_changes}
\renewcommand{\arraystretch}{1.5}
\begin{tabular}{lcccc}
\hline \hline 
  & \textbf{Pantheon+}& \textbf{DES-SN5YR} & \textbf{DES-Dovekie} & \textbf{Unite} \\ \hline
\textbf{Section~\ref{sec:calibration}: Calibration} & Fragilistic & Fragilistic & Dovekie & Dovekie-like \\
\textbf{Section~\ref{sec:lcfit}: SALT model} & \texttt{SALT2.B22}& \texttt{SALT3.DES5YR} & \texttt{SALT3.DOVEKIE} & \texttt{SALT3.UNITE} \\
\textbf{Section~\ref{sec:intrinsic_scatter}: Intrinsic scatter model} & BS21 & P23 & P23 & P23 \\
\textbf{Section~\ref{sec:hostmass}: Host galaxy masses} & Inconsistent & Consistent & Consistent & Consistent \\
\textbf{Section~\ref{sec:bbc}: \texttt{SNANA} version} & \texttt{v11\_03e} & \texttt{v11\_05c} & \texttt{v11\_06b} & \texttt{v12\_01c} \\

\textbf{Section~\ref{sec:samplecuts}: Quality cuts} & \multicolumn{4}{c}{A summary of the quality cut changes are given in Table~\ref{tab:cuts}}  \\
\textbf{Section~\ref{sec:samplecuts}: $P_{\rm Ia}$ cut} & - & No & No & Yes \\
\textbf{Section~\ref{sec:lens}: Lensing pdfs for sims} & Underestimated & Underestimated & Underestimated & Updated \\
\textbf{Section~\ref{sec:lens}: $\Delta m_{\mathrm{lens}}(\vec{r}, z)$ correction} & No & No  & No & Yes \\
\textbf{Number of SNe in the cosmology sample} & 1701 & 1829 & 1820 & \sntot \\ \hline
\end{tabular}

\end{table*}

\subsection{Survey cross-calibration}\label{sec:calibration}
One of the largest systematic uncertainties in measuring cosmological parameters with SNe~Ia from multiple surveys is the calibration of the differing telescopes and filters used to make observations. Both the \ppdes~analyses used the `Supercal-Fragilistic' photometric cross-calibration values presented in \cite{Brout_2022scal}. Recently, \cite{p25a} released `Dovekie,' an improved cross-calibration using new data from HST \citep{narayan2019,axelrod2023,Boyd25} and Gaia \citep{GAIAONE,GAIATWO,GAIATHREE}. The software is open-source,\footnote{\url{https://github.com/bap37/Dovekie/}} allowing additional surveys to be easily added or removed.

\begin{figure}
    \centering \includegraphics[width=\linewidth]{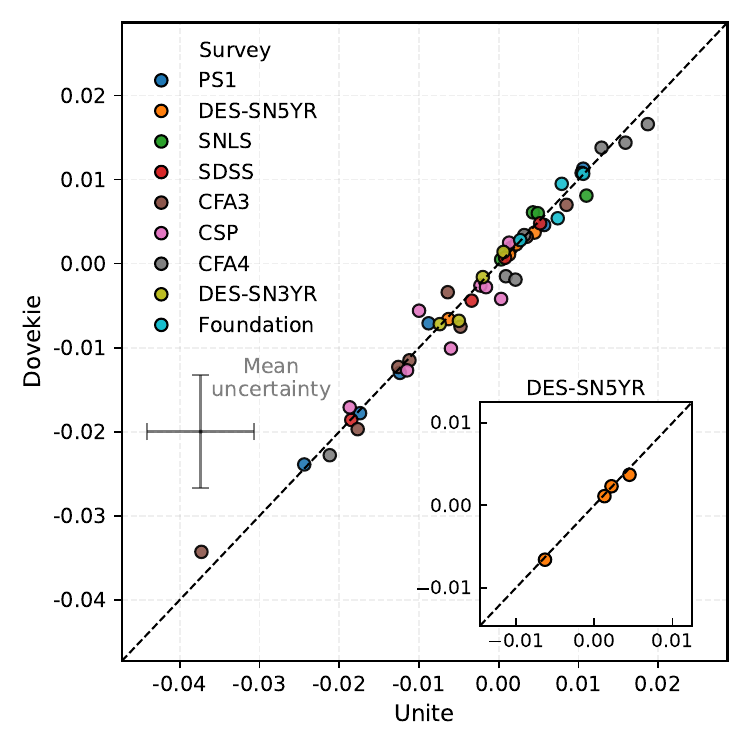}
    \caption{A comparison of the Unite and Dovekie derived magnitude zero-point offsets for common surveys, showing the offsets applied to the survey-specific photometric bands. The mean uncertainty in the Unite estimated zero point offsets is shown for reference.}
    \label{fig:comparezps}
\end{figure}

Dovekie's fiducial release did not include all of the photometric systems included in Supercal-Fragilistic, such as LOSS1, LOSS2 or SOUSA\footnote{We do not include CfA1, CfA2 and various $\mathcal{O}(1)$ SN light-curves in our cross-calibration following the same reasoning outlined in sec. 2.6 of \cite{Brout_2022scal}. We likewise do not include HST observations in the cross-calibration, as the HST CALSPEC standards define the absolute flux calibration.} which we require for Unite. We therefore take advantage of the public Dovekie code (which allows us to include the additional surveys we require), and perform a cross-calibration with the addition of these missing surveys. 
When comparing common surveys, we observe a high degree of consistency between Dovekie and Unite zero-points offsets that are applied to the survey-specific photometric bands; see Figure~\ref{fig:comparezps}. The derived zero-point offsets for all surveys are provided in the public data release.\footnote{The data will be publicly available upon acceptance of the paper.}

\subsection{Light-curve fitting}\label{sec:lcfit}
To convert sparse light-curve observations to the SN standardisation parameters, stretch $x_1$, colour $c$, and an amplitude term $m_x$, we use the SALT model framework \citep{guy05}. These models require a training sample of SNe~Ia and determine three components. First, $M_0$, which is a spectral time-series describing the mean energy distribution (SED) of a SN Ia (for $x_1=0$ and $c=0$). Second, a spectral time-series describing the first-order deviation around $M_0$, which is correlated with light-curve width. Third, $CL$, a wavelength-dependent colour law capturing the phase-independent colour variation from combined intrinsic SN colour and host-galaxy dust extinction.

\pplus~trained the \texttt{SALT2} model \citep{guy_2007, guy_2010} which was released in \citet[][hereafter \texttt{SALT2.B22}]{Brout_2022scal}. \texttt{SALT2.B22} was trained on a compilation of 1083 SNe with 1207 spectra from \citet[][hereafter K21]{Kenworthy_2021}, and used the `Supercal-Fragilistic' photometric cross-calibration.

In \des, the model was upgraded to \texttt{SALT3} \citep{Kenworthy_2021}, which includes an extended central passband wavelength range ($2800-7000$\AA~for \texttt{SALT2} vs $2800-8000$\AA~for \texttt{SALT3}) and improved uncertainty estimation as a function of phase and colour. For more details on the differences between \texttt{SALT2} and \texttt{SALT3}, see \cite{2023MNRAS.tmp..345T}. 
The \des~analysis trained their own \texttt{SALT3} model presented in \citet[][hereafter \texttt{SALT3.DES5YR}]{vincenzi24} using the same K21 training set (with 1083 SNe with 1207 spectra) and Supercal-Fragilistic cross-calibration as in \texttt{SALT2.B22} however, the observer-frame U-band data was removed from the training set due to the challenging nature of calibrating the near UV ground-based data. 

For the Unite analysis, we follow the same approach as \des: we use the K21 training set \citep[for more details on the training set and its constituent surveys, see section~4 of][]{Kenworthy_2021} and exclude observer-frame U-band data. Note this differs from the DES-Dovekie \citep{desdov} analysis, which uses a smaller training set. However, we adopt the improved photometric cross-calibration (discussed in Section~\ref{sec:calibration}) and update the magnitude zero-point offset applied to the training set, necessitating a retraining of the model. Therefore, in this analysis, we train our own \texttt{SALT3} model, \texttt{\saltunite}. 

\subsection{SN Ia Intrinsic scatter model}\label{sec:intrinsic_scatter}
\cite{Brout_2021dust} showed that the SNe host galaxy mass correlates with the intrinsic scatter of SN Ia distance modulus residuals after standardisation. With this understanding, they presented a model of the intrinsic scatter with 12 free parameters that characterise intrinsic colour variations, as well as dust properties and their dependency on the host galaxy mass.

Both \ppdes~use the same dust model framework presented in \cite{Brout_2021dust} to describe the intrinsic scatter, although they differ in the value of the dust parameters and how they were determined. \pplus~used dust parameters that were originally presented by \cite{Brout_2021dust} and found by tuning, such that \texttt{SNANA} simulations reproduced trends in observed data. This iteration is commonly referred to as BS21. In \des~these parameters were instead measured using a forward-modelling approach presented in \cite{P21_dust2dust} and this iteration we refer to as P23.\footnote{In the \des~analysis this model was referred to as P21.}

We do not re-fit dust parameters in this work. For our nominal analysis, we use the P23 intrinsic scatter model used in both the \des~anlaysis \citep{vincenzi24} and DES-Dovekie \citep{desdov}. For systematic variations, we follow the previous analysis by using the same three sets of dust parameters previously drawn from the MCMC chains and retaining the BS21 intrinsic scatter model. Improvements to the dust model and the Dust2dust method are ongoing in separate work \citep[][in prep.]{Popovic26, Carreres26}, and in lieu of re-fitting the dust parameters we perform additional validation tests of the best fit P23 model in Appendix~\ref{appendix2}.

\subsection{Host galaxy stellar masses}\label{sec:hostmass}

SNe~Ia in higher stellar mass galaxies tend to be brighter than those in lower mass galaxies by about $0.05$ to $0.10$ mag after standard corrections for SN colour and light curve width have been applied \citep{kelly2010_mass-correlation,sullivan2010_mass-step,2010ApJ...722..566L}. Therefore, it has become customary to account for these differences using a host-galaxy stellar `mass step' \citep{2022_pantheon_analysis,Rubin_2025,DES-SN5YR}.  Typically, the mass step is assumed to be located around~$10^{10} M_{\odot}$. 
For DES-SN5YR, \citet{Wiseman20} updated the host-galaxy stellar masses by using deeper coadd photometry with an updated galaxy spectral energy distribution (SED) fitting method based on the \verb|PÉGASE.2| spectral synthesis templates \citep{PEGASE1997,le_borgne2002_ZPEG}. To ensure consistency with the high-$z$ sample, the host masses for the low-$z$ external samples (Foundation, CfA3, CfA4 and CSP) were also recomputed. 

\citet{vincenzi25} found that different host-galaxy mass measurements could lead to $0.01$ mag shifts between the low-$z$ and high-$z$ distance moduli and hence impact cosmological constraints (especially for beyond $\Lambda$CDM). Therefore, we re-derive nearly all (> 98\%) of the host-galaxy masses for both the DES-SN5YR and the Pantheon+ SNe~Ia subsamples in our Hubble diagram starting from the photometry. 

We perform aperture photometry using the \verb|HostPhot|\footnote{\url{https://github.com/temuller/hostphot}} \citep{muller2022hostphot} package using coadded images from DES  \citep{abbott2021DES_DR2,Qu_hostMismatch}, LegacySurvey Data Release 10 \citep{DESI_LegacySurvey2019}, or PanSTARRS \citep{magnier2020PS} for the optical wavelengths, and from the \textit{Galaxy Evolution Explorer} \citep[\textit{GALEX},][]{martin2005GALEX} for the near-ultraviolet wavelengths. Although it is possible to include the near-infrared photometry from the \textit{2 Micron All Sky Survey} \citep[\textit{2MASS},][]{skrutskie2006_2MASS} \citep[as done in][]{Wiseman20}, we choose to not include them in our default photometry as its addition typically degrades the quality of the SED fit, likely due to the $0.8$ mag scatter we see in 2MASS photometry. We note that using the same near-ultraviolet and optical bands for SED fitting across the full sample results in a different rest-frame wavelength coverage depending on the redshift. This however, only causes small shifts in cosmological parameters well below the current level of uncertainties. See Section 2.1 of \citet{lee25mass} for more details.

We visually inspect each of the multi-band coadded host-galaxy images and apply masking to foreground stars to ensure that the aperture is appropriately chosen with minimal contamination and use the same photometry pipeline throughout all redshifts for consistency. We find that our new photometry is consistent with the DES-SN5YR host-galaxy photometry for the overlapping sample, at the 0.01 mag level. We then use the SED fitting code CIGALE \citep{boquien2019cigale} assuming a Kroupa initial mass function (IMF) \citep{kroupa2001} and the \verb|m2005| \citep{maraston2005} library of single stellar populations (SSPs) to obtain host-galaxy properties including the host stellar mass. Additionally, we compare each of the multi-band host-galaxy fluxes from the CIGALE SED fit with the input flux from \verb|HostPhot| to identify \textit{individual} bands with issues. For $ < 3$\% of the host-galaxies, the image cutouts for one or two specific band(s) have issues such as bad pixels, in which case we remove the specific band(s) and rerun the SED fitting.

We find that $\langle \Delta \log M_{\rm stellar} \rangle = \langle\log M_{\rm stellar}^{\mathrm{Previous}} - \log M_{\rm stellar}^{\mathrm{Unite}}\rangle$ between the DES-SN5YR subsample of our re-derived masses (denoted as `Unite') and the DES-SN5YR data-release (denoted as `Previous') is  $0.160 \pm 0.009$ (rms: 0.114) at $z < 0.1$ while being more consistent with DES-SN5YR at $z \ge 0.1$ with $ \langle \Delta \log M_{\rm stellar} \rangle = 0.017 \pm 0.004$ (rms: 0.162) with a $5\sigma$ clip based on the median absolute deviation of $\Delta \log M_{\rm stellar}$ to prevent outliers from dominating the quoted statistics ($<2$\% when all redshifts are included). The difference between the Unite re-derived masses and DES-SN5YR masses can be attributed to the use of a different SSP, as well as including a systematic error budget in the SED fitting for the Unite measurements. Updating the Unite-DES-SN5YR subsample masses from DES-SN5YR data-release masses to Unite masses does not result in a shift in cosmology. 

When we compare the Unite-Pantheon+ subsample and Pantheon+ masses we see there is an offset in the mean stellar mass between the low and high redshift samples, $\langle \Delta \log M_{\rm stellar} \rangle = -0.028 \pm 0.019$ (rms: 0.471) at $z < 0.15$ and $0.283 \pm 0.015$ (rms: 0.385) at $z \ge 0.15$, excluding the $\sim$$8$\% catastrophic outliers. The discrepancy between the $z \ge 0.15$ and $z < 0.15$ measurements for the Pantheon+ sample, first shown in \citet{Hoyt2026}, is in part due to Pantheon+ assuming the Chabrier IMF \citep{scolnic2022pantheon+_DR} when redetermining host stellar masses for $z < 0.15$, which prefers lower masses compared to the historical values Pantheon+ chose to adopt for $z \ge 0.15$. Updating the Unite-Pantheon+ subsample masses from Pantheon+ data-release masses to Unite masses \textit{does} result in a shift in cosmology. See Section 5 of \citet{lee25mass} for details. In \citet{lee25mass}, we present the details of our host-galaxy stellar mass measurements, the sensitivity of cosmological results to the host galaxy mass measurements, as well as how much offsetting the high-$z$/low-$z$ discrepancy in the Pantheon+ data-release masses changes Pantheon+ subsample results. We will release the Unite host-galaxy stellar mass measurements and photometry in an accompanying repository.\footnote{The host-galaxy stellar masses and photometry will be available upon acceptance of the paper.}
\begin{figure}
    \centering \includegraphics[width=\linewidth]{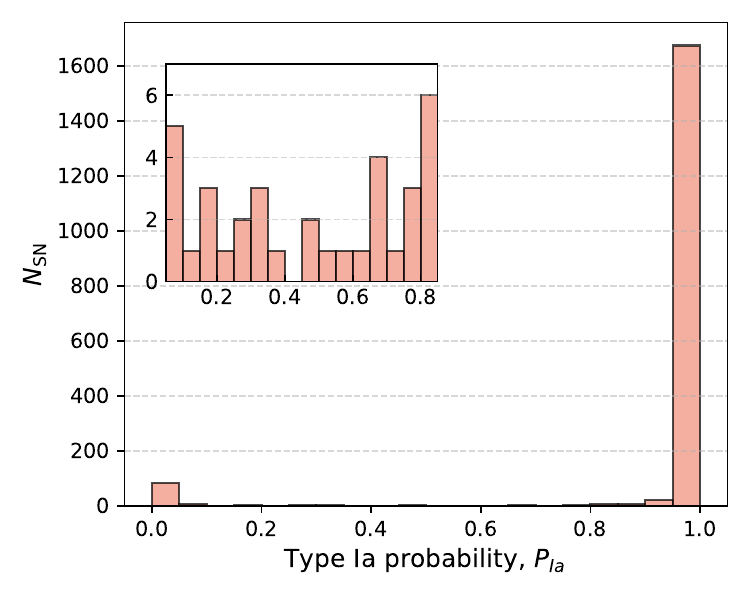}
    \caption{Distribution of the probability of being a SN Ia, $P_{\mathrm{Ia}}$ for the publicly released DES-Dovekie sample. The inset axes show the $ 0.05 < P_{\mathrm{Ia}} < 0.85$ region in greater detail. We see that the distribution is essentially two delta functions at $\sim$$0$ and $\sim$$1$.}
    \label{fig:Pia}
\end{figure} 
\subsection{BEAMS with Bias Corrections (BBC)}\label{sec:bbc}
Both \ppdes~used the BBC framework \citep{2017_BiasCor}, which returns a bias corrected Hubble diagram from a photometrically identified sample of SNe (which can equivalently be applied to a spectroscopic SNe~Ia sample, by setting the probability of each SN event being type Ia to $1$). In order to properly account for contaminants in the cosmology analysis, BBC incorporates the Bayesian Estimation Applied to Multiple Species (BEAMS) \citep{kunz07,Hlozek_2012, kunz2013} likelihood which includes two terms, one that models the SN Ia population and another that models a population of contaminants. To estimate bias corrections, the BBC approach uses simulations that model the survey detection efficiency, Malmquist bias as well as other biases introduced in the analysis. Simulations are generated using the \texttt{ SNANA}\footnote{\url{https://github.com/RickKessler/SNANA}} \citep{2019_DESbias} software.

There have been continual improvements and minor bug fixes since the release of the \ppdes~Hubble diagrams. The most significant update was related to the incorrect treatment of the $\beta$ parameter in \pplus, where the incorrect intrinsic $\beta$ parameter was used in Equation~\ref{eq:TrippMOD} to estimate bias corrections. This error was corrected prior to the \des~analysis. The different BBC\footnote{The current \texttt{SNANA} version is \texttt{v12\_01c-36-g125e61f} with BBC version 4 and specific updates to the BBC code detailed in \url{https://github.com/RickKessler/SNANA/blob/master/src/SALT2mu.c}} versions were shown to have negligible effect on the computed distances \citep{vincenzi25}.

\subsection{Sample cuts}\label{sec:samplecuts}
\begin{table*}

\centering
\caption{Summary of the Unite cosmology sample cuts \& a comparison to cuts used in \ppdes}
\label{tab:cuts}
\renewcommand{\arraystretch}{1.5}
\begin{tabular}{ccccc}
\hline \hline
& \multicolumn{2}{c}{\textbf{Unite}} &  \multicolumn{2}{c}{Requirement difference from:} \\ \cline{2-3} 
Requirement &  \# Discarded & \# Remaining & \pplus & \des    \\ \hline
Spec-$z$ available \& SALT3 fit converged & - & 5271 & SALT2 fit converged$^{*}$ & $z > 0.025$ \\
`Normal~SNe~Ia' ($|x_1|<3$~\&~$|c|<0.3$) & 1220 & 4051 &  &  \\
`Well~constrained' ($\sigma_{x_1}<1.15$,~$\sigma_{t_{\rm peak}}<2$) & 472 & 3579 & $\sigma_{x_1}<1.5$ & $\sigma_{x_1}<1$ \\
Fit probability (fitprob $>0.001$) & 50 & 3529 & no min. for SNLS &  \\
Milky Way reddening ($E(B-V)_{\mathrm{MW}}<0.2$) & 22 & 3507 &  & $E(B-V)_{\mathrm{MW}}<0.25$ \\
Detected host galaxy & 58 & 3349 &  &  \\
Spec-$z$ from the host galaxy emission lines (not SN spectrum) & 40 & 3309 & &  \\
`Likely~SNe~Ia' ($P_{\text{Ia, SCONE}}>0.8$) & 313 & 2996 &  & $P_{\text{Ia}}>0$ \\
Chauvenets criterion & 22 & 2974 &  &  \\ 
Valid bias correction $^{\S}$ & 30 & 2944 &  &  \\
Common SNe across all systematic variants $^{\dag}$ & 60 & \sntot &  &  \\
\hline
\textbf{Cosmology Sample} & & \textbf{\sntot} &  &  \\ 
\hline
\end{tabular}
\begin{tabnote}
\raggedright
{$^{\S}$Some events are lost as their light-curve properties fall within a region of parameter space that is too sparsely populated in the BBC simulations.}\\
{$^{\dag}$In order to build the systematic covariance matrix, we require the same SNe across all systematic variants.}\\
{$^{*}$While \pplus\ did not apply a redshift cut, a $z>0.01$ threshold was imposed during cosmological fitting, which we do not adopt here.}\\
\end{tabnote}
\end{table*}
Here we discuss the selection cuts applied to our combined sample. These cuts are summarised in Table~\ref{tab:cuts}, which highlight any differences from prior analyses. For the most part, we follow the \des~analysis that, in general, applied stricter cuts compared to \pplus. 

We place an initial cut on our sample, prior to the \texttt{SALT3} fit based on the quality of the light-curve. We require SNe with at least two bands that each have at least one detection with SNR $> 5$ and require at least one observation before phase $5$ days after $B-$band peak (we do not require observations before light-curve peak). With these conditions applied and following the convergence of the \texttt{SALT3} fit, we start with an initial SN count of 5271. The loss of SNe due to additional cuts is shown sequentially in Table~\ref{tab:cuts}.

We apply a `Normal SNe~Ia' cut, on both the stretch, $|x_1|<3$ and colour, $|c|<0.3$ and ensure the light-curves are `well constrained' in both the uncertainty on the stretch,  $\sigma_{x_1}<1.15$ and the fitted peak-date of the light-curve, $\sigma_{t_{\rm peak}}<2$. Note that the cut on $\sigma_{x_1}$ is more stringent than \pplus~(requiring $\sigma_{x_1}<1.5$) and more lenient than \des~(requiring $\sigma_{x_1}<1$). Following \citet{desdov}, we relax the $\sigma_{x_1}$ cut because the median $\sigma_{x_1}$ of the new SALT model is 0.15 higher than in \texttt{SALT3.DES5YR}. This adjustment roughly preserves the selection of SNe based on $\sigma_{x_1}$ used in DES-SN5YR.

We apply a uniform minimum on the fit probability of $0.001$ to all high-redshift samples (in \pplus, SNLS had no minimum requirement; SNLS was treated this way to remain consistent with previous work, as \cite{2014JLA} found no impact on light-curve fit accuracy). For Milky Way reddening, we use the stricter cut applied in \pplus~and exclude SNe with reddening larger than $0.20$ (versus $0.25$ used in \des). Following the \des~method we do not include events without a detected host galaxy. Furthermore, we do not include DES SNe with redshift information that could only be determined from the SN spectrum instead of host galaxy features.\footnote{Host galaxy redshifts are more precise than those from supernovae due to sharper spectral lines, and the location of the features in SN spectra change in time due to changes in the velocity of the photosphere.} We iteratively apply a $4\sigma$ cut on the Hubble diagram residuals. 

As \des~is a photometrically classified sample, each event is assigned a probability of being type Ia, $P_{\text{Ia}}$. We perform machine learning light curve classification on the sample following \citet{maria_2021} and \citet{ 10.1093/mnras/stac1691} using two advanced machine learning classifiers, \texttt{SuperNNova} \citep{2020MNRAS.491.4277M} and \texttt{SCONE} \citep{QU21}. For our final cut, we keep only those SNe for which the \texttt{SCONE} classifier assigns a $P_{\text{Ia}}>0.8$. This is in part motivated by Figure~\ref{fig:Pia}, which shows the distribution of $P_{\text{Ia}}$ for the DES-Dovekie sample which essentially consists of two delta functions at $\sim$$0$ and $\sim$$1$ where SNe with $P_{\rm Ia}\sim$$0$ have extremely large uncertainties. When testing classifiers on simulations, algorithms like \texttt{SuperNNova} and \texttt{SCONE} can achieve levels of purity and efficiency $\gtrsim 98\%$ \citep[based on a threshold for binary classification of 0.5; see results reported in][]{10.1093/mnras/stac1691, maria_2021, QU21}, which is comparable to SNe samples confirmed spectroscopically \citep{UNITY_2015}. Therefore, we adopt a threshold of \(P_{\mathrm{Ia}}>0.8\) to obtain a higher-purity sample.

With this conservative cut applied, we then treat the DES-SN as an effectively pure sample throughout the BBC framework. The decision to apply this probability cut does not significantly impact inferred cosmological parameters as the SNe that are cut are the ones that would have been given very large uncertainties had they remained within the cosmology sample. We compare the inferred cosmological parameters on blinded data, with the $P_{\text{Ia}}>0.8$ cut and treating the DES sample as pure vs. without the $P_{\text{Ia}}>0.8$ cut and using BEAMS to deal with the impure sample. We find shifts of $\Delta \Omega_m = 0.001 \pm 0.016$ ($0.06\sigma$) for Flat-$\Lambda$CDM and $\Delta\Omega_m = 0.0007 \pm 0.011$ ($0.06\sigma$) and  $\Delta w = -0.001 \pm 0.033$ ($0.03\sigma$) for Flat-$w$CDM (with a CMB prior; see Section~\ref{sec:cosmology}). 

We also tested applying the $P_{\text{Ia}}>0.8$ cut while simultaneously modelling residual contamination with BEAMS; however, this approach introduced small biases on our best fit cosmological parameters when fit on simulated data, and we therefore do not adopt it. The reason for this bias was investigated in \citet[][see Appendix A]{maria_2021} who also show that probability-based selection can, counter-intuitively, lead to equal or higher biases on cosmological parameters. They find that as the classifier becomes more accurate, the contamination decreases (after a probability-based cut), but the modelling of contamination in BEAMS becomes more uncertain. Essentially, there is so little contamination that trying to model it does more harm than good.

The last two rows of Table~\ref{tab:cuts} detail SN losses owing to the need to implement bias corrections and systematics. To determine valid bias corrections, the simulations of SN Ia are required to populate the entire \texttt{SALT3} parameter space of the data. Furthermore, to build the systematic covariance matrix \citep[see section~$3.6$ of][for details]{vincenzi24}, we require a common set of SNe across all systematic variants. For a small fraction of SNe, these requirements are not met.

\subsection{Lensing}\label{sec:lens}
It has been shown that SN magnitudes are weakly lensed by foreground matter \citep{Shah2022}, which adds skewness and an observed scatter of $\sigma_{\rm lens} = (0.052 \pm 0.009) \; (D_M(z)/D_M(z=1)) ^{3/2}$ to the distance moduli $\mu$, where $D_M = D_L/(1+z_{\rm obs})$ is the perpendicular comoving distance. 

The \texttt{SNANA} simulations used by both \ppdes~incorporated a redshift-dependent Malmquist bias correction derived from lensing probability density functions (pdfs) generated using ray tracing in N-body simulations. However, due to resolution limitations originating from the particle mass of the simulation, $\sigma_{\rm lens}$ of these pdfs was too low by approximately $30\%$. They have been updated in this paper using pdfs generated from the code package \texttt{TurboGL} \citep{Kainulainen2009} which are consistent with the observed value of $\sigma_{\rm lens}$.

Going further, it is possible to reduce the scatter by constructing an estimator of weak lensing magnification on a per SN basis. This uses as input data the photometric magnitude and redshifts of foreground galaxies around the line of sight, which is easily obtainable from survey catalogues. We use the SDSS D16 catalogue \citep[for SN Ia within the \pplus\ footprint;][]{Eisenstein2011} and DES Y6 Gold \citep[for SN Ia within the \des\ footprint;][]{DESY6Gold}. The lensing estimator has two free parameters, a mass-to-observable scaling and halo shape parameter, which are calibrated using the data itself \citep{Shah2023}. The estimate may then be used to `de-lens' SN distances \citep[equation 20 of][]{shah_weaklens}. We include this term in the Unite analysis, applying it after the bias corrections and neglecting any second-order impact on the bias-correction procedure. While \des\ presented de-lensing results as a comparison to their baseline analysis, this work is the first supernova analysis to incorporate the de-lensing correction term into the primary analysis ($\Delta m_{\mathrm{lens}}(\vec{r}, z)$ in Equation~\ref{eq:TrippMOD}).

\subsection{The \des~re-analysis}
\cite{desdov} present updated cosmological constraints from a re-analysis of \des~using the Dovekie cross-calibration. To ensure that the change in cosmology results was entirely due to Dovekie, a full recovery of the original \des~cosmology parameters was included as part of the unblinding criteria and led to the discovery of two errors in the \des~analysis. \cite{desdov} corrected for these issues in their analysis. In this work, we implement their corrections and describe them below.

First, the \cite{Fitzpatrick99} colour law implemented in \des~used a polynomial expansion about $R_V = 3.1$, which is accurate to within 1\% for that value. However, this approximation introduces larger errors in bias-correction simulations that sample a range of $R_V$ values. The impact of this change on the \des~analysis, which shifted the best fit matter density higher in Flat-$\Lambda$CDM by $1\sigma$, is discussed in appendix B of \cite{desdov}.

Second, the photometric uncertainty was under weighted by $\sim$$20\%$. This arose because the individual weight assigned to each of the 9 \texttt{SALT3} calibration variants mistakenly rounded to 0.3 rather than the more accurate value of $1/\sqrt{9} \approx 0.333$.

\subsection{Covariance matrix and systematics}
The $N_{\mathrm{SN}}\times N_{\mathrm{SN}}$ covariance matrix, $\mathcal{C}_{\mathrm{SN}}$ is defined as the sum of statistical, $\mathcal{C}_{\mathrm{stat}}$ and systematic, $\mathcal{C}_{\mathrm{syst}}$ terms.\footnote{For details on the construction of the systematic term, $\mathcal{C}_{\mathrm{syst}}$ see section~3.6 of \citet{vincenzi24}.} In the \des~analysis $\mathcal{C}_{\mathrm{stat}}$ is a diagonal matrix, however duplicate SNe~Ia observed by multiple \pplus~contributing surveys are also included in the Unite analysis. Therefore, we follow \citet[][section~2.2]{2022_pantheon_analysis}  and compute $\mathcal{C}_{\mathrm{stat}}$ including off-diagonal terms that account for correlations between light curves observations of the same SN. The term $\sigma_{\rm lens} = 0.055z$ added to the statistical error in \ppdes~is replaced here with the uncertainty in the lensing estimator described in Section \ref{sec:lens}. This is calculated for each SN, with the dominant contribution being foreground photo-$z$ uncertainty. 

In line with the methodology of \cite{vincenzi24, desdov}, $\mathcal{C}_{\mathrm{syst}}$ is derived from the set of systematic uncertainty sources summarised in \citet[][table 6]{vincenzi24}, with some exceptions. First, we exclude the contamination systematics associated with variations of the BEAMS formalism, and instead apply a conservative probability cut as discussed in Section~\ref{sec:samplecuts}. Second, we exclude the `P21($u-r$)' systematic. For this systematic, the best fit dust parameters within the P23 intrinsic scatter model are determined when splitting SNe by host-galaxy $u-r$ rest-frame colour rather than by host stellar mass. Further investigation \cite[see section~7.1 of][]{vincenzi24} determined that the P21($u-r$) model fails to adequately describe the Hubble residuals and we note that both the \des~analysis and the subsequent Dovekie reanalysis find the size of this systematic has no impact on $\sigma_{w,\mathrm{syst}}$, while having a moderate impact on $w$ (when only including this systematic, \des~find $\delta w=0.048$ and the Dovekie reanalysis find $\delta w=0.029$). Therefore we discard the P21($u-r$) variation of the dust model as it has been superseded by newer models that give a superior fit. Third, we also apply a shift of $\pm 0.2$ mag to the efficiency curves of \citep[with sources provided in table 1 of][]{2022_pantheon_analysis} of SDSS, SNLS and PS1MD. This extends the corresponding systematic applied to DES in \des\ to the additional surveys included in Unite, ensuring a consistent treatment across the sample.

\subsection{Host galaxy mass uncertainties}
\cite{efstathiou25} noted that one source of systematic uncertainty not included in the \des~analysis (nor \pplus) is related to the host galaxy stellar mass uncertainty estimates. This affects the SNe close to the mass step location, meaning that host-galaxies could be assigned to the incorrect log mass bin due to uncertainties. Upon further investigation, \cite{vincenzi25} estimated the missing systematic uncertainty related to host stellar mass would increase the systematic error by less than 3\%. In this work, we use the derived host mass uncertainties determined in Section~\ref{sec:hostmass} to randomly draw 10 realisations of host masses for the Unite sample. We then include each realisation as an analysis variant, which we use to estimate the systematic uncertainty due to host mass uncertainties.

\section{Hubble diagram}\label{sec:hubble_diagram}

The Unite Hubble diagram is shown in Figure~\ref{fig:hubble_diagram} and includes \sntot~likely SNe~Ia. In order to populate a Hubble diagram with the distance moduli of SNe, the observed light-curves are first converted into SN standardisation parameters, $\{m_x,x_1,c\}_i$ for each event using \texttt{SALT3} (see Section~\ref{sec:lcfit}). The distance modulus of each SN, $ \mu_{{\rm obs, i}}$ can then be calculated using the modified Tripp equation \citep{1998_tripp},
\begin{equation} \label{eq:TrippMOD}
\begin{aligned}
\mu_{{\rm obs, }i} = m_{x, i} +\alpha x_{1,i}-\beta c_i + \gamma G_{{\rm host},i}  &- M_x  - \Delta \mu_{{\rm bias,} i} \\ &-\Delta m_{\mathrm{lens},i}(\vec{r}, z)
\end{aligned}
\end{equation}
which includes a mass-step correction, where $G_{\rm{host}} =\pm 1/2$ is a step function that describes the magnitude offset observed between SNe found in high ($M_{\rm host}>S$) and low ($M_{\rm host}<S$) stellar mass galaxies, where $S$ is typically taken to be $10^{10 }M_{\odot}$. $M_x$ is the SN Ia peak absolute magnitude and is completely degenerate with $H_0$, however can be calibrated by setting an absolute distance scale with primary distance anchors. The global nuisance parameters, $\alpha$, $\beta$ and $\gamma$ correct for luminosity correlations with stretch, colour and host stellar mass, respectively. $\Delta \mu_{{\rm bias,} i}$, corrects each SN distance for selection effects and analysis biases and is determined using large simulations of the Unite sample. The simulations are created using the \texttt{SNANA} software and $\Delta \mu_{{\rm bias,} i}$, along with the nuisance parameters are fit for using the BBC framework. The final term in Equation~\ref{eq:TrippMOD} $\Delta m_{\mathrm{lens}}(\vec{r}, z)$ is a novel term introduced by \citet[][see Section~\ref{sec:lens}]{shah_weaklens} to correct for weak lensing magnification by foreground matter and is done after the standarisation process.

The distance modulus uncertainties, $\sigma_{\mu, i}$ are also calculated within the BBC framework as
\begin{equation}\label{eq:mu_uncer}
\begin{aligned}
\sigma_{\mu, i}^2=f\left(z_i, c_i, M_{*, i}\right) \sigma_{\text {S3fit }, i}^2 & +\sigma_{\text {floor }}^2\left(z_i, c_i, M_{*, i}\right) +\sigma_{z, i}^2 \\ 
& +\sigma_{\text {vpec }, i}^2+\sigma_{\text {lens }, i}^2
\end{aligned}
\end{equation}
where $\sigma_{\text {S3fit }, i}$ is computed from the \texttt{SALT3} light-curve fit parameters. The terms $f\left(z_i, c_i, M_{*, i}\right)$ and $\sigma_{\text {floor }}\left(z_i, c_i, M_{*, i}\right)$ are survey specific scaling and additive factors that are estimated from the same simulations used for bias corrections to ensure that the reduced $\chi^2$ in each cell of a $\{z_i, c_i, M_{*, i}\}$ grid is close to unity. Additionally, $\sigma_{\text {floor}}$ accounts for any additional scatter not captured by $\sigma_{\text {S3fit }}$ and consists of a grey term and a term that depends on redshift, colour and host mass, 
\begin{equation}
\sigma_{\text {floor }}^2\left(z_i, c_i, M_{*, i}\right)=\sigma_{\text {scat }}^2\left(z_i, c_i, M_{*, i}\right)+\sigma_{\text {grey }}^2 .
\end{equation}
The final terms in Equation~\ref{eq:mu_uncer}, $\sigma_{z, i}$, $\sigma_{\text {vpec }, i}$ and $\sigma_{\text {lens }, i}$ are uncertainties associated with spectroscopic redshifts, peculiar velocities and weak lensing, respectively.  

The sources of systematic uncertainties are summarised in Table~\ref{tab:systsources} and discussed further in Section~\ref{sec:systbudget}. In Table~\ref{tab:nuisance_params}, we present our fits for the nuisance parameters, $\alpha$, $\beta$, $\gamma$ and $\sigma_{\text{grey}}$ as well as fit diagnostics for our baseline analysis and systematic variants related to SN Ia astrophysics. For the Unite Hubble diagram, we find consistent results with \des~and DES-Dovekie and observe a residual mass step, $\gamma=0.040\pm0.006$ and residual intrinsic scatter $\sigma_{\rm grey}=0.037$. We discuss the results from systematic variants further in Section~\ref{sec:systbudget}.

\subsection{Per-survey Hubble residuals}\label{sec:survey_residuals}
\begin{figure}
    \centering \includegraphics[width=\linewidth]{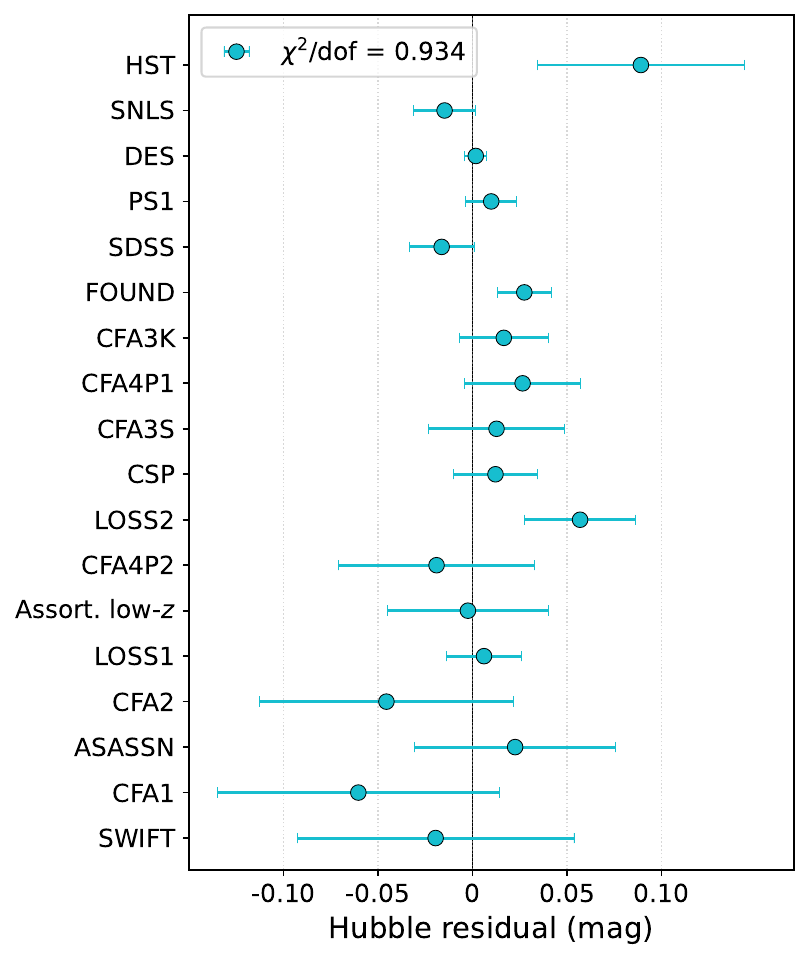}
    \caption{Per-survey inverse-covariance weighted mean Hubble residual relative to the best fit Flat-$\Lambda$CDM cosmology. Surveys are ordered by mean survey redshift.}
    \label{fig:surveyresiduals}
\end{figure}

In Figure~\ref{fig:surveyresiduals}, we compare per-survey offsets in the Hubble diagram, with surveys ordered by mean survey redshift. We calculate the inverse-covariance weighted mean for the entire redshift range relative to the best fit Flat-$\Lambda$CDM cosmology and find $\chi^2/\mathrm{dof}=0.934$.

\subsection{Unblinding procedure}\label{sec:unblind}
To mitigate confirmation bias in our analysis, we blinded the inferred cosmological parameters while the analysis pipeline was being finalised and only unblinded them after predefined criteria had been met.

First, we test the validity of our pipeline on simulated light-curves generated using \texttt{SNANA} and confirm that our input cosmology is recovered. In particular, we generate 25 realisations of the Unite sample assuming Flat-$\Lambda$CDM with $\Omega_{\mathrm{m}}=0.315$. We run these simulations through the entire pipeline and fit each Hubble diagram to Flat-$\Lambda$CDM (with no prior) and Flat-$w$CDM (see Section~\ref{sec:cosmology}). For Flat-$\Lambda$CDM fits, we find a mean bias of $\Omega_{\mathrm{m}} - \Omega_\mathrm{m, true} = 0.002 \pm0.002$. For Flat-$w$CDM fits we find a mean bias of $\Omega_\mathrm{m} - \Omega_\mathrm{m, true} = 0.005 \pm0.010$ and $w - w_\mathrm{true} = 0.01 \pm0.03$. 

\begin{figure}
    \centering \includegraphics[width=\linewidth]{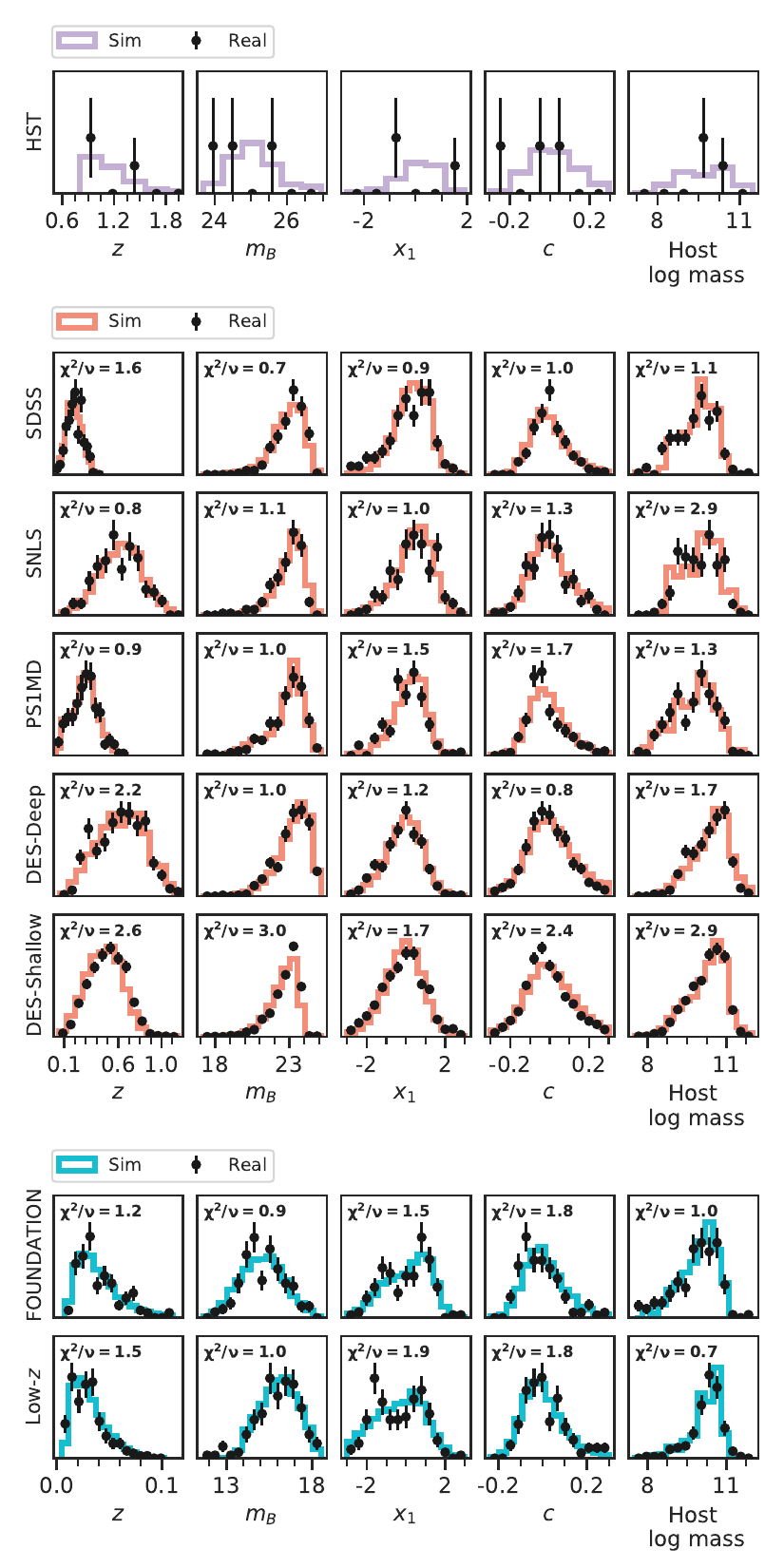}
    \caption{Comparison between observed and simulated SN and host galaxy properties for each survey. Note that the DES-SN has been separated into shallow and deep fields. We impose a reduced $\chi^2$ criterion of $0.7 - 3.0$ for all surveys. The HST sample is excluded from this requirement due to the limited number of SNe.}
    \label{fig:sims}
\end{figure}

Second, we examine the accuracy of our simulated redshift, \texttt{SALT3} parameters, and host log mass distributions. Following \cite{vincenzi24, desdov}, we impose a reduced $\chi^2$ criterion of $0.7 - 3.0$ for all surveys. The HST sample is excluded from this requirement due to the limited number of SNe. In Figure~\ref{fig:sims}, we present the comparisons. We find good agreement between data and simulations with all simulations meeting our criteria.

Following the initial unblinding, as part of the review process, a small number of methodological improvements were identified that enhanced the analysis. These updates have been included in the current version, presented in this paper. The changes include:
\begin{enumerate}
    \item \textbf{Duplicate observations:} 
    In the initial iteration of Unite, we inherited the \pplus~off-diagonal covariance entries. These have since been replaced with a dedicated Unite implementation that consistently accounts for correlations between multiple observations of the same supernova. This increases the uncertainty contribution to $\sigma_{Q_H}$ from $0.002$ to $0.003$, while $\Delta Q_H$ also increases from $0.004$ to $0.005$ (Here, $Q_H$ is a parameter designed to assess the cosmological shift perpendicular to the degeneracy direction in the $\Omega_m$-$w$ plane introduced in detail in Section~\ref{sec:systbudget}).

    \item \textbf{Additional systematics:} The spectroscopic efficiency systematic was updated to apply consistently across all high-redshift surveys. In the initial analysis, this variation was included for DES only. This change had a negligible impact on $\sigma_{Q_H}$, while $\Delta Q_H$ reduced from $0.002$ to $0.001$.
    
    \item \textbf{low-$z$ bias-corrections:} In preparation of the companion $H_0$ analysis, we identified and removed an unintended $\chi^2/\mathrm{dof}$ selection applied to high signal-to-noise low-redshift data. This change resulted in the recovery of five calibrator light-curves. 

    \item \textbf{SNLS SN-dependent filter transmission:} As part of post-unblinding validation, we examined per-survey Hubble residuals (Figure~\ref{fig:surveyresiduals}) and found a $\sim$$2\sigma$ offset for SNLS relative to the best-fit cosmology. While such an offset is not in itself statistically significant, we investigated further and identified an omission in the treatment of the SNLS filter transmission. \citet{2009A&A...506..999R} reported that the SNLS filter transmission varies with radial position on the detector. Therefore, the effective filter transmission can be unique to each SN, and we incorporated this dependence into the updated analysis, reducing the SNLS offset to $\sim$$0.9\sigma$.

\end{enumerate}

Overall, these changes had a negligible impact on the inferred cosmological parameters. The largest shift was $0.3\sigma$ for the SN-only Flat-$\Lambda$CDM fit. For the remaining models, the best fits shifted by $\sim$$0.1\sigma$ or less, predominantly along the existing parameter degeneracy directions.

\section{Systematic uncertainty budget}\label{sec:systbudget}

\begin{table*}
\caption{Sources of uncertainty}
\label{tab:systsources}
\renewcommand{\arraystretch}{1.5}
\begin{tabular}{llll}
\hline \hline Baseline & Size & Systematic & Label \\
\hline \multicolumn{4}{l}{\textbf{Calibration and LC model}} \\
 SALT3 surfaces \& ZP & 1/10 & 10 covariance realisations & `SALT3+calibration' \\
HST Calspec 2020 Update & 1 & $5 \mathrm{mmag} / 7000$ \AA & `HST calspec' \\[2mm]
\multicolumn{4}{l}{\textbf{SN Ia properties and astrophysics}} \\
\cite{P21_dust2dust} dust-based model & 1/4 & 3 realizations of the dust model & `P23 dust pop 1/2/3' \\
& 1/4 & Original BS21 dust parameters & `BS21 dust params' \\
Empirical modelling of $x_1-\mathrm{M}_{\star}$ correlations & 1 & modelling SN age as in \cite{Wiseman22} & `Model SN age' \\
No $\alpha$ evolution & 1 & $\alpha(z)=\alpha_0+\alpha_1 \times z$ & `$\alpha$ evolution' \\
No $\beta$ evolution & 1 & $\beta(z)=\beta_0+\beta_1 \times z$ & `$\beta$ evolution' \\
No $\gamma$ evolution & 1 & $\gamma(z)=\gamma_0+\gamma_1 \times z$ & `$\gamma$ evolution' \\
$\alpha=0.15$ and $\beta=2.87$ initial estimate & 1 & Change initial estimates to $\alpha=0.16$ and  $\beta=3.1$ &  `Change initial $\alpha\beta$' \\
Mass step location at $10^{10} M_{\odot}$ & 1 & Mass step location at $10^{10.2} M_{\odot}$ & `Mass step location' \\
$\sigma_{\text {int }}$ modelling with scaling+additive scatter terms (Equation~\ref{eq:mu_uncer}) & 1 & Scaling term only & `$\sigma_{\text {int }}$ modelling' \\
Host masses determined from \citet{lee25mass} & 1/10 & 10 random realisations & `Host mass uncertainties' \\[2mm]
\multicolumn{4}{l}{\textbf{Milky Way extinction}} \\
MW scaling \citet{Schlafly11} & 1 & 5\% scaling & `MW 5\% scaling' \\
MW colour law $R_V=3.1$ and \citetalias{Fitzpatrick99} & 1/3 & $R_V=3.0$ and \citetalias{Cardelli89} & `MW colour law' \\[2mm]
\multicolumn{4}{l}{\textbf{Survey modelling}} \\
SN host catalogue by \citet{Qu_hostMismatch} & 1 & SN hosts from the DES-SVA galaxy catalogue & `DES SV catalogue' \\
Efficiency $\epsilon_z^{\text {spec }}$ presented by \citetalias{Vincenzi_2021_} (DES) and sources in & \multirow[c]{2}{*}{1} & \multirow[c]{2}{*}{Shift of $\pm 0.2 \mathrm{mag}$ in the efficiency curves} & \multirow[c]{2}{*}{`Shift host spec eff.'} \\ table 1 of \citet[][SDSS, SNLS, PS1MD]{2022_pantheon_analysis} & & &\\[2mm]
\multicolumn{4}{l}{\textbf{Redshift}} \\
2M++ Peculiar velocities \citep{Peterson_2022, Carr_2022}  & 1 & Integrated line-of-sight 2M++ \citep{2015MNRAS.450..317C} & `Peculiar velocities' \\
& & or 2MRS \citep{2021MNRAS.507.1557L} & \\
No redshift shift & 1/6$^{\dag}$ & $\Delta z=1 \times 10^{-4}$ & `Redshift shift' \\[2mm]
\multicolumn{4}{l}{\textbf{Additional statistical uncertainty}} \\
Measurement noise from multi-survey SN observations & N/A & N/A & `Duplicate SNe' \\
\hline
\end{tabular}
\begin{tabnote}
{$^{\dag}$~The scaling is applied such that \(\sqrt{1/6}\times10^{-4}\approx4\times10^{-5}\).}\\
\end{tabnote}
\end{table*}

\begin{table*}
\centering
\caption{Global nuisance parameters and fit diagnostics for the Unite analysis and systematic variants related to SN Ia astrophysics. }
\label{tab:nuisance_params}
\renewcommand{\arraystretch}{1.5}
\begin{tabular}{lcccccccccc}
\hline\hline
 & $\alpha_0$ & $\alpha_1$ & $\beta_0$ & $\beta_1$ & $\gamma_0$ &  $\gamma_1$ & $\sigma_{\rm grey}$ & RMS & $\Delta \chi^2$ & $\Delta Q_{H, \rm syst}$ \\
\hline
\textbf{Unite} & \textbf{0.159(3)} & - & \textbf{3.08(3)} & - & \textbf{0.040(6)} & - & \textbf{0.037} & \textbf{0.177} & \textbf{0.0} & - \\
\hline
\multicolumn{5}{l}{\textbf{Systematic variants from SN Ia astrophysics}} & & & \\
BS21 dust params & 0.159(3) & - & 3.14(3) & - & 0.028(6) & - & 0.054 & 0.178 & 2.7 & -0.003 \\
P23 dust pop 1 & 0.160(3) & - & 3.15(3) & - & 0.045(6) & - & 0.050 & 0.177 & 4.2 & 0.000 \\
P23 dust pop 2 & 0.159(3) & - & 3.02(3) & - & 0.035(6) & - & 0.035 & 0.177 & 0.2 & -0.002 \\
P23 dust pop 3 & 0.159(3) & - & 3.00(3) & - & 0.044(6) & - & 0.055 & 0.178 & 49.2 & -0.001 \\
Model SN Age & 0.162(3) & - & 3.00(3) & - & 0.037(6) & - & 0.043 & 0.177 & 23.6 & -0.001 \\

$\alpha$ Evolution & 0.151(5) & 0.03(1)  & 3.03(3) & - & 0.040(6) & - & 0.035 & 0.177 & -4.7 & 0.001 \\

$\beta$ Evolution & 0.159(3) & - & 2.96(5) & 0.3(1) & 0.039(6) & - & 0.035 & 0.177 & -9.5 & -0.016\\

$\gamma$ Evolution & 0.159(3) & - & 3.08(3) & - & 0.039(10) & 0.01(2) & 0.037 & 0.177 & -0.4 & -0.000 \\

$\sigma_{\rm int}$ model & 0.159(3) & - & 3.06(3) & - & 0.038(6) & - & 0.116 & 0.177 & -1.2 & 0.001 \\
Change initial $\alpha\beta$ & 0.169(3) & - & 3.31(3) & - & 0.040(6) & - & 0.039 & 0.177 & 0.6 & 0.000 \\
Mass step location & 0.160(3) & - & 3.08(3) & - & 0.043(6) & - & 0.036 & 0.177 & -8.4 & -0.004 \\
\hline
\end{tabular}
\end{table*}
The sources of systematic uncertainties described in \citet{vincenzi24, desdov} and summarised in Table~\ref{tab:systsources} are used as the basis for constructing the covariance matrix $\mathcal{C}$. Using the Hubble diagram and $\mathcal{C}$, we perform a cosmological fit to constrain the parameters $w$ and $\Omega_{\mathrm{m}}$ in a Flat-$w$CDM model and examine the sensitivity of the parameters to each systematic variant. We note that in previous analyses, the systematic uncertainty budget and parameter shifts have been shown for $w$. However, it is difficult to assess systematics within generalised dark energy models due to a  degeneracy between the equation of state of dark energy and the matter content of the universe. Therefore, in this work we use the parameter $Q_H(z=0.2)$ (hereafter we drop the $z=0.2$ for conciseness) defined in sec. 3 of \citet{camilleri24} as 
\begin{equation}
    Q_H(z) = \frac{1}{2}\left[ \Omega_{\mathrm{m}} a^{-3} + \Omega_{\rm de} \left(1+3w\right) a^{-3(1+w)} \right],
\end{equation}
where $Q_H \equiv -\ddot{a}/(aH_0^2)\equiv q(H/H_0)^2$ and $a = (1+z)^{-1}$. The $Q_H$ parameter allows us to present a single non-degenerate number summarising a constraint in the $w-\Omega_{\mathrm{m}}$ plane. For completeness, we present the separate systematic uncertainty budgets on $\Omega_m$ and $w$ in Appendix~\ref{appendix1}.

\begin{figure}
    \centering \includegraphics[width=\linewidth]{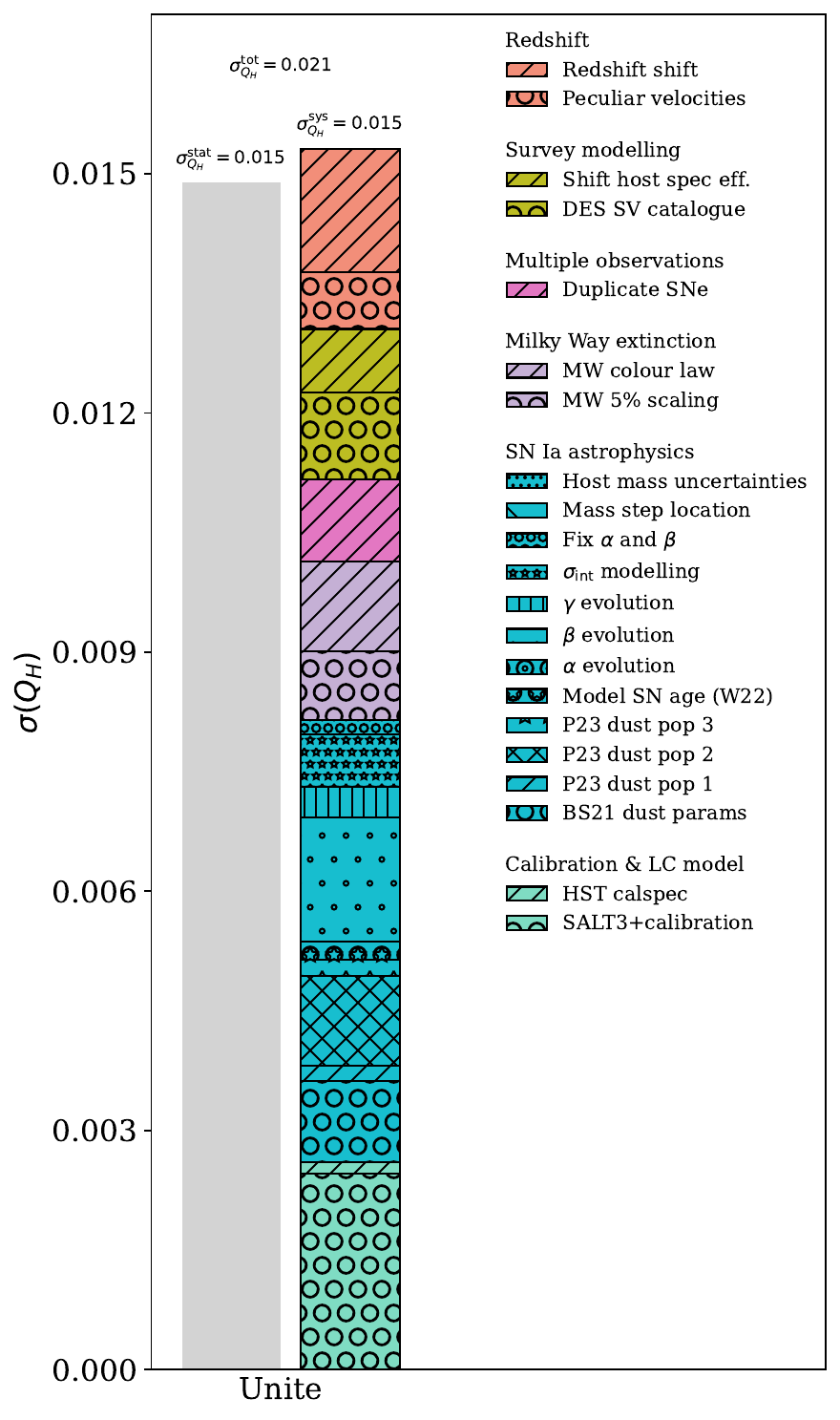}
    \caption{Systematic error budget on $Q_H(z=0.2)$ from the Unite SN sample alone. The statistical uncertainty is also shown for comparison. The various sources of systematic uncertainty in this analysis and labelling conventions are presented in Table~\ref{tab:systsources} and we detail individual contributions in Table~\ref{tab:uncertainty_budget}.}
    \label{fig:unc_budget}
\end{figure}

The total uncertainty budget on $Q_H$, for the Unite analysis, is presented in Figure~\ref{fig:unc_budget} and Table~\ref{tab:uncertainty_budget} summarises the individual uncertainty contributions. The contribution from systematic uncertainties is evaluated as 
\begin{equation}\label{eq:systcontr}
    \sigma_{Q_H,\mathrm{syst}} = \sqrt{\sigma_{Q_H,\mathrm{tot}}^2 - \sigma_{Q_H,\mathrm{stat}}^2},
\end{equation}
where $\sigma_{Q_H,\mathrm{tot}}$ is the total uncertainty resulting from the covariance matrix $\mathcal{C}_\mathrm{SN}$ including both statistical and systematic errors. To evaluate individual systematic contributions, $\sigma_{Q_H,\mathrm{tot}}$ in Equation~\ref{eq:systcontr} is then defined as the total uncertainty resulting $\mathcal{C}_{\mathrm{stat}}+\mathcal{C}_{\mathrm{syst}, i}$ for the $i$th systematic. We note that, although the increased uncertainty due to correlations between duplicate SNe is included in the statistical covariance matrix, we additionally report the resulting shifts in the best fit parameters and their contribution to the uncertainty budget as a systematic. This is to ensure that $\sigma_{Q_H,\mathrm{syst}}$ in Equation~\ref{eq:systcontr} is purely from the $i$th systematic and off diagonal terms from the duplicate SNe are not included in $\sigma_{Q_H,\mathrm{stat}}$, which could result in internal correlations and some systematics to partly cancel out.

We find that the total systematic uncertainty, is approximately equal to the statistical uncertainty, $\sigma^{\mathrm{sys}}_{Q_H}=\sigma^{\mathrm{stat}}_{Q_H}=0.015$. We find trends consistent with previous analyses, such as calibration and SN Ia intrinsic properties contributing heavily to the uncertainty budget.

\subsection{Calibration and LC modelling}
We find a strong contribution from the SALT3+calibration to our total systematic uncertainty budget. The SALT3+calibration systematic produces a shift in $Q_H$ of $-0.009$, moving the contour perpendicular to the $\om-w$ degeneracy direction towards higher values.

\subsection{SN Ia properties and astrophysics}
Table~\ref{tab:nuisance_params} presents the BBC-fitted nuisance parameters, $\Delta\chi^2$ of the cosmology fit relative to the nominal analysis, RMS of the Hubble residual, and shift in best fit $Q_H$ for different systematic variants related to SN Ia intrinsic properties. 

When testing for redshift evolution of the standardisation parameters $\alpha$, $\beta$ and $\gamma$, we define $\alpha(z) = \alpha_0 + \alpha_1\times z$, and similarly for $\beta(z)$ and $\gamma(z)$. Notably, we detect a $\sim$$3\sigma$ redshift evolution in the $\alpha$ and $\beta$ nuisance parameters and find $\alpha_1 = 0.03\pm0.01$ and $\beta_1=0.3\pm0.1$. The best fit parameters suggest that $\alpha$ and $\beta$ increase with redshift. The significance for beta evolution is greater in Unite than in the \des~and DES-Dovekie analyses, and we find that the $\beta$ systematic contributes strongly to the systematic error budget. This is in part due to the construction of the error budget using the $Q_H$ parameter, and when we look at the systematic error budget on $w$ for Unite, we see a negligible contribution from these two systematics. The $\Delta \chi^2$ values given in Table~\ref{tab:nuisance_params} correspond to a moderate and strong preference ($\Delta$AIC $=-2.7$ and $-9.5$) for $\alpha$ and $\beta$ evolution. To investigate this further, we also performed a fit that simultaneously allowed for evolution in the $\alpha$, $\beta$ and $\gamma$ nuisance parameters, and still find a $\sim$$2\sigma$ preference for redshift evolution in both $\alpha$ and $\beta$.

We find a strong preference, $\Delta \chi^2=-8.4$, for a mass step at $10^{10.2}M_{\odot}$ over the baseline value of $10^{10}M_{\odot}$. Consequently, we considered adjusting the baseline analysis to fit for the location of the mass step. However, this choice induced a bias on inferred cosmological parameters when fitting to simulated data (with a mass step of $10^{10}M_{\odot}$), and the implementation required significantly more testing, which we leave for future work.

\citet{vincenzi25} estimated a 3\% increase to the systematic error on $w$ if host galaxy stellar mass uncertainties were accounted for in \des. In this work, we find a stronger contribution of $\sim$$10\%$ to the systematic error on $w$, which is in part due to assuming larger (and therefore more conservative) uncertainties for host-galaxy stellar masses in Unite \citep{lee25mass}; however, it gives a negligible contribution to the systematic error on $Q_H$. Even though the contribution to $\sigma_{Q_H}$ is negligible, we see a shift in $Q_H$ of $\Delta Q_H = -0.004$.

\subsection{Other systematics}
For the remaining systematics, we find a moderate contribution due to Milky Way extinction, where we follow DES and Pantheon+ in adopting a 5\% scaling systematic for Milky Way dust extinction. Future work by \citet[][in prep.]{Elliott26} intends to extend this treatment to account for uncertainties in the Galactic dust map and incorporate these systematics directly into SALT model training. We find a small contribution from shifts in the spectroscopic efficiency curves. Interestingly, we find a moderate contribution from the DES SV catalogue systematic, which arises from replacing the nominal galaxy catalogue derived from DES co-added images \citep{Wiseman20,Qu_hostMismatch} with a shallower `SVA Gold' catalogue \citep{Smith20,Kessler19} in the \texttt{SNANA} simulations used to model SN–host correlations. Finally, the redshift systematics contribute heavily to the total systematic error budget, which is driven by the larger low-redshift sample compared to \des~or DES-Dovekie.
\begin{table}
\renewcommand{\arraystretch}{1.4}
\centering
\caption{Size of systematic uncertainty (SN-only). A description of the various systematics and labelling conventions are presented in Table~\ref{tab:systsources}.}
\label{tab:uncertainty_budget}
\begin{tabular}{lccc}
\hline\textbf{Systematic} & $\boldsymbol{\sigma_{Q_H,\mathrm{syst}}^*}$ & \textbf{\% $\boldsymbol{\sigma_{\mathrm{stat+sys}}}$} & $\boldsymbol{\Delta Q_{H,\mathrm{syst}}^{\dag}}$ \\[1mm] \hline
\textbf{Total Systematic} & 0.015 & 72 & -0.019 \\
\hline
\textbf{Calibration \& LC model} &  &  &  \\
\hspace{2mm} SALT3+calibration & 0.008 & 35 & -0.009 \\
\hspace{2mm} HST calspec & 0.000 & 2 & -0.000 \\
[2mm]
\textbf{SN Ia astrophysics} &  &  &  \\
\hspace{2mm} BS21 dust params & 0.003 & 15 & -0.003 \\
\hspace{2mm} P23 dust pop 1 & 0.001 & 2 & 0.000 \\
\hspace{2mm} P23 dust pop 2 & 0.003 & 16 & -0.002 \\
\hspace{2mm} P23 dust pop 3 & 0.001 & 3 & -0.001 \\
\hspace{2mm} Model SN age (W22) & 0.001 & 3 & -0.001 \\
\hspace{2mm} $\alpha$ evolution & 0.000 & 0 & 0.001 \\
\hspace{2mm} $\beta$ evolution & 0.005 & 22 & -0.016 \\
\hspace{2mm} $\gamma$ evolution & 0.001 & 5 & -0.000 \\
\hspace{2mm} $\sigma_{\rm int}$ modelling & 0.002 & 9 & 0.001 \\
\hspace{2mm} Fix $\alpha$ and $\beta$ & 0.001 & 2 & 0.000 \\
\hspace{2mm} Mass step location & 0.000 & 0 & -0.004 \\
\hspace{2mm} Host mass uncertainties & 0.000 & 0 & -0.004 \\
[2mm]
\textbf{Milky Way extinction} &  &  &  \\
\hspace{2mm} MW 5\% scaling & 0.003 & 12 & 0.000 \\
\hspace{2mm} MW colour law & 0.004 & 16 & -0.002 \\
[2mm]
\textbf{Survey modelling} &  &  &  \\
\hspace{2mm} DES SV catalogue & 0.003 & 15 & 0.012 \\
\hspace{2mm} Shift host spec eff. & 0.002 & 11 & 0.001 \\
[2mm]
\textbf{Redshift} &  &  &  \\
\hspace{2mm} Peculiar velocities & 0.002 & 10 & 0.002 \\
\hspace{2mm} Redshift shift & 0.005 & 22 & 0.001 \\
[2mm]
\multicolumn{2}{l}{\textbf{Additional statistical uncertainty}} &  &  \\
\hspace{2mm} Duplicate SNe & 0.003 & 15 & 0.005 \\
\hline
\end{tabular}
\begin{tabnote}
\raggedright
{$^{\dag}$Shift in $Q_H$ when including \emph{only} this systematic.}\\
{$^{*}$The quadrature sum of systematic uncertainties is larger than the total systematic uncertainty. Internal correlations in the sample cause some systematics to partially cancel when considering the full covariance matrix.}\\    
\end{tabnote}
\end{table}

\section{Cosmology constraints}\label{sec:results}
\begin{table}
\centering
\caption{Parameters and the associated priors used in this analysis. We list the complete parameter set and priors for the Flat-$\Lambda$CDM model and only the additional parameters invoked by the extended models. Background-only refers to constraints using SN and/or BAO data only. CMB refers to any data combinations that include CMB likelihoods.}
\label{tab:priors}
\renewcommand{\arraystretch}{1.5}
\begin{tabular}{lccc}
\hline \hline 
 & Parameter & Background-only & CMB \\ \hline
\textbf{Baseline} & $\Omega_{\mathrm{m}}$ & $\mathcal{U}[0.1, 0.5]$ & $\mathcal{U}[0.1, 0.5]$ \\
& $H_0~\left[\frac{\mathrm{km/s}}{\mathrm{Mpc}}\right]$

 & $\mathcal{U}[55, 91]$ & $\mathcal{U}[55, 91]$ \\
& $\Omega_{\rm b}^{\dag}$  & $\mathcal{U}[0.03, 0.07]$ & $\mathcal{U}[0.03, 0.07]$ \\
& ln$(10^{10}A_s)^{\dag}$ & 2.0 & $\mathcal{U}[0.5, 5.0]$ \\
& $n_s^{\dag}$ & 0.965 & $\mathcal{U}[0.87, 1.07]$ \\
& $\tau^{\dag}$ & 0.07 & $\mathcal{N}(0.067, 0.023)$ \\
\hline
\textbf{Extended} & $\ok$ & $\mathcal{U}[-0.15, 0.15]$ & $\mathcal{U}[-0.15, 0.15]$ \\
\textbf{models} & $w/w_0$ & $\mathcal{U}[-3, -0.4]$ & $\mathcal{U}[-3, -0.4]$\\ 
& $w_a$ & $\mathcal{U}[-3, 2]$ & $\mathcal{U}[-3, 2]$ \\
\hline
\textbf{CMB} & $A_{\rm Planck}$ & - & $\mathcal{N}[1.0, 0.0025]$ \\
 \textbf{nuisance} & $P_{\rm ACT}$ & - & $\mathcal{N}[1.0, 0.003]$\\ 
\textbf{parameters}& $E_{\rm cal}$ & - & $\mathcal{U}[0.8, 1.2]$ \\
& $T_{\rm cal}$ & - & $\mathcal{N}[1.0, 0.0036]$ \\
& $A_{\rm foregorund}$ & - & $\mathcal{U}[0.0, 2.0]$ \\
\hline
\end{tabular}
\begin{tabnote}
\raggedright
{
$^{\dag}$$\Omega_{\rm b}$ is the baryon fraction of the critical density, $A_s$ is the amplitude of primordial fluctuations, $n_s$ is the scalar spectral index, and $\tau$ is the optical depth to reionization.
}\\
\end{tabnote}
\end{table}
We estimate cosmological constraints and combine the Unite sample with external probes using the \texttt{CosmoSIS} \citep{ZUNTZ201545} framework. 
We use the nested sampler, \texttt{Nautilus}\footnote{\url{https://github.com/johannesulf/nautilus}} \citep{nautilus} to determine our best fit parameters, as well as tension metrics and model comparison statistics.
The priors used on the fitted parameters and CMB nuisance parameters are given in Table~\ref{tab:priors}. These priors are consistently adopted for all tension metrics and model comparison statistics. However, our best fit constraints for Unite-only on the $\Lambda$CDM and Flat-$w_0 w_a$CDM models have been determined with wider priors to encompass the full posterior. For all fits, we present the median of the marginalised posterior and cumulative 68.27\% confidence intervals.  In the following sections we summarise the most notable results; we provide a comprehensive list of the results for all combinations in Table~\ref{tab:cosmo_results}.

\subsubsection{Flat-$\Lambda$CDM}\label{sec:flcdm_results}
\begin{figure}
    \centering
    \hspace*{-0cm}
    \includegraphics[width=\linewidth]{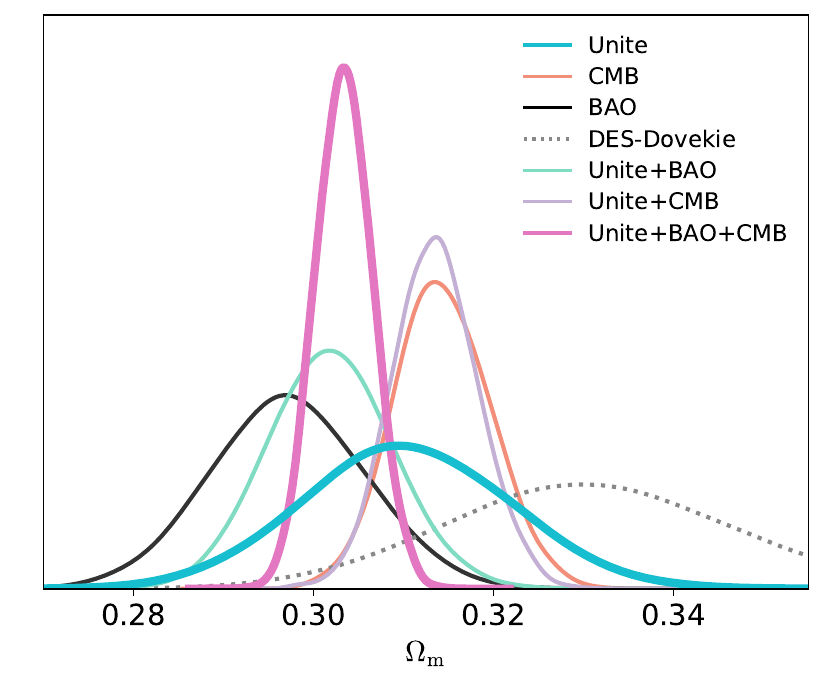}
    \caption{The posteriors for the various constraints on $\om$ in the Flat-$\Lambda$CDM model from Unite only (blue), from Unite combined with BAO (cyan), from Unite combined with CMB (purple), and Unite combined with all external probes (pink). We also show the CMB only, the BAO only and the DES-Dovekie posteriors for comparison (orange, black, and light grey respectively).}
    \label{fig:flcdm_contours}
\end{figure}

Figure~\ref{fig:flcdm_contours} shows the probability density function for the constraints on $\om$ in the Flat-$\Lambda$CDM model. Using the Unite sample alone, we obtain $\om = \mflcdmS$, which is lower than the corresponding best fit values from both DES-Dovekie and \pplus, making it more consistent with the CMB data. We discuss why Unite prefers a lower matter density in Section~\ref{sec:comparisons}. 
When combining all external probes we obtain $\om = \mflcdmSBC$.

\subsubsection{$\Lambda$CDM}

\begin{figure}
    \centering \includegraphics[width=\linewidth]{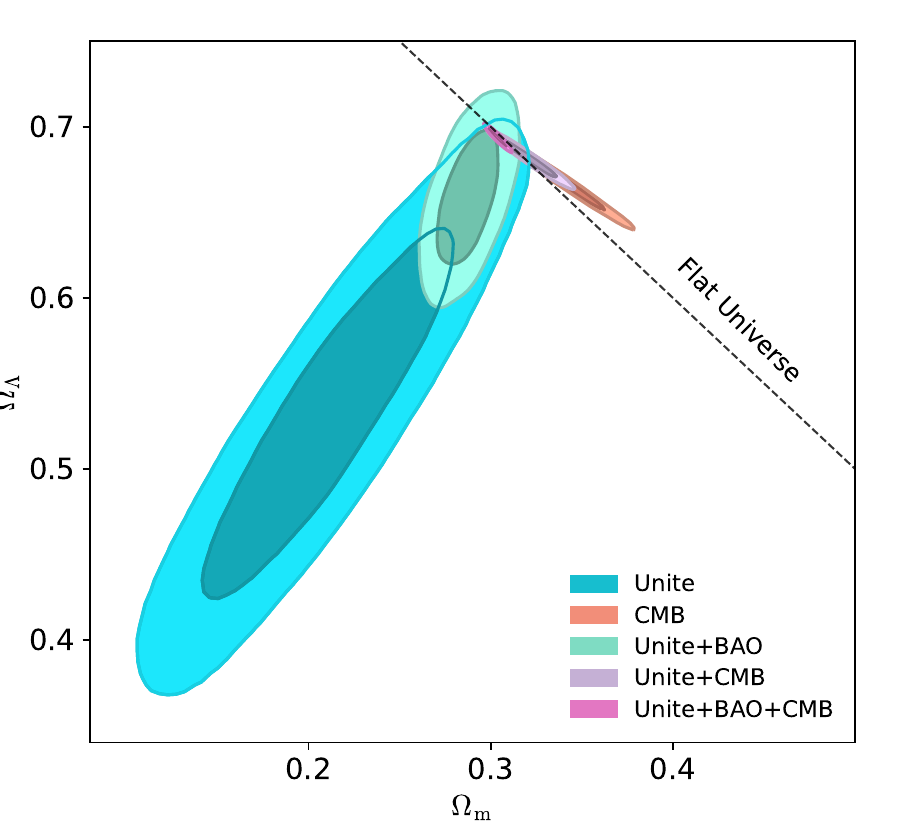}
    \caption{Constraints on the $\Lambda$CDM model from Unite only (blue), from Unite combined with BAO (cyan), from Unite combined with CMB (purple), and Unite combined with all external probes (pink). We also show the CMB only constraints for comparison (orange). The dashed line marks the parameter values corresponding to a spatially flat universe.}
    \label{fig:lcdm_contours}
\end{figure}

We show the posteriors resulting from the fits to the $\Lambda$CDM model in Figure~\ref{fig:lcdm_contours}. Using the Unite sample alone, we obtain $(\om, \ok) = (\mlcdmS, \klcdmS)$, approximately $2\sigma$ from a flat universe. 
When combining all external probes we obtain $(\om, \ok) = (\mlcdmSBC, \klcdmSBC)$. This is $2.6\sigma$ (MAP) away from flatness, and that deviation is driven by the combined low-redshift probes (Unite+BAO), but nevertheless indicates a Universe that is very close to spatially flat. 

\subsubsection{Flat-$w$CDM}
\begin{figure}
    \centering \includegraphics[width=\linewidth]{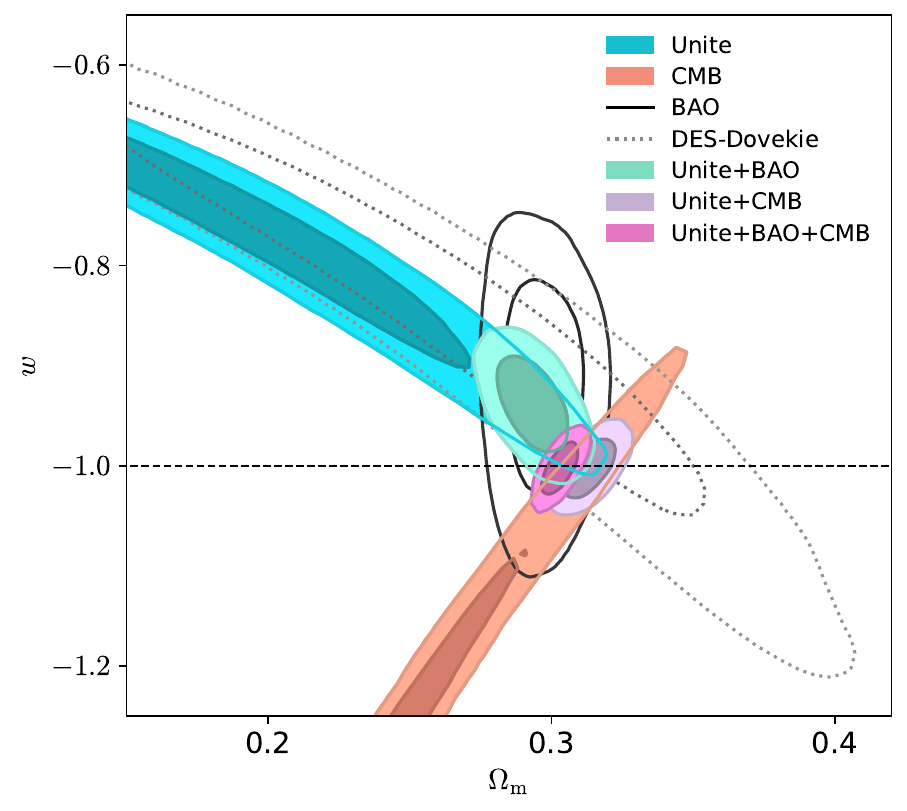}
    \caption{Constraints on the Flat-$w$CDM model from Unite only (blue), from Unite combined with BAO (cyan), from Unite combined with CMB (purple), and Unite combined with all external probes (pink). We also show the CMB only (orange), DES-Dovekie (light grey), and BAO only constraints for comparison. The contour from the combined Unite+BAO+CMB constraint lies outside the $1\sigma$ regions of the Unite, Unite+BAO, and CMB contours, indicating a tension between datasets (see Section~\ref{sec:tensions}). The dashed line marks $w=-1$, corresponding to the cosmological constant.}
    \label{fig:fwcdm_contours}
\end{figure}

In Figure~\ref{fig:fwcdm_contours} we present the posterior distributions of $w$ and $\om$ resulting from fits of the Flat-$w$CDM model. For the Unite sample alone, we obtain $(\om, w) = (\mfwcdmS, \wfwcdmS)$, approximately $2\sigma$ from the cosmological constant equation of state value, $w=-1$. We also overplot the DES-Dovekie constraints for comparison and observe that the Unite contours shift primarily along the degeneracy direction, toward lower $\om$ and higher $w$, with an additional smaller shift perpendicular to the degeneracy direction, toward lower values in the parameter plane. 

Due to the $\om-w$ degeneracy, it is useful to compare supernova constraints in the context of the $Q_H$ parameter defined in Section~\ref{sec:systbudget}. We find $Q_H=-0.418\pm0.021$ for the Unite sample. In the Flat-$\Lambda$CDM analysis (see Section~\ref{sec:flcdm_results}), we find $\om=\mflcdmS$, which corresponds to $Q_H=-0.422^{+0.022}_{-0.021}$, consistent with the Flat-$w$CDM constraint. Therefore, the degeneracy line crosses $w=-1$ at $\om\simeq0.310$ and the perpendicular shift we observe with respect to DES-Dovekie reflects Unite's preference for a lower $\om$ than DES-Dovekie in Flat-$\Lambda$CDM.

Combining SNe with external probes breaks the $\om-w$ degeneracy. 
Combining Unite with all external probes results in $(\om, w) = (\mfwcdmSBC, \wfwcdmSBC$). We have greyed out these constraints, which may have compressed error bars, as the contour from the combined Unite+BAO+CMB constraint lies outside the $1\sigma$ regions of the Unite and CMB contours, and we find a significant tension between the combined low-redshift probes (Unite+BAO) and the CMB; we quantify this tension in Section~\ref{sec:tensions}.

\subsubsection{Flat-$w_0 w_a$CDM}

\begin{figure*}
    \centering \includegraphics[width=1\linewidth]{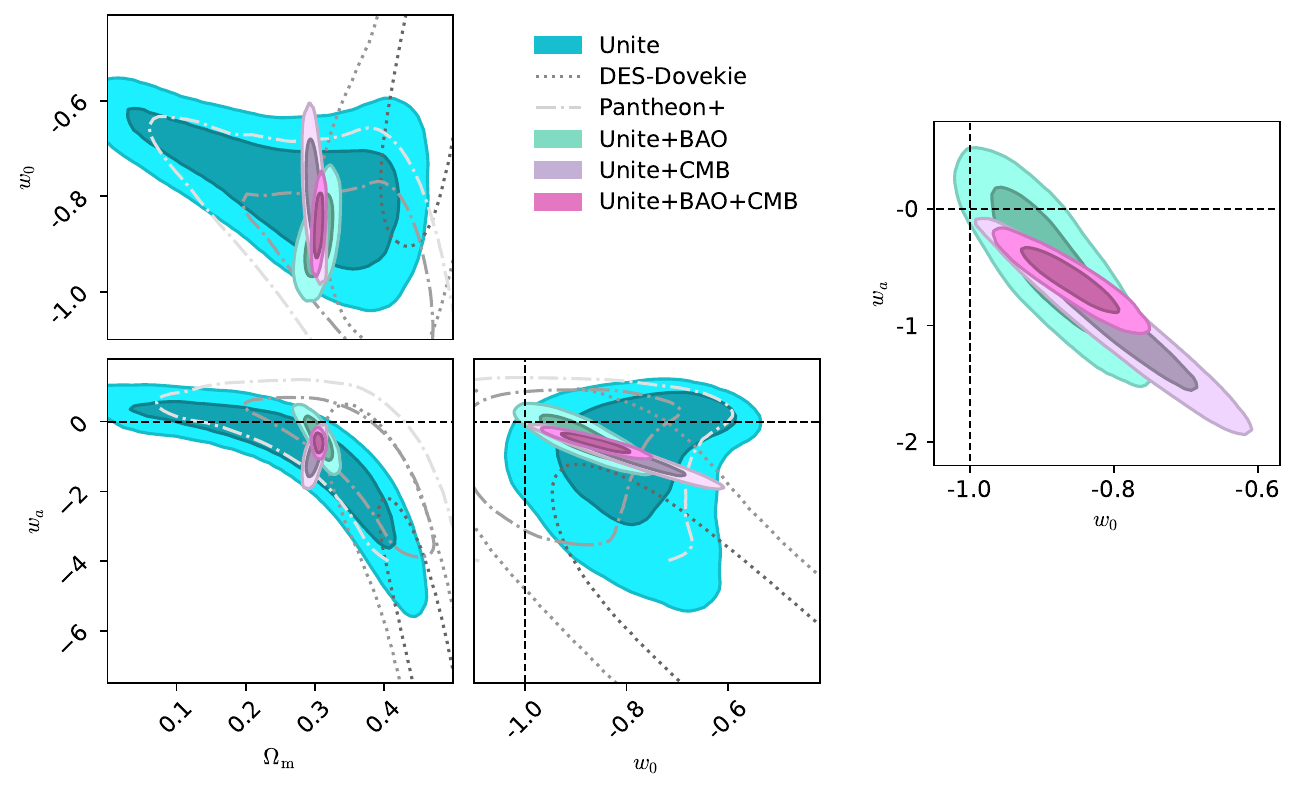}
    \caption{Constraints on the Flat-$w_0 w_a$CDM model from Unite only (blue), from Unite combined with BAO (cyan), from Unite combined with CMB (purple), and Unite combined with all external probes (pink). We also show the DES-Dovekie (dotted grey line) and \pplus~(dash-dotted light grey line) constraints for comparison. The dashed lines mark $(w_0, w_a)=(-1, 0)$, corresponding to the cosmological constant.}
    \label{fig:fw0wacdm_contours}
\end{figure*}

The posterior distributions for the constraints on time-varying dark energy are shown in Figure~\ref{fig:fw0wacdm_contours}. When fitting to Unite alone, we obtain $(\om, w_0, w_a) = (\mfwacdmS, \wfwacdmS, \afwacdmS)$, consistent with a dark energy scenario without time evolution. In Figure~\ref{fig:fw0wacdm_contours}, we also compare the Unite-only posteriors with those from DES-Dovekie and \pplus. The differences between the Unite-only posteriors and their DES-Dovekie counterparts are driven by parameter degeneracies and the differing redshift coverage of the two samples. This is evident from the closer alignment of the Unite contours with those of \pplus, and further highlights the role of parameter degeneracies in setting the overall posterior geometry. 
%

\begin{figure}
    \centering \includegraphics[width=1\linewidth]{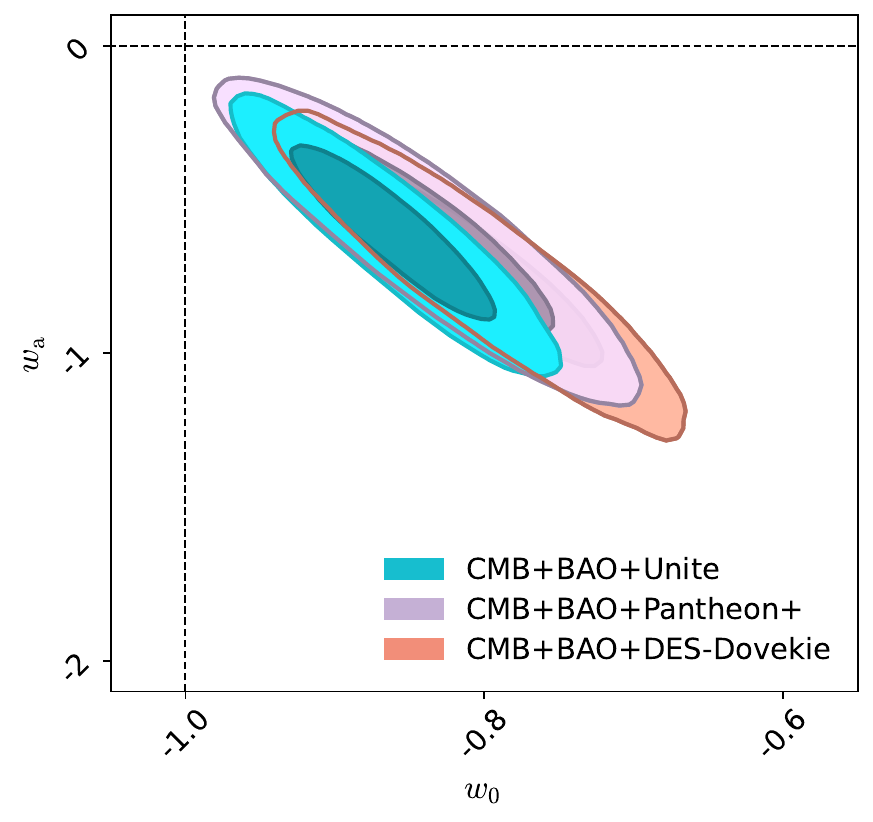}
    \caption{Constraints in the $w_0-w_a$ plane for the \pplus, DES-Dovekie and Unite samples when combined with BAO and CMB data (using the respective published versions of \pplus\ and DES-Dovekie). The 68\% and 95\% confidence regions are shown using the corresponding results from DESI-DR2 and the Dovekie reanalysis and do not in general use identical BAO and CMB data. The dashed lines mark $(w_0, w_a)=(-1, 0)$, corresponding to the cosmological constant.}
    \label{fig:fw0wacdm_compare_contours}
\end{figure}

When combining all external probes we obtain $(w_0, w_a) = (\wfwacdmSBC, \afwacdmSBC)$. This result is consistent with the latest results on dark energy evolution from the Dark Energy Surveys multiple probe analysis \citep{DESCOMBINED}, while improving on their most constraining data combination, which yielded a dark energy Figure of Merit of 222. Our combined constraints achieve $\mathrm{FoM} \equiv 1/\sqrt{\det(\mathrm{Cov}_{w_0w_a})}=315$, corresponding to a $\sim$$30\%$ reduction in the area of the $w_0-w_a$ confidence region and providing the tightest constraints to date. We compare the $w_0-w_a$ constraints obtained from combining Unite with CMB and BAO data to previous constraints obtained using different SN samples in Figure~\ref{fig:fw0wacdm_compare_contours}.

In the $w_0 - w_a$ plane, the degeneracy direction has a sharp turn to the right near $w_a=0$. Unite has moved the best fit up this degeneracy direction, making the SN-alone fit consistent with $w_a=0$.  However, the deviation in $w_0$ remains strong, preferring $w_0>-1$. Thus when combined with BAO and CMB the preference for time-varying dark energy persists. We assess the significance of any deviation from the cosmological constant in Section~\ref{sec:modelpref}. 

\begin{table*}
    \centering
    \caption{Summary of cosmological parameter constraints for four models fitted to the Unite sample and in combination with external datasets. Reported values correspond to the medians of the marginalised posterior distributions, with 68.27\% credible intervals. For each fit we report the $\chi^2$ at the maximum likelihood, $\chi^2_{\rm ML}$. We also present various model preference statistics using frequentist (based on both the $\chi^2$ at maximum likelihood, $\chi^2_{\rm ML}$, and at maximum \textit{a posteriori} probability, $\chi^2_{\rm MAP}$) and Bayesian (comparing Bayesian evidence) approaches. We have greyed out the Flat-$w$CDM constraints when combining all data because this result is subject to significant tensions between the datasets (see Section~\ref{sec:tensions}).}
    \renewcommand{\arraystretch}{1.5}
    \label{tab:cosmo_results}
    \begin{tabular}{lcccccccccccc}
        \cline{1-6} \cline{8-12}
        & $\om$ & $\ok$ & $w_0$ & $w_a$ & $\chi^2_{\rm ML}$ & & \multicolumn{5}{c}{\textbf{Model preference}} \\ 
        
        \cline{1-6} \cline{8-12} 
        \multicolumn{6}{l}{\textbf{Unite}} & & $\Delta \chi^2_{\rm ML}$ & $n\sigma_{\rm ML}$ & $\Delta \chi^2_{\rm MAP}$ & $n\sigma_{\rm MAP}$ & $\Delta {\rm log}~\mathcal{Z}$ \\   \cline{1-6} \cline{8-12}
		Flat-$\Lambda$CDM & $\mflcdmS$  &  -- & -- & -- & 2793 & & 0 & 0 & 0 & 0 & 0\\ 
		$\Lambda$CDM & $\mlcdmS$  &  $\klcdmS$ & -- & -- & 2787 & & -4.6 & 2.2 & -4.6 & 2.2 & -0.89\\ 
        Flat-$w$CDM & $\mfwcdmS$  &  -- & $\wfwcdmS$ & -- & 2787 & & -5.5 & 2.3 & -5.5 & 2.3 & -0.42\\
        Flat-$w_0w_a$CDM & $\mfwacdmS$  &  -- & $\wfwacdmS$ & $\afwacdmS$ & 2787 & & -5.5 & 1.9 & -5.5 & 1.9 & 0.36\\
        \cline{1-6}  \cline{8-12} 
        \multicolumn{6}{l}{\textbf{Unite + CMB}}  & &  $\Delta \chi^2_{\rm ML}$ & $n\sigma_{\rm ML}$ & $\Delta \chi^2_{\rm MAP}$ & $n\sigma_{\rm MAP}$ & $\Delta {\rm log}~\mathcal{Z}$ \\  \cline{1-6} \cline{8-12}
		Flat-$\Lambda$CDM & $\mflcdmSC$  &  -- & -- & -- & 3371 & & 0 & 0 & 0 & 0 & 0\\  
		$\Lambda$CDM & $\mlcdmSC$  &  $\klcdmSC$ & -- & -- & 3368 & & -2.1 & 1.5 & -2.3 & 1.5 & 2.48\\ 
        Flat-$w$CDM & $\mfwcdmSC$  &  -- & $\wfwcdmSC$ & -- & 3371 & & 0 & 0 & 0 & 0 & 4.02\\ 
        Flat-$w_0w_a$CDM & $\mfwacdmSC$  &  -- & $\wfwacdmSC$ & $\afwacdmSC$ & 3363 &  & -7.4 & 2.2 & -7.6 & 2.3 & 1.83 \\ 
		\cline{1-6} \cline{8-12}  
        \multicolumn{6}{l}{\textbf{Unite + BAO}} & & $\Delta \chi^2_{\rm ML}$ & $n\sigma_{\rm ML}$ & $\Delta \chi^2_{\rm MAP}$ & $n\sigma_{\rm MAP}$ & $\Delta {\rm log}~\mathcal{Z}$ \\  \cline{1-6} \cline{8-12}
		Flat-$\Lambda$CDM & $\mflcdmSB$  &  -- & -- & -- & 2887 & & 0 & 0 & 0 & 0 & 0\\   
		$\Lambda$CDM & $\mlcdmSB$  &  $\klcdmSB$ & -- & -- & 2885 & & -2.6 & 1.6 & -2.6 & 1.6 & -0.12\\   
        Flat-$w$CDM & $\mfwcdmSB$  &  -- & $\wfwcdmSB$ & -- & 2884 & & -3.8 & 1.9 & -3.8 & 1.9 & 1.55 \\  
        Flat-$w_0w_a$CDM & $\mfwacdmSB$  &  -- & $\wfwacdmSB$ & $\afwacdmSB$ & 2882 & & -5.4 & 1.8 & -5.4 & 1.8 & 2.46 \\  
        \cline{1-6} \cline{8-12}
        \multicolumn{6}{l}{\textbf{Unite + BAO + CMB}} & & $\Delta \chi^2_{\rm ML}$ & $n\sigma_{\rm ML}$ & $\Delta \chi^2_{\rm MAP}$ & $n\sigma_{\rm MAP}$ & $\Delta {\rm log}~\mathcal{Z}$ \\ \cline{1-6} \cline{8-12}
		Flat-$\Lambda$CDM & $\mflcdmSBC$  &  -- & -- & -- & 3472 & & 0 & 0 & 0 & 0 & 0\\  
		$\Lambda$CDM & $\mlcdmSBC$  &  $\klcdmSBC$ & -- & -- & 3466 & & -5.7 & 2.4 & -6.7 & 2.6 & 1.30\\ 
        Flat-$w$CDM & $\mfwcdmSBC$  &  -- & $\wfwcdmSBC$ & -- & 3472 & & 0 & 0 & 0 & 0 & 4.14 \\
        Flat-$w_0w_a$CDM & $\mfwacdmSBC$  &  -- & $\wfwacdmSBC$ & $\afwacdmSBC$ & 3459 & & -12.4 & 3.1 & -13.6 & \fwasig & -0.44\\
        \cline{1-6} \cline{8-12}
    \end{tabular}
\end{table*}

\subsection{Delensing}
Since this is the first major supernova cosmology analysis to implement the lensing correction as default, we report on its impact. Delensing improves the $\chi^2$ in Flat-$\Lambda$CDM by $-35$ (as measured from the covariance matrix before adjusting for lensing), which corresponds to a $5.6 \sigma$ detection of lensing. It also reduces the scatter of residuals around the Hubble diagram by $\sim$$5\%$ for $z > 0.5$. 

The significance of the lensing detection when going from \des\ to Unite might seem less than expected from the increase in the number of SN Ia alone (\des\ detected lensing at $5\sigma$ significance). However, this is because the lensing signal is largest at high-redshift, and the majority of high-redshift supernovae were already in the \des\ sample.

The impact on median values of cosmological parameters is $< 0.2 \sigma$ for the range of models we consider in this paper, and we find that de-lensing reduces parameter uncertainties by $\sim$$5\%$ on average. We compare our cosmological fits and posteriors with and without the lensing correction term in Appendix~\ref{appendix3}. 

\section{Model assessment and tension}\label{sec:modelcomp_tensions}

\subsection{Tension metrics}\label{sec:tensions}
We quantify the tension between different datasets and their combinations using the Suspiciousness, $S$, \citep{handley19}. This statistic is closely related to the Bayes ratio, $R$, with $\ln S\equiv \ln R - \ln I$, where $I$ is the Bayesian information; however, unlike $R$, Suspiciousness is not dependent on the choice of priors. To asses the significance of the results, we adopt the criteria described in \citet{Trotta08}: ln$~S < -5$ indicates strong tension, $-5 < $ ln$~S < -2.5$ indicates moderate tension, and ln$~S > -2.5$ indicates that the datasets agree.

\begin{figure}
    \centering \includegraphics[width=\linewidth]{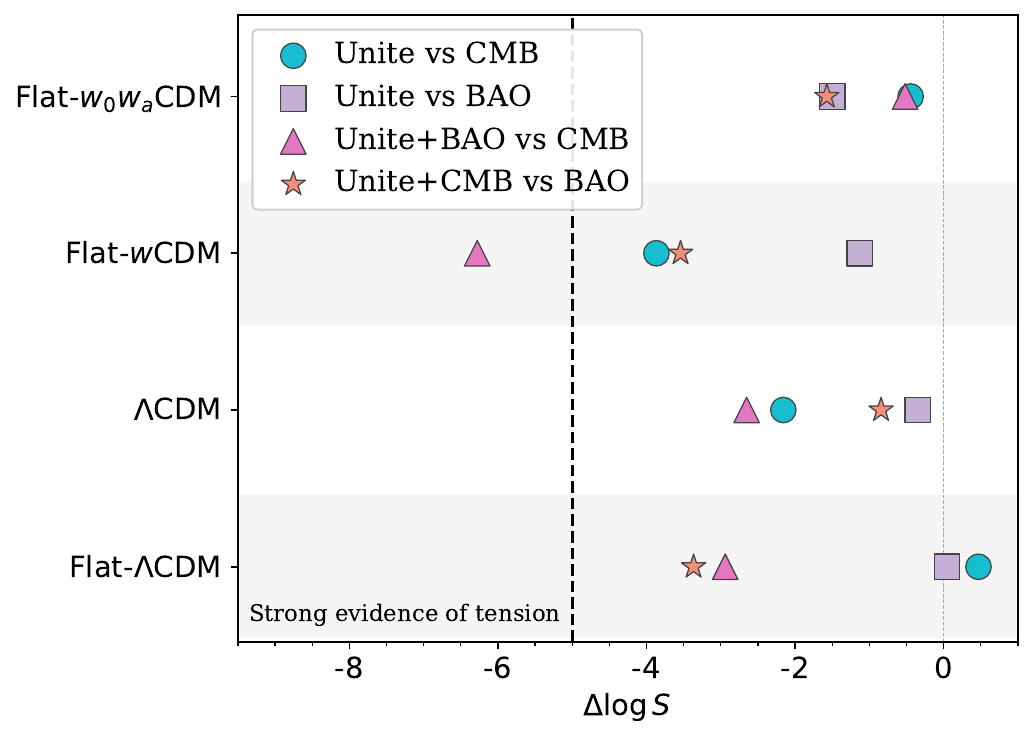}
\caption{Difference in Suspiciousness, $\Delta{\rm log}~S$, between Unite and the CMB, Unite and the BAO, the combined Unite+BAO datasets and the CMB, and lastly, the combined Unite+CMB datasets and the BAO. The region to the left of the dashed line indicates strong evidence of tension according to \citet{Trotta08}. We find strong evidence of tension between the Unite+BAO datasets and the CMB within the Flat-$w$CDM model.}
    \label{fig:tensions}
\end{figure}

Examining all tension metrics (see Figure~\ref{fig:tensions}), we find no strong evidence for tension between the datasets, with the exception of the Flat-$w$CDM model comparing the Unite+BAO and CMB datasets. This becomes apparent when considering Figure~\ref{fig:fwcdm_contours}, where the combined Unite+BAO+CMB posterior lies outside the $1\sigma$ regions of the Unite, Unite+BAO, and CMB confidence intervals.

This is the behaviour expected when a model is not flexible enough to simultaneously fit all datasets.  It indicates that Flat-$w$CDM cannot provide a good fit to all data simultaneously, and the best fit parameters in that model should be treated with caution. It has long been known that the CMB is inconsistent with $w=-1$ at $>1\sigma$ level, and now the supernovae are finding a similar sized discrepancy but in the opposite direction. Although it remains important to consider whether the disagreements could arise from systematic errors in one or more of the analyses, we also take seriously the possibility that a more complex model may be needed.

\subsection{Model preference}\label{sec:modelpref}
In this section, we assess the preference for extended models beyond Flat-$\Lambda$CDM. We use various metrics to compare results within our study, which ensures our findings can also be directly compared with numerous other studies. These metrics are presented in Table~\ref{tab:cosmo_results} and described below.

First, we compute frequentist significance measures based on both the $\chi^2$ at maximum likelihood, $\chi^2_{\rm ML}$, and at maximum \textit{a posteriori} probability, $\chi^2_{\rm MAP}$. In particular we use $\Delta \chi^2$ values between an extended model and Flat-$\Lambda$CDM. As Flat-$\Lambda$CDM is nested within all models considered in this work (with corresponding free parameters, $\ok=0$, $w_0=-1$ and $w_a=0$), Wilks' theorem \citep{wilks1938} implies that  $\Delta\chi^2$ is $\chi^2$-distributed with one (for $\Lambda$CDM and Flat-$w$CDM) or two (for Flat-$w_0 w_a$CDM) additional degrees of freedom (dof), assuming the null hypothesis holds and that the errors are Gaussian and correctly estimated. Therefore, we can convert $\Delta\chi^2$ into a p-value and the corresponding significance $n\sigma$ for a 1D Gaussian distribution,
\begin{equation}
\label{eq:cdf}
{\rm CDF_{\chi^2} (\Delta\chi^2 \; | \; dof)} = \frac{1}{\sqrt{2\pi}} \int_{-N}^{N} e^{-t^2/2} dt~,
\end{equation}
where CDF$_{\chi^2}$ is the cumulative distribution of $\chi^2$.

Second, we use Bayesian quantities that depend on the posterior distribution over the full parameter space. To quantify the preference between models, we use the logarithm of the Bayes Ratio, $R$, defined as the ratio of the Bayesian evidence, $\mathcal{Z}$ evaluated for each model so that
\begin{equation}
    {\rm log}~R = \Delta{\rm log}~\mathcal{Z }~.
\end{equation}
${\rm log}~\mathcal{Z}$ is computed by \texttt{Nautilus} as part of the sampling process and we adopt the criteria described in \citet{Trotta08} to evaluate the significance of the results.

Examining the Bayesian evidence, we find that none of the dataset combinations considered in this analysis strongly favours models beyond Flat-$\Lambda$CDM. The same trend is seen when we examine both the $\Delta\chi^2_{\rm ML}$ and $\Delta\chi^2_{\rm MAP}$ values converted into an equivalent $\sigma$, with the exception of time-varying dark energy, which is preferred over Flat-$\Lambda$CDM when fit to the combined Unite+BAO+CMB datasets. We obtain $3.3\sigma$ significance of time-varying dark energy when using $\Delta\chi^2_{\rm MAP}$ (or $3.1 \sigma$ using the $\Delta\chi^2_{\rm ML}$).

We also note a mild preference for spatial curvature. For $\Lambda$CDM, we obtain $2.6\sigma$ significance of curvature when using $\Delta\chi^2_{\rm MAP}$ (or $2.4 \sigma$ using the $\Delta\chi^2_{\rm ML}$). Historically, low-redshift observations such as weak lensing, \pplus, and BAO measurements from the Sloan Digital Sky Survey, generally favour a spatially flat Universe when analysed alongside CMB data. In contrast, analyses based on the Planck PR3 dataset alone indicate a preference for non-zero curvature at $\sim$$2\sigma$ level, and this preference is maintained when the CMB measurements are paired with the new DESI-BAO data. Therefore, when considering CMB and DESI-BAO data, the preference for a non-spatially flat universe has been noted as a viable alternative to evolving dark energy \citep{Chen2025}. \cite{desdov}, also find a $2.3\sigma$ (using $\Delta\chi^2_{\rm ML}$)  preference\footnote{Using Bayesian evidence, \cite{desdov} find that non-zero spatial curvature is weakly not preferred over Flat-$\Lambda$CDM.} for non-zero spatial curvature when combining DES-Dovekie with DESI-BAO and the same CMB data we consider here. They partially attribute the preference for spatial curvature to DESI-BAO favouring a lower matter density, which in turn causes a slight internal tension between DES-Dovekie and DESI-BAO in $\Lambda$CDM. Interestingly however, Unite prefers a lower matter density than DES-Dovekie (we comment on this further in Section~\ref{sec:comparisons}) and therefore the degeneracy direction of the Unite posterior is more consistent with the intersection of the CMB and BAO data (see Figure~\ref{fig:lcdm_contours}). We note however that using Bayesian evidence, we find non-zero spatial curvature is weakly \textit{not} preferred over Flat-$\Lambda$CDM, consistent with \cite{desdov}.

The strength of the preference for the Flat-$w_0w_a$CDM model remains consistent with DES-Dovekie, which is interesting given that Unite finds $w_a$ consistent with zero. The reason for this is seen in the bottom right panel of Figure~\ref{fig:fw0wacdm_contours}, where the $2\sigma$ contour from Unite alone just hits the crosshairs of the Flat-$\Lambda$CDM parameters. Rather than being $2\sigma$ away from Flat-$\Lambda$CDM towards the bottom right as in DES-SN5YR, Unite is $2\sigma$ away in the direction of larger $w_0$. Projection effects have been shown to occur in parameter spaces with long degeneracy directions, and when the posterior exhibits strong non-Gaussianity \citep{2025JCAP...07..028A}. This is the case for supernova constraints in the $w_0-w_a$ plane, where the posterior extends along a broad degeneracy direction from strongly negative values of $w_0$ and $w_a$ toward the $\Lambda$CDM point $(w_0, w_a) = (-1, 0)$, before exhibiting a strong hook to the right when approaching $w_a=0$. Future work will address this by adopting a different time-varying parameterisation of $w$ that is less susceptible to projection effects arising from the strong non-linear parameter degeneracy \citep[][in prep.]{leo_bigW}.

\section{Comparing Hubble diagrams}\label{sec:comparisons}
\begin{figure*}
    \centering \includegraphics[width=\linewidth]{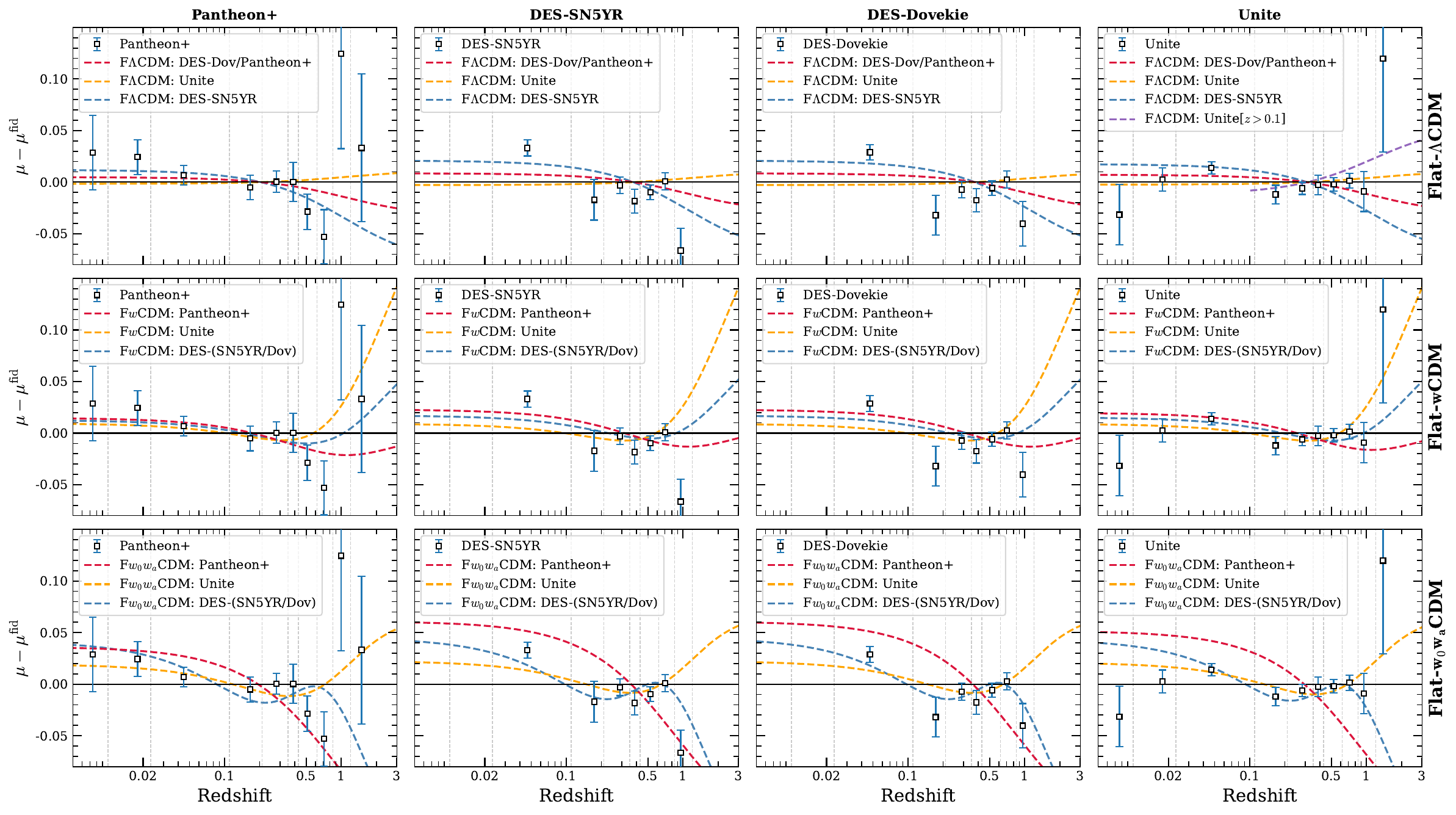}\vspace{-4mm}
    \caption{Comparison of binned Hubble residuals of the Unite and constituent Hubble diagrams (using the respective published versions) that are combined in the Unite analysis, calculated with respect to $\mu^{\rm fid}$, a Flat-$\Lambda$CDM cosmology with $\Omega_{\rm m}=0.315$. In each subplot, we show the weighted mean of the Hubble residuals (using statistical only uncertainties), plotted at the weighted mean of the redshifts. In the top, middle, and bottom rows, we overlay the SN-only best fit Flat-$\Lambda$CDM, Flat-$w$CDM and Flat-$w_0 w_a$CDM models respectively from Unite, DES-Dovekie, \des~and \pplus. We group fits that are sufficiently similar and, in such cases, adopt the best fit result the most recent analysis, which is DES-Dovekie in all cases considered here (e.g., for Flat-$\Lambda$CDM, \pplus\ finds $\Omega_{\rm m}=0.334$, while DES-Dovekie finds $\Omega_{\rm m}=0.330$). In the top-right panel, we also plot in purple the best fit Flat-$\Lambda$CDM model obtained from Unite with a $z>0.1$ redshift cut, which corresponds to removing the first three bins. Note that the difference in the lowest bin between \pplus\ and Unite is {\em almost entirely} because they sit at slightly {\em different redshifts}, due to the slightly different number of supernova in that bin. Since the Hubble diagram is so steep at low redshift, the difference in $\mu^{\rm fid}$ at $z=0.00739$ (\pplus) and $z=0.00761$ (Unite) is $\Delta \mu^{\rm fid}=0.064$.  
    }
    \label{fig:HDs_compare}
\end{figure*}
\begin{figure*}
    \centering \includegraphics[width=\linewidth]{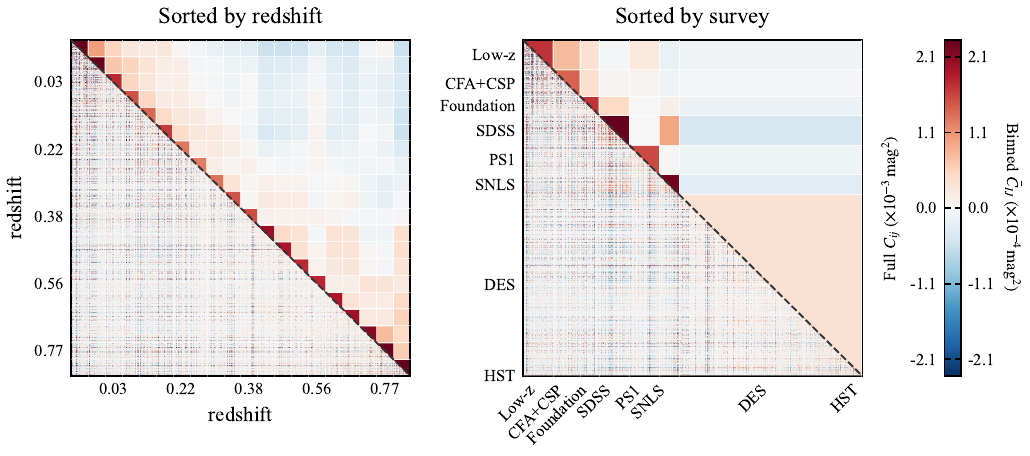}
\caption{Covariance matrix for the supernovae.  The lower-left half of each diagram shows the covariance between individual supernovae, while the upper-right half shows the mean covariance in bins. The left plot orders and bins the supernovae by redshift.  The right plot orders and bins the supernovae by survey. One can see enhanced covariance at low-$z$ and high-$z$ relative to intermediate redshifts, explaining why these extreme redshift ranges only have a weak impact on the cosmology fits. }
    \label{fig:plot_cov}
\end{figure*}

\subsection{General observations} \label{sec:general_obs}
In Figure~\ref{fig:HDs_compare}, we compare binned Hubble residuals of the Unite and constituent Hubble diagrams (using the respective published versions of \pplus, \des, and DES-Dovekie) that are combined in the Unite analysis, calculated with respect to $\mu^{\rm fid}$, a Flat-$\Lambda$CDM cosmology with $\Omega_{\rm m}=0.315$. The residuals are shown as weighted means in redshift bins using statistical uncertainties only, and are overlaid with SN-only best fit cosmological models for Flat-$\Lambda$CDM, Flat-$w$CDM, and Flat-$w_0 w_a$CDM. For clarity, we group fits that yield broadly similar predictions (e.g., for Flat-$\Lambda$CDM, \pplus\ finds $\Omega_{\rm m}=0.334$, while DES-Dovekie finds $\Omega_{\rm m}=0.330$) and, in such cases, adopt the best fit from the most recent analysis, which is DES-Dovekie in all cases considered here.

The error bars in Figure~\ref{fig:HDs_compare} are only the diagonal elements of the covariance matrix.  To complement this, Figure~\ref{fig:plot_cov} shows the covariance between supernova, sorted by redshift, and sorted by survey. Data points with high covariance have lower weight in the cosmology fit than their diagonal errors would suggest. 

These plots facilitate comparisons between analyses without the complications introduced by parameter degeneracies. From these plots, we make the following observations:

\begin{enumerate}
    \item {\bf Reduced scatter:}
The most notable feature of the binned Unite Hubble Diagram is its reduced scatter relative to previous datasets. The high-$z$ Unite bins clearly demonstrate the impact of the combined sample, with substantially reduced uncertainties and noticeably tighter scatter compared to either \pplus~or the DES datasets alone.

\item {\bf Impact of redshift ranges:} 
At both the low- and high-redshift ends of the Hubble diagram, the uncertainties increase due to different systematic and statistical limitations. At low-redshifts ($z \lesssim 0.023$), the sensitivity of peculiar velocities is large, and due to volumetric effects (Eddington bias), more objects are scattered to lower redshifts than the reverse.  At high-redshifts ($z \gtrsim 0.85$), the number of available SNe decreases substantially, leading to larger statistical uncertainties and increased sensitivity to individual objects. The uncertainties quoted are intended to encapsulate these effects, although residual biases may still remain. We therefore test the sensitivity of our results by excluding the first two bins\footnote{We fit to the unbinned data, so excluding a bin means excluding all the supernovae that are in that bin.} (retaining $z>0.0233$), excluding the first three bins (retaining $z>0.1$), or removing the two highest-redshift bins (retaining $z<0.85$) used in Figure~\ref{fig:HDs_compare}. 

The only cut that causes a substantial shift in the best fit cosmology is cutting all of the $z<0.1$ supernovae, corresponding to 687 events ($24\%$ of the full sample) and resulting in a $\sim$2$\sigma$ shift in the best fit.  This further motivates obtaining modern low-$z$ data from ongoing surveys, including the Dark Energy Bedrock All-Sky Supernova Program \citep[DEBASS;][]{Sherman2025,Acevedo2026}, which has measured $>400$ supernovae at $z<0.08$ using the same telescope and filters as DES, as well as the Asteroid Terrestrial-impact Last Alert System \citep{ATLASSURVEY, 2026ApJ..1004..173M}, the Zwicky Transient Factory \citep{Rigault25}, and the Young Supernova Experiment \citep{jones21,aleo23}.
\begin{figure}
    \centering \includegraphics[width=\linewidth]{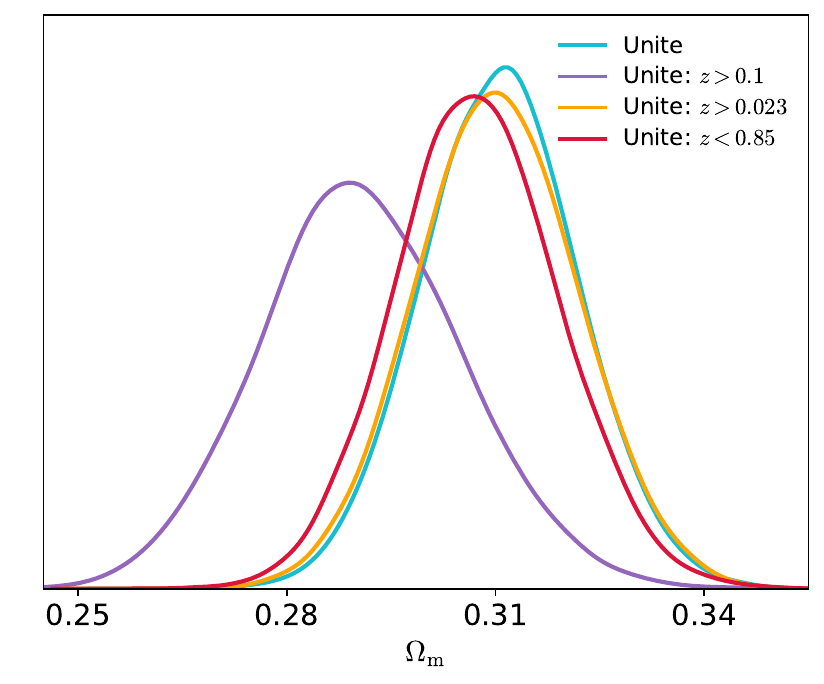}
\caption{The posteriors for constraints on $\om$ in the Flat-$\Lambda$CDM model from Unite (blue) and Unite with redshift cuts that correspond to removing the first two bins ($z>0.0233$; orange), removing the first three bins ($z>0.1$; purple), and removing the two highest-redshift bins ($z<0.85$; red) from Figure~\ref{fig:HDs_compare}. }
    \label{fig:flcdm_cuts}
\end{figure}

\item  {\bf Lower matter density:} One might think that the preference for a lower matter density than previous analyses is driven by the inclusion of the lowest redshift bins, and the apparent lower value of those bins relative to \pplus.\footnote{The magnitudes have not actually decreased on average for the supernovae in the lowest redshift bin, it is just that the average redshift of that bin has increased by $\Delta z=0.00022$, which corresponds to a $\Delta \mu^{\rm fid}=0.064$.  So it is the fiducial distance modulus that has changed, not the measured ones.}  However, as we show in (ii) our cosmology fits are relatively insensitive to low-redshift cuts. We find that the lower matter density is driven primarily by the SNe in the range $0.023 \leq z \leq 0.85$. Within this interval, the binned measurements exhibit an upward trend relative to $\mu^{\rm fid}$, which pulls the inferred matter density toward lower values. When excluding the $z<0.1$ SNe from Unite, we obtain $\om=0.290\pm0.015$ when fit to the Flat-$\Lambda$CDM model. We plot the corresponding best fit model in the top-right panel of Figure~\ref{fig:HDs_compare} (purple dashed line), which highlights the upward trend relative to $\mu^{\rm fid}$.

\end{enumerate}

\subsection{Specific Hubble diagram comparisons}
In the remainder of this section, we analyse the changes in distance moduli of individual SNe that appear in both Unite and the \pplus~and/or DES-Dovekie Hubble diagrams. 

Since we use SNe~Ia as relative distance indicators, any mean offset between the various Hubble diagrams is absorbed by a combination of the Hubble constant $H_0$ and the peak absolute magnitude $M_x$. Therefore, we follow the approach of \citet{vincenzi25} and consider a distance modulus offset between low-$z$ and high-$z$ parts of the diagram. This quantity is important because any such redshift-dependent change in the mean distance moduli would manifest as a change in best fit cosmology. 

We generalise the \citet{vincenzi25} definition to include all surveys by splitting the sample into high ($z>0.1$) and low-redshift ($z\leq0.1$) regimes rather than comparing two distinct surveys,
\begin{equation}
\Delta \mu_{\text {offset }}=\left\langle\mu_{\text {HD}}-\mu_{\text {Unite}}\right\rangle_{\text {high}-z}-
\left\langle\mu_{\text{HD}}-\mu_{\text{Unite}}\right\rangle_{\text {low}-z}
\end{equation}
where $\mu_{\rm HD}$ are the distances from the \pplus~or DES-Dovekie Hubble diagrams and $\left\langle\mu_{\text {HD}}-\mu_{\text {Unite}}\right\rangle$ is the mean difference between $\mu_{\rm HD}$ and Unite distance moduli, computed from the inverse-variance weighted average over the common high-$z$ or low$-z$ SNe.

\subsubsection{Unite and DES-Dovekie}
Unite and DES-Dovekie have 1555 common SNe (182 with $z<0.1$ and 1373 with $z>0.1)$. The high-$z$ subset of common SNe, consists exclusively of supernovae from DES, and the difference in the number of SNe is entirely driven by the probability cut ($P_{\rm Ia}>0.8$) applied in the Unite analysis. 

We observe a high degree of consistency between Unite and DES-Dovekie ($\Delta \mu_{\text {offset }}=-0.0031$ mag between the low-$z$ and high-$z$ regimes), as expected given the methodological similarity between the analyses. Motivated by the fact that \citet{vincenzi25} found different host-galaxy mass measurements could lead to $0.01$ mag shifts between the low-$z$ and high-$z$ distance moduli, and hence impact cosmological constraints, we consistently re-measured host-galaxy stellar masses for over 98\% of the Unite sample (as was done in DES-SN5YR). When reverting to the host-galaxy masses determined in DES-SN5YR and adopted in DES-Dovekie, the consistency between Unite and DES-Dovekie slightly increases ($\Delta \mu_{\text{offset}} = -0.0012$ mag).

\subsubsection{Unite and \pplus}
Unite and \pplus~ have 1375 (642 with $z<0.1$ and 733 with $z>0.1)$ common SNe. As expected, we find less consistency between Unite and \pplus~than DES-Dovekie, reflecting the cumulative impact of successive analysis improvements relative to this older work ($\Delta\mu_{\text {offset }}=-0.0199$). As noted above, in Unite, the host-galaxy stellar masses are uniformly re-measured across the full sample. If instead, we revert to the host-galaxy masses used in \pplus, the consistency increases ($\Delta \mu_{\text {offset }}=-0.0077$). Furthermore, \pplus~used the original BS21 dust parameters as the nominal intrinsic scatter model (see Section~\ref{sec:intrinsic_scatter}). When we additionally revert to these BS21 dust parameters (on top of adopting the original \pplus~host-galaxy masses), we find a further reduction in the offset, to $\Delta \mu_{\text {offset }}=0.0002$. 

While this residual offset is small, there remains variation between the analyses, and we have not reverted all of the analysis choices that differ between the samples (see Table~\ref{tab:analysis_changes}). A notable example is the adoption of the latest photometric calibration, Dovekie \citep{p25a}, in contrast to \pplus\ and DES-SN5YR, which use an earlier calibration presented in \cite{Brout_2022scal}.

\subsubsection{General comments on consistency}
\pplus, DES-SN5YR/DES-Dovekie, Union3, and Unite are all consistent with each other to well within $1\sigma$. 
When discussing differences in the supernova cosmology contours it is important to note that approximately half of the error bar in each of the modern analyses comes from systematic uncertainties.  These systematic uncertainties include model choices, and therefore different methods (such as Unity and BBC) are expected to result in different best fits, even on identical data.  It is standard in the field to acknowledge systematic errors by estimating their size and incorporating them in the error bars. Nevertheless it is important to acknowledge that systematic errors are not statistical.  Therefore using statistical techniques to estimate the significance of a deviation from $\Lambda$CDM is prone to over interpretation, and should be done with caution.  This is not isolated to supernova analyses, other techniques such as BAO and CMB also incorporate systematic uncertainties in their error bars (such as due to modelling choices).

\section{Conclusions}\label{sec:conclusion}
We present the Unite Hubble diagram, which is the combination of the \ppdes\ supernova samples (including the DES-Dovekie updates). Unite features consistent SN Ia modelling, sample selection,
and bias corrections.  We also rederived host-galaxy stellar masses for over 98\% of the sample. As a result, Unite represents the most comprehensive and internally consistent SN Ia dataset currently available with \sntot\ likely SNe~Ia.

The Unite Hubble diagram can be used in both anchored and unanchored forms, with both implementations provided in the public likelihood code. In this work, we present cosmological constraints from the unanchored Hubble diagram. In a companion paper, \citet[][in prep.]{lee25h0} uses the anchored Unite Hubble diagram to infer the sound horizon scale and $H_0$.

\subsection{Cosmological constraints}
We present cosmological constraints on the standard cosmological model, a spatially flat universe with cold dark matter and a cosmological constant, and three basic extensions using Unite alone and after combining Unite with external BAO and CMB datasets. These results are summarised in Table~\ref{tab:cosmo_results}.

\begin{itemize}
\item  In the standard cosmological model, Unite prefers a lower matter density than previous analyses. We discuss how this is driven by  the SNe in the range $0.023 \leq z \leq 0.85$ in Section~\ref{sec:comparisons}. 

\item When fitting Unite alone to a cosmological model assuming a constant dark energy equation of state ($w$), we obtain $w=\wfwcdmS$. This result lies approximately $\sim$$2\sigma$ from the cosmological constant value, $w=-1$, and in the opposite direction to the best fit value inferred by the CMB, which itself is inconsistent with the cosmological constant at $>1\sigma$ level. Consequently, we find a tension between CMB measurements and the combined supernova and BAO datasets.

\item Using Unite alone to constrain time-varying dark energy, parameterised as $w=w_0 + w_a(1-a)$, we find consistency with a dark energy scenario without time evolution, $w_a=0$, however the results remain approximately $\sim$$2\sigma$ from $w_0=-1$. Allowing $w$ to evolve with time alleviates the tension between the low-$z$ probes and the CMB data, and after combining Unite with the BAO and CMB datasets, we find a preference for $w_a<0$. 

\item According to model comparison statistics, Bayesian evidence indicates only weak preference for time evolving dark energy, whereas frequentist analyses yield a preference over Flat-$\Lambda$CDM at a significance of $\fwasig\sigma$ when using the maximum \textit{a posteriori} probability, or $3.1\sigma$ when using the maximum likelihood. 

\item 
Unite is also the first dataset to include as default a delensing term in the distance modulus estimate. This delensing is calculated based on the density of galaxies along the line of sight to the supernova, and reduces the scatter in the Hubble diagram by $\sim$$5$\% for $z>0.5$. In DES-SN5YR we found delensing pushed the cosmology constraints very slightly toward $\Lambda$CDM, but the opposite is true for Unite. Nonetheless, the difference between lensed and de-lensed best fits are small and shift along degeneracy directions (see Appendix~\ref{appendix3}).   
\end{itemize}

\subsection{Comparison between supernova datasets}
There has been much discussion in the literature about the pros and cons of the different supernova datasets. Our intention with this dataset is to satisfy the community’s desire for a single supernova dataset to use for cosmological fits. The Unite sample contains the largest number of supernovae, homogeneously analysed, and results in the tightest constraints of any dataset to date. The current analysis employs the BBC framework, while application of the alternative {\em UNITY} Bayesian Hierarchical Model will be presented in future work.

Some notable features compared to previous datasets are:
\begin{itemize}
\item  Unite contains all of the Pantheon+ surveys but modernises the \pplus\ analysis including all the improvements made by the Dovekie recalibration, and additionally putting all the host galaxies on the same mass scale.  
\item For applications that need broad sky coverage and/or many nearby supernovae, one advantage of \pplus\ over DES-SN5YR/DES-Dovekie was the inclusion of a much larger low-$z$ sample.  Unite reinstates this low-$z$ sample with improved distances and combines it with DES, enabling applications such as peculiar velocity analyses and $H_0$ measurements.

\item DES-Dovekie remains the cleanest sample that avoids some of the systematic uncertainties inherent in combining many datasets from different telescopes and filter systems, at the price of reduced precision. That was the original motivation for excluding some of the \pplus\ sample from the DES cosmology analysis. However, with the improvements in methodology since \pplus\ was analysed, we deemed it worth reincorporating them into the analysis.   
\item The Union3 sample uses most of the same supernovae as Pantheon+ and adds some extra surveys, but derives cosmological parameters using the `UNITY’ Bayesian Hierarchical Model rather than BBC.  Having two different pipelines is an important robustness check for our supernova models, and future work will compare UNITY and BBC applied to the Unite sample.  
\end{itemize}

\subsection{The low-redshift sample}
The low-$z$ sample is assembled from numerous surveys with differing instruments, calibrations, and selection functions.  It also has a strong impact on the cosmology fit both because it provides a strong lever arm on time variation and because it comes from the epoch of the Universe in which dark energy is most dominant.  We observe a noticeable shift in the best fit cosmology when excluding SNe below $z=0.1$ (discussed in Section~\ref{sec:general_obs}). This sensitivity underscores the importance of next-generation low-redshift samples for SN cosmology. 

Fortunately, the Asteroid Terrestrial-impact Last Alert System \citep{ATLASSURVEY, 2026ApJ..1004..173M}, the Zwicky Transient Factory \citep{Rigault25}, Young Supernova Experiment \citep{jones21, aleo23}, and Dark Energy Bedrock All-sky Supernova Survey  \citep[DEBASS;][]{Acevedo2026, Sherman2025} are all expected to produce new low-$z$ results in the near future. In particular, DEBASS would provide a valuable cross-check of our results, having measured more than $400$ supernovae at $z<0.08$ using the same telescope and filters as DES. Since the high-redshift regime of Unite is dominated by DES observations obtained with DECam, DEBASS would enable a fully homogeneous sample across low and high redshift. Therefore, there is opportunity for future work to either replace the current low-redshift or incorporate them directly into the Unite compilation.

\section*{Acknowledgements}
We thank Sesh Nadathur for a question that prompted us to revisit and clarify our MAP values. In the first version submitted to arXiv, we inadvertently used the MAP posterior, which includes a contribution from the prior volume, rather than the likelihood evaluated at the MAP.

RC is grateful for the support provided by the Big Questions Institute and its commitment to fostering fundamental research. RC, TMD and KG acknowledge support by the Australian Research Council Centre of Excellence for Gravitational Wave Discovery (project number CE230100016) funded by the Australian Government.

J.L. and K.B. are supported by the US Department of Energy grant DE‐SC0022950. AM is supported by Australian Research Council award (DE230100055). LG acknowledges financial support from CSIC, MCIN and AEI 10.13039/501100011033 under projects PID2023-151307NB-I00, PIE 20215AT016, CEX2020-001058-M, and by the MaX-CSIC Excellence Award MaX4-SOMMA-ICE. M.S. was partially supported by DOE grant DE-FOA-0003177.

\section*{Data Availability}
\newcommand{\URLpippin}{\url{https://github.com/dessn/Pippin}}
\newcommand{\URLSNANA}{\url{https://github.com/RickKessler/SNANA}}
\newcommand{\URLSNN}{\url{https://github.com/supernnova/SuperNNova}}
\newcommand{\URLSCONE}{\url{https://github.com/helenqu/scone}}
\newcommand{\URLSMP}{\url{https://smp???}}
\newcommand{\URLDUST}{\url{https://github.com/djbrout/dustdriver}}
\newcommand{\URLSALT}{\url{https://github.com/djones1040/SALTShaker}}
\newcommand{\URLCOSMOSIS}{\url{https://github.com/joezuntz/cosmosis}}
\newcommand{\URLSNDATAROOT}{\url{https://zenodo.org/records/4015325}}
The key data, code, and likelihoods along with the host-galaxy stellar mass measurements and photometry will be publicly available upon acceptance of the paper.
The Unite analysis was run using the \texttt{PIPPIN} pipeline framework
\citep{Hinton2020}\footnote{\URLpippin}
that incorporated \texttt{SNANA} \citep{kessler09}\footnote{\URLSNANA} codes for simulations, light curve
fitting, BBC, and covariance matrix computation.

\bibliographystyle{pasa-mnras}
\bibliography{mybib}

@ARTICLE{DES-SN5YR,
       author = {{DES Collaboration} and {Abbott}, T.~M.~C. and {Acevedo}, M. and {Aguena}, M. and {Alarcon}, A. and {Allam}, S. and {Alves}, O. and {Amon}, A. and {Andrade-Oliveira}, F. and {Annis}, J. and {Armstrong}, P. and {Asorey}, J. and {Avila}, S. and {Bacon}, D. and {Bassett}, B.~A. and {Bechtol}, K. and {Bernardinelli}, P.~H. and {Bernstein}, G.~M. and {Bertin}, E. and {Blazek}, J. and {Bocquet}, S. and {Brooks}, D. and {Brout}, D. and {Buckley-Geer}, E. and {Burke}, D.~L. and {Camacho}, H. and {Camilleri}, R. and {Campos}, A. and {Carnero Rosell}, A. and {Carollo}, D. and {Carr}, A. and {Carretero}, J. and {Castander}, F.~J. and {Cawthon}, R. and {Chang}, C. and {Chen}, R. and {Choi}, A. and {Conselice}, C. and {Costanzi}, M. and {da Costa}, L.~N. and {Crocce}, M. and {Davis}, T.~M. and {DePoy}, D.~L. and {Desai}, S. and {Diehl}, H.~T. and {Dixon}, M. and {Dodelson}, S. and {Doel}, P. and {Doux}, C. and {Drlica-Wagner}, A. and {Elvin-Poole}, J. and {Everett}, S. and {Ferrero}, I. and {Fert{\'e}}, A. and {Flaugher}, B. and {Foley}, R.~J. and {Fosalba}, P. and {Friedel}, D. and {Frieman}, J. and {Frohmaier}, C. and {Galbany}, L. and {Garc{\'\i}a-Bellido}, J. and {Gatti}, M. and {Gaztanaga}, E. and {Giannini}, G. and {Glazebrook}, K. and {Graur}, O. and {Gruen}, D. and {Gruendl}, R.~A. and {Gutierrez}, G. and {Hartley}, W.~G. and {Herner}, K. and {Hinton}, S.~R. and {Hollowood}, D.~L. and {Honscheid}, K. and {Huterer}, D. and {Jain}, B. and {James}, D.~J. and {Jeffrey}, N. and {Kasai}, E. and {Kelsey}, L. and {Kent}, S. and {Kessler}, R. and {Kim}, A.~G. and {Kirshner}, R.~P. and {Kovacs}, E. and {Kuehn}, K. and {Lahav}, O. and {Lee}, J. and {Lee}, S. and {Lewis}, G.~F. and {Li}, T.~S. and {Lidman}, C. and {Lin}, H. and {Malik}, U. and {Marshall}, J.~L. and {Martini}, P. and {Mena-Fern{\'a}ndez}, J. and {Menanteau}, F. and {Miquel}, R. and {Mohr}, J.~J. and {Mould}, J. and {Muir}, J. and {M{\"o}ller}, A. and {Neilsen}, E. and {Nichol}, R.~C. and {Nugent}, P. and {Ogando}, R.~L.~C. and {Palmese}, A. and {Pan}, Y.-C. and {Paterno}, M. and {Percival}, W.~J. and {Pereira}, M.~E.~S. and {Pieres}, A. and {Malag{\'o}n}, A.~A. Plazas and {Popovic}, B. and {Porredon}, A. and {Prat}, J. and {Qu}, H. and {Raveri}, M. and {Rodr{\'\i}guez-Monroy}, M. and {Romer}, A.~K. and {Roodman}, A. and {Rose}, B. and {Sako}, M. and {Sanchez}, E. and {Sanchez Cid}, D. and {Schubnell}, M. and {Scolnic}, D. and {Sevilla-Noarbe}, I. and {Shah}, P. and {Smith}, J. Allyn. and {Smith}, M. and {Soares-Santos}, M. and {Suchyta}, E. and {Sullivan}, M. and {Suntzeff}, N. and {Swanson}, M.~E.~C. and {S{\'a}nchez}, B.~O. and {Tarle}, G. and {Taylor}, G. and {Thomas}, D. and {To}, C. and {Toy}, M. and {Troxel}, M.~A. and {Tucker}, B.~E. and {Tucker}, D.~L. and {Uddin}, S.~A. and {Vincenzi}, M. and {Walker}, A.~R. and {Weaverdyck}, N. and {Wechsler}, R.~H. and {Weller}, J. and {Wester}, W. and {Wiseman}, P. and {Yamamoto}, M. and {Yuan}, F. and {Zhang}, B. and {Zhang}, Y.},
        title = "{The Dark Energy Survey: Cosmology Results with {\ensuremath{\sim}}1500 New High-redshift Type Ia Supernovae Using the Full 5 yr Data Set}",
      journal = {\apjl},
         year = 2024,
        month = sep,
       volume = {973},
       number = {1},
          eid = {L14},
        pages = {L14},
          doi = {10.3847/2041-8213/ad6f9f},
archivePrefix = {arXiv},
       eprint = {2401.02929},
 primaryClass = {astro-ph.CO},
       adsurl = {https://ui.adsabs.harvard.edu/abs/2024ApJ...973L..14D}
}

@ARTICLE{Acevedo2026,
       author = {{Acevedo}, Maria and {Sherman}, Nora F. and {Brout}, Dillon and {Carreres}, Bastien and {Scolnic}, Daniel and {Popovic}, Brodie and {Armstrong}, Patrick and {Cao}, Dingyuan and {Chen}, Rebecca C. and {Drlica-Wagner}, Alex and {Ferguson}, Peter S. and {Lidman}, Christopher and {Martin}, Bailey and {Peterson}, Erik R. and {Riess}, Adam G.},
        title = "{The Dark Energy Bedrock All-sky Supernova Program: Cross Calibration, Simulations, and Cosmology Forecasts}",
      journal = {\apj},
         year = 2026,
        month = jan,
       volume = {996},
       number = {1},
          eid = {7},
        pages = {7},
          doi = {10.3847/1538-4357/ae1e78},
archivePrefix = {arXiv},
       eprint = {2508.10877},
 primaryClass = {astro-ph.CO},
       adsurl = {https://ui.adsabs.harvard.edu/abs/2026ApJ...996....7A}
}

@ARTICLE{Kessler2009SDSS,
       author = {{Kessler}, Richard and {Becker}, Andrew C. and {Cinabro}, David and {Vanderplas}, Jake and {Frieman}, Joshua A. and {Marriner}, John and {Davis}, Tamara M. and {Dilday}, Benjamin and {Holtzman}, Jon and {Jha}, Saurabh W. and {Lampeitl}, Hubert and {Sako}, Masao and {Smith}, Mathew and {Zheng}, Chen and {Nichol}, Robert C. and {Bassett}, Bruce and {Bender}, Ralf and {Depoy}, Darren L. and {Doi}, Mamoru and {Elson}, Ed and {Filippenko}, Alexei V. and {Foley}, Ryan J. and {Garnavich}, Peter M. and {Hopp}, Ulrich and {Ihara}, Yutaka and {Ketzeback}, William and {Kollatschny}, W. and {Konishi}, Kohki and {Marshall}, Jennifer L. and {McMillan}, Russet J. and {Miknaitis}, Gajus and {Morokuma}, Tomoki and {M{\"o}rtsell}, Edvard and {Pan}, Kaike and {Prieto}, Jose Luis and {Richmond}, Michael W. and {Riess}, Adam G. and {Romani}, Roger and {Schneider}, Donald P. and {Sollerman}, Jesper and {Takanashi}, Naohiro and {Tokita}, Kouichi and {van der Heyden}, Kurt and {Wheeler}, J.~C. and {Yasuda}, Naoki and {York}, Donald},
        title = "{First-Year Sloan Digital Sky Survey-II Supernova Results: Hubble Diagram and Cosmological Parameters}",
      journal = {\apjs},
         year = 2009,
        month = nov,
       volume = {185},
       number = {1},
        pages = {32-84},
          doi = {10.1088/0067-0049/185/1/32},
archivePrefix = {arXiv},
       eprint = {0908.4274},
 primaryClass = {astro-ph.CO},
       adsurl = {https://ui.adsabs.harvard.edu/abs/2009ApJS..185...32K}
}

@ARTICLE{Sherman2025,
       author = {{Sherman}, Nora F. and {Acevedo}, Maria and {Brout}, Dillon and {Martin}, Bailey and {Scolnic}, Daniel and {Cao}, Dingyuan and {Lidman}, Christopher and {Ali}, Noor and {Armstrong}, Patrick and {Auchettl}, K. and {Chen}, Rebecca C. and {Drlica-Wagner}, Alex and {Ferguson}, Peter S. and {Herner}, Kenneth and {Narayan}, Gautham and {Peterson}, Erik R. and {Rauf}, Liana and {Rest}, Armin and {Riess}, Adam G. and {Sako}, Masao and {Schmidt}, Brian and {Tang}, Xianzhe TZ and {Tucker}, Brad E.},
        title = "{The Dark Energy Bedrock All-sky Supernova Program: Motivation, Design, Implementation, and Preliminary Data Release}",
      journal = {\apj},
         year = 2026,
        month = may,
       volume = {1002},
       number = {2},
          eid = {146},
        pages = {146},
          doi = {10.3847/1538-4357/ae48ff},
archivePrefix = {arXiv},
       eprint = {2508.10878},
 primaryClass = {astro-ph.CO},
       adsurl = {https://ui.adsabs.harvard.edu/abs/2026ApJ..1002..146S}
}

@article{Goliath_2001,
   title={Supernovae and the nature of the dark energy},
   volume={380},
   ISSN={1432-0746},
   url={http://dx.doi.org/10.1051/0004-6361:20011398},
   DOI={10.1051/0004-6361:20011398},
   number={1},
   journal={\aap},
   publisher={EDP Sciences},
   author={Goliath, M. and Amanullah, R. and Astier, P. and Goobar, A. and Pain, R.},
   year={2001},
   month={Dec},
   pages={6–18}
}

@article{Riess_1998,
   title={Observational Evidence from Supernovae for an Accelerating Universe and a Cosmological Constant},
   volume={116},
   ISSN={0004-6256},
   url={http://dx.doi.org/10.1086/300499},
   DOI={10.1086/300499},
   number={3},
   journal = {\aj},
   publisher={American Astronomical Society},
   author={Riess, Adam G. and Filippenko, Alexei V. and Challis, Peter and Clocchiatti, Alejandro and Diercks, Alan and Garnavich, Peter M. and Gilliland, Ron L. and Hogan, Craig J. and Jha, Saurabh and Kirshner, Robert P. and et al.},
   year={1998},
   month={Sep},
   pages={1009–1038}
}

@article{Perlmutter_1999,
   title={Measurements of $\Omega$ and $\Lambda$ from 42 High‐Redshift Supernovae},
   volume={517},
   ISSN={1538-4357},
   url={http://dx.doi.org/10.1086/307221},
   DOI={10.1086/307221},
   number={2},
   journal = {\apj},
   publisher={American Astronomical Society},
   author={Perlmutter, S. and Aldering, G. and Goldhaber, G. and Knop, R. A. and Nugent, P. and Castro, P. G. and Deustua, S. and Fabbro, S. and Goobar, A. and Groom, D. E. and et al.},
   year={1999},
   month={Jun},
   pages={565–586}
}

@article{Riess_2007,
   title={NewHubble Space TelescopeDiscoveries of Type Ia Supernovae atz≥ 1: Narrowing Constraints on the Early Behavior of Dark Energy},
   volume={659},
   ISSN={1538-4357},
   url={http://dx.doi.org/10.1086/510378},
   DOI={10.1086/510378},
   number={1},
   journal = {\apj},
   publisher={American Astronomical Society},
   author={Riess, Adam G. and Strolger, Louis‐Gregory and Casertano, Stefano and Ferguson, Henry C. and Mobasher, Bahram and Gold, Ben and Challis, Peter J. and Filippenko, Alexei V. and Jha, Saurabh and Li, Weidong and et al.},
   year={2007},
   month={Apr},
   pages={98–121}
}

@article{scolnic2021pantheon,
doi = {10.3847/1538-4357/ac8b7a},
url = {https://dx.doi.org/10.3847/1538-4357/ac8b7a},
year = {2022},
month = {oct},
publisher = {The American Astronomical Society},
volume = {938},
number = {2},
pages = {113},
author = {Dan Scolnic and Dillon Brout and Anthony Carr and Adam G. Riess and Tamara M. Davis and Arianna Dwomoh and David O. Jones and Noor Ali and Pranav Charvu and Rebecca Chen and Erik R. Peterson and Brodie Popovic and Benjamin M. Rose and Charlotte M. Wood and Peter J. Brown and Ken Chambers and David A. Coulter and Kyle G. Dettman and Georgios Dimitriadis and Alexei V. Filippenko and Ryan J. Foley and Saurabh W. Jha and Charles D. Kilpatrick and Robert P. Kirshner and Yen-Chen Pan and Armin Rest and Cesar Rojas-Bravo and Matthew R. Siebert and Benjamin E. Stahl and WeiKang Zheng},
title = {The Pantheon+ Analysis: The Full Data Set and Light-curve Release},
journal = {\apj}
}

@article{2014JLA,
   title={Improved cosmological constraints from a joint analysis of the SDSS-II and SNLS supernova samples},
   volume={568},
   ISSN={1432-0746},
   url={http://dx.doi.org/10.1051/0004-6361/201423413},
   DOI={10.1051/0004-6361/201423413},
   journal={\aap},
   publisher={EDP Sciences},
   author={Betoule, M. and Kessler, R. and Guy, J. and Mosher, J. and Hardin, D. and Biswas, R. and Astier, P. and El-Hage, P. and Konig, M. and Kuhlmann, S. and Marriner, J. and Pain, R. and Regnault, N. and et al.},
   year={2014},
   month={Aug},
   pages={A22} }

@article{Brout_2019,
	doi = {10.3847/1538-4357/ab06c1},
	url = {https://doi.org/10.3847/1538-4357/ab06c1},
	year = 2019,
	month = {mar},
	publisher = {American Astronomical Society},
	volume = {874},
	number = {1},
	pages = {106},
	author = {D. Brout and M. Sako and D. Scolnic and R. Kessler and C. B. D'Andrea and T. M. Davis and S. R. Hinton and A. G. Kim and J. Lasker and E. Macaulay and A. Möller and R. C. Nichol and M. Smith and M. Sullivan and R. C. Wolf and S. Allam and B. A. Bassett and P. Brown and F. J. Castander and M. Childress and R. J. Foley and L. Galbany and K. Herner and E. Kasai and M. March and E. Morganson and P. Nugent and Y.-C. Pan and R. C. Thomas and B. E. Tucker and W. Wester and T. M. C. Abbott and J. Annis and S. Avila and E. Bertin and D. Brooks and D. L. Burke and A. Carnero Rosell and M. Carrasco Kind and J. Carretero and M. Crocce and C. E. Cunha and L. N. da Costa and C. Davis and J. De Vicente and S. Desai and H. T. Diehl and P. Doel and T. F. Eifler and B. Flaugher and P. Fosalba and J. Frieman and J. Garc{\'{\i}}a-Bellido and E. Gaztanaga and D. W. Gerdes and D. A. Goldstein and D. Gruen and R. A. Gruendl and J. Gschwend and G. Gutierrez and W. G. Hartley and D. L. Hollowood and K. Honscheid and D. J. James and K. Kuehn and N. Kuropatkin and O. Lahav and T. S. Li and M. Lima and J. L. Marshall and P. Martini and R. Miquel and B. Nord and A. A. Plazas and A. Roodman and E. S. Rykoff and E. Sanchez and V. Scarpine and R. Schindler and M. Schubnell and S. Serrano and I. Sevilla-Noarbe and M. Soares-Santos and F. Sobreira and E. Suchyta and M. E. C. Swanson and G. Tarle and D. Thomas and D. L. Tucker and A. R. Walker and B. Yanny and Y. Zhang and},
	title = {First Cosmology Results Using Type Ia Supernovae from the Dark Energy Survey: Photometric Pipeline and Light-curve Data Release},
	journal = {\apj}
}

@ARTICLE{guy_2010,
       author = {{Guy}, J. and {Sullivan}, M. and {Conley}, A. and {Regnault}, N. and {Astier}, P. and {Balland}, C. and {Basa}, S. and {Carlberg}, R.~G. and {Fouchez}, D. and {Hardin}, D. and {Hook}, I.~M. and {Howell}, D.~A. and {Pain}, R. and {Palanque-Delabrouille}, N. and {Perrett}, K.~M. and {Pritchet}, C.~J. and {Rich}, J. and {Ruhlmann-Kleider}, V. and {Balam}, D. and {Baumont}, S. and {Ellis}, R.~S. and {Fabbro}, S. and {Fakhouri}, H.~K. and {Fourmanoit}, N. and {Gonz{\'a}lez-Gait{\'a}n}, S. and {Graham}, M.~L. and {Hsiao}, E. and {Kronborg}, T. and {Lidman}, C. and {Mourao}, A.~M. and {Perlmutter}, S. and {Ripoche}, P. and {Suzuki}, N. and {Walker}, E.~S.},
        title = "{The Supernova Legacy Survey 3-year sample: Type Ia supernovae photometric distances and cosmological constraints}",
      journal = {\aap},
         year = 2010,
        month = nov,
       volume = {523},
          eid = {A7},
        pages = {A7},
          doi = {10.1051/0004-6361/201014468},
archivePrefix = {arXiv},
       eprint = {1010.4743},
 primaryClass = {astro-ph.CO},
       adsurl = {https://ui.adsabs.harvard.edu/abs/2010A&A...523A...7G}
}

@article{2022_pantheon_analysis,
doi = {10.3847/1538-4357/ac8e04},
url = {https://dx.doi.org/10.3847/1538-4357/ac8e04},
year = {2022},
month = {oct},
publisher = {The American Astronomical Society},
volume = {938},
number = {2},
pages = {110},
author = {Dillon Brout and Dan Scolnic and Brodie Popovic and Adam G. Riess and Anthony Carr and Joe Zuntz and Rick Kessler and Tamara M. Davis and Samuel Hinton and David Jones and W. D’Arcy Kenworthy and Erik R. Peterson and Khaled Said and Georgie Taylor and Noor Ali and Patrick Armstrong and Pranav Charvu and Arianna Dwomoh and Cole Meldorf and Antonella Palmese and Helen Qu and Benjamin M. Rose and Bruno Sanchez and Christopher W. Stubbs and Maria Vincenzi and Charlotte M. Wood and Peter J. Brown and Rebecca Chen and Ken Chambers and David A. Coulter and Mi Dai and Georgios Dimitriadis and Alexei V. Filippenko and Ryan J. Foley and Saurabh W. Jha and Lisa Kelsey and Robert P. Kirshner and Anais Möller and Jessie Muir and Seshadri Nadathur and Yen-Chen Pan and Armin Rest and Cesar Rojas-Bravo and Masao Sako and Matthew R. Siebert and Mat Smith and Benjamin E. Stahl and Phil Wiseman},
title = {The Pantheon+ Analysis: Cosmological Constraints},
journal = {\apj}
}

@article{P21_dust2dust,
doi = {10.3847/1538-4357/aca273},
url = {https://dx.doi.org/10.3847/1538-4357/aca273},
year = {2023},
month = {mar},
publisher = {The American Astronomical Society},
volume = {945},
number = {1},
pages = {84},
author = {Brodie Popovic and Dillon Brout and Richard Kessler and Daniel Scolnic},
title = {The Pantheon+ Analysis: Forward Modeling the Dust and Intrinsic Color Distributions of Type Ia Supernovae, and Quantifying Their Impact on Cosmological Inferences},
journal = {\apj}
}

@article{2020_planck,
   title={Planck 2018 results},
   volume={641},
   ISSN={1432-0746},
   url={http://dx.doi.org/10.1051/0004-6361/201833910},
   DOI={10.1051/0004-6361/201833910},
   journal={\aap},
   publisher={EDP Sciences},
   author={{Planck Collaboration}, Aghanim, N. and Akrami, Y. and Ashdown, M. and Aumont, J. and Baccigalupi, C. and Ballardini, M. and Banday, A. J. and Barreiro, R. B. and Bartolo, N. and Basak, S. and Battye, R. and Benabed, K. and Bernard, J.-P. and Bersanelli, M. and Bielewicz, P. and Bock, J. J. and Bond, J. R. and Borrill, J. and Bouchet, F. R. and Boulanger, F. and Bucher, M. and Burigana, C. and Butler, R. C. and Calabrese, E. and Cardoso, J.-F. and Carron, J. and Challinor, A. and Chiang, H. C. and Chluba, J. and Colombo, L. P. L. and Combet, C. and Contreras, D. and Crill, B. P. and Cuttaia, F. and de Bernardis, P. and de Zotti, G. and Delabrouille, J. and Delouis, J.-M. and Di Valentino, E. and Diego, J. M. and Doré, O. and Douspis, M. and Ducout, A. and Dupac, X. and Dusini, S. and Efstathiou, G. and Elsner, F. and Enßlin, T. A. and Eriksen, H. K. and Fantaye, Y. and Farhang, M. and Fergusson, J. and Fernandez-Cobos, R. and Finelli, F. and Forastieri, F. and Frailis, M. and Fraisse, A. A. and Franceschi, E. and Frolov, A. and Galeotta, S. and Galli, S. and Ganga, K. and Génova-Santos, R. T. and Gerbino, M. and Ghosh, T. and González-Nuevo, J. and Górski, K. M. and Gratton, S. and Gruppuso, A. and Gudmundsson, J. E. and Hamann, J. and Handley, W. and Hansen, F. K. and Herranz, D. and Hildebrandt, S. R. and Hivon, E. and Huang, Z. and Jaffe, A. H. and Jones, W. C. and Karakci, A. and Keihänen, E. and Keskitalo, R. and Kiiveri, K. and Kim, J. and Kisner, T. S. and Knox, L. and Krachmalnicoff, N. and Kunz, M. and Kurki-Suonio, H. and Lagache, G. and Lamarre, J.-M. and Lasenby, A. and Lattanzi, M. and Lawrence, C. R. and Le Jeune, M. and Lemos, P. and Lesgourgues, J. and Levrier, F. and Lewis, A. and Liguori, M. and Lilje, P. B. and Lilley, M. and Lindholm, V. and López-Caniego, M. and Lubin, P. M. and Ma, Y.-Z. and Macías-Pérez, J. F. and Maggio, G. and Maino, D. and Mandolesi, N. and Mangilli, A. and Marcos-Caballero, A. and Maris, M. and Martin, P. G. and Martinelli, M. and Martínez-González, E. and Matarrese, S. and Mauri, N. and McEwen, J. D. and Meinhold, P. R. and Melchiorri, A. and Mennella, A. and Migliaccio, M. and Millea, M. and Mitra, S. and Miville-Deschênes, M.-A. and Molinari, D. and Montier, L. and Morgante, G. and Moss, A. and Natoli, P. and Nørgaard-Nielsen, H. U. and Pagano, L. and Paoletti, D. and Partridge, B. and Patanchon, G. and Peiris, H. V. and Perrotta, F. and Pettorino, V. and Piacentini, F. and Polastri, L. and Polenta, G. and Puget, J.-L. and Rachen, J. P. and Reinecke, M. and Remazeilles, M. and Renzi, A. and Rocha, G. and Rosset, C. and Roudier, G. and Rubiño-Martín, J. A. and Ruiz-Granados, B. and Salvati, L. and Sandri, M. and Savelainen, M. and Scott, D. and Shellard, E. P. S. and Sirignano, C. and Sirri, G. and Spencer, L. D. and Sunyaev, R. and Suur-Uski, A.-S. and Tauber, J. A. and Tavagnacco, D. and Tenti, M. and Toffolatti, L. and Tomasi, M. and Trombetti, T. and Valenziano, L. and Valiviita, J. and Van Tent, B. and Vibert, L. and Vielva, P. and Villa, F. and Vittorio, N. and Wandelt, B. D. and Wehus, I. K. and White, M. and White, S. D. M. and Zacchei, A. and Zonca, A.},
   year={2020},
   month={Sep},
   pages={A6} }

@article{2019_DESbias,
   title={First cosmology results using Type Ia supernova from the Dark Energy Survey: simulations to correct supernova distance biases},
   volume={485},
   ISSN={1365-2966},
   url={http://dx.doi.org/10.1093/mnras/stz463},
   DOI={10.1093/mnras/stz463},
   number={1},
   journal = {\mnras},
   publisher={Oxford University Press (OUP)},
   author={Kessler, R and Brout, D and D’Andrea, C B and Davis, T M and Hinton, S R and Kim, A G and Lasker, J and Lidman, C and Macaulay, E and Möller, A and Sako, M and Scolnic, D and Smith, M and Sullivan, M and Zhang, B and Andersen, P and Asorey, J and Avelino, A and Calcino, J and Carollo, D and Challis, P and Childress, M and Clocchiatti, A and Crawford, S and Filippenko, A V and Foley, R J and Glazebrook, K and Hoormann, J K and Kasai, E and Kirshner, R P and Lewis, G F and Mandel, K S and March, M and Morganson, E and Muthukrishna, D and Nugent, P and Pan, Y-C and Sommer, N E and Swann, E and Thomas, R C and Tucker, B E and Uddin, S A and Abbott, T M C and Allam, S and Annis, J and Avila, S and Banerji, M and Bechtol, K and Bertin, E and Brooks, D and Buckley-Geer, E and Burke, D L and Carnero Rosell, A and Carrasco Kind, M and Carretero, J and Castander, F J and Crocce, M and da Costa, L N and Davis, C and De Vicente, J and Desai, S and Diehl, H T and Doel, P and Eifler, T F and Flaugher, B and Fosalba, P and Frieman, J and García-Bellido, J and Gaztanaga, E and Gerdes, D W and Gruen, D and Gruendl, R A and Gutierrez, G and Hartley, W G and Hollowood, D L and Honscheid, K and James, D J and Johnson, M W G and Johnson, M D and Krause, E and Kuehn, K and Kuropatkin, N and Lahav, O and Li, T S and Lima, M and Marshall, J L and Martini, P and Menanteau, F and Miller, C J and Miquel, R and Nord, B and Plazas, A A and Roodman, A and Sanchez, E and Scarpine, V and Schindler, R and Schubnell, M and Serrano, S and Sevilla-Noarbe, I and Soares-Santos, M and Sobreira, F and Suchyta, E and Tarle, G and Thomas, D and Walker, A R and Zhang, Y},
   year={2019},
   month={Feb},
   pages={1171–1187} }

@article{guy_2007,
   title={SALT2: using distant supernovae to improve the use   of type Ia 
supernovae as distance indicators},
   volume={466},
   ISSN={1432-0746},
   url={http://dx.doi.org/10.1051/0004-6361:20066930},
   DOI={10.1051/0004-6361:20066930},
   number={1},
   journal={\aap},
   publisher={EDP Sciences},
   author={Guy, J. and Astier, P. and Baumont, S. and Hardin, D. and Pain, R. and Regnault, N. and Basa, S. and Carlberg, R. G. and Conley, A. and Fabbro, S. and Fouchez, D. and Hook, I. M. and Howell, D. A. and Perrett, K. and Pritchet, C. J. and Rich, J. and Sullivan, M. and Antilogus, P. and Aubourg, E. and Bazin, G. and Bronder, J. and Filiol, M. and Palanque-Delabrouille, N. and Ripoche, P. and Ruhlmann-Kleider, V.},
   year={2007},
   month={Feb},
   pages={11–21} }

@article{1998_tripp,
doi = {10.1086/307883},
url = {https://dx.doi.org/10.1086/307883},
year = {1999},
month = {nov},
publisher = {},
volume = {525},
number = {1},
pages = {209},
author = {Robert Tripp and David Branch},
title = {Determination of the Hubble Constant Using a Two-Parameter Luminosity Correction for Type Ia Supernovae},
journal = {\aj}
}

@article{Vincenzi_2021_,
	doi = {10.1093/mnras/stab1353},
  
	url = {https://doi.org/10.1093%2Fmnras%2Fstab1353},
  
	year = 2021,
	month = {may},
  
	publisher = {Oxford University Press ({OUP})},
  
	volume = {505},
  
	number = {2},
  
	pages = {2819--2839},
  
	author = {M Vincenzi and M Sullivan and O Graur and D Brout and T M Davis and C Frohmaier and L Galbany and C P Guti{\'{e}
}rrez and S R Hinton and R Hounsell and L Kelsey and R Kessler and E Kovacs and S Kuhlmann and J Lasker and C Lidman and A Möller and R C Nichol and M Sako and D Scolnic and M Smith and E Swann and P Wiseman and J Asorey and G F Lewis and R Sharp and B E Tucker and M Aguena and S Allam and S Avila and E Bertin and D Brooks and D L Burke and A Carnero~Rosell and M Carrasco~Kind and J Carretero and F J Castander and A Choi and M Costanzi and L N da~Costa and M E S Pereira and J De~Vicente and S Desai and H T Diehl and P Doel and S Everett and I Ferrero and P Fosalba and J Frieman and J Garc{\'{\i}}a-Bellido and E Gaztanaga and D W Gerdes and D Gruen and R A Gruendl and G Gutierrez and D L Hollowood and K Honscheid and B Hoyle and D J James and K Kuehn and N Kuropatkin and M A G Maia and P Martini and F Menanteau and R Miquel and R Morgan and A Palmese and F Paz-Chinch{\'{o}}n and A A Plazas and A K Romer and E Sanchez and V Scarpine and S Serrano and I Sevilla-Noarbe and M Soares-Santos and E Suchyta and G Tarle and D Thomas and C To and T N Varga and A R Walker and R D Wilkinson and  (DES~Collaboration)},
  
	title = {The Dark Energy Survey supernova programme: modelling selection efficiency and observed core-collapse supernova contamination},
  
	journal = {\mnras}
}

@article{2017_BiasCor,
   title={Correcting Type Ia Supernova Distances for Selection Biases and Contamination in Photometrically Identified Samples},
   volume={836},
   ISSN={1538-4357},
   url={http://dx.doi.org/10.3847/1538-4357/836/1/56},
   DOI={10.3847/1538-4357/836/1/56},
   number={1},
   journal = {\apj},
   publisher={American Astronomical Society},
   author={Kessler, R. and Scolnic, D.},
   year={2017},
   month={Feb},
   pages={56} }

@article{Hinton2020,
  doi = {10.21105/joss.02122},
  url = {https://doi.org/10.21105/joss.02122},
  year = {2020},
  publisher = {The Open Journal},
  volume = {5},
  number = {47},
  pages = {2122},
  author = {Samuel Hinton and Dillon Brout},
  title = {Pippin: A pipeline for supernova cosmology},
  journal={J. Open Source Softw.}
}

@ARTICLE{2018_Foley_foundation,
       author = {{Foley}, Ryan J. and {Scolnic}, Daniel and {Rest}, Armin and {Jha}, S.~W. and {Pan}, Y. -C. and {Riess}, A.~G. and {Challis}, P. and {Chambers}, K.~C. and {Coulter}, D.~A. and {Dettman}, K.~G. and {Foley}, M.~M. and {Fox}, O.~D. and {Huber}, M.~E. and {Jones}, D.~O. and {Kilpatrick}, C.~D. and {Kirshner}, R.~P. and {Schultz}, A.~S.~B. and {Siebert}, M.~R. and {Flewelling}, H.~A. and {Gibson}, B. and {Magnier}, E.~A. and {Miller}, J.~A. and {Primak}, N. and {Smartt}, S.~J. and {Smith}, K.~W. and {Wainscoat}, R.~J. and {Waters}, C. and {Willman}, M.},
        title = "{The Foundation Supernova Survey: motivation, design, implementation, and first data release}",
      journal = {\mnras},
         year = 2018,
        month = mar,
       volume = {475},
       number = {1},
        pages = {193-219},
          doi = {10.1093/mnras/stx3136},
archivePrefix = {arXiv},
       eprint = {1711.02474},
 primaryClass = {astro-ph.HE},
       adsurl = {https://ui.adsabs.harvard.edu/abs/2018MNRAS.475..193F}
}

@article{maria_2021,
    author = {Vincenzi, M and Sullivan, M and Möller, A and Armstrong, P and Bassett, B A and Brout, D and Carollo, D and Carr, A and Davis, T M and Frohmaier, C and Galbany, L and Glazebrook, K and Graur, O and Kelsey, L and Kessler, R and Kovacs, E and Lewis, G F and Lidman, C and Malik, U and Nichol, R C and Popovic, B and Sako, M and Scolnic, D and Smith, M and Taylor, G and Tucker, B E and Wiseman, P and Aguena, M and Allam, S and Annis, J and Asorey, J and Bacon, D and Bertin, E and Brooks, D and Burke, D L and Rosell, A Carnero and Carretero, J and Castander, F J and Costanzi, M and da Costa, L N and Pereira, M E S and De Vicente, J and Desai, S and Diehl, H T and Doel, P and Everett, S and Ferrero, I and Flaugher, B and Fosalba, P and Frieman, J and García-Bellido, J and Gerdes, D W and Gruen, D and Gutierrez, G and Hinton, S R and Hollowood, D L and Honscheid, K and James, D J and Kuehn, K and Kuropatkin, N and Lahav, O and Li, T S and Lima, M and Maia, M A G and Marshall, J L and Miquel, R and Morgan, R and Ogando, R L C and Palmese, A and Paz-Chinchón, F and Pieres, A and Malagón, A A Plazas and Reil, K and Roodman, A and Sanchez, E and Schubnell, M and Serrano, S and Sevilla-Noarbe, I and Suchyta, E and Tarle, G and To, C and Varga, T N and Weller, J and Wilkinson, R D and (DES Collaboration)},
    title = "{The Dark Energy Survey supernova program: cosmological biases from supernova photometric classification}",
    journal = {\mnras},
    volume = {518},
    number = {1},
    pages = {1106-1127},
    year = {2022},
    month = {06},
    issn = {0035-8711},
    doi = {10.1093/mnras/stac1404},
    url = {https://doi.org/10.1093/mnras/stac1404},
    eprint = {https://academic.oup.com/mnras/article-pdf/518/1/1106/47161313/stac1404.pdf},
}

@article{Smith_2020,
	doi = {10.3847/1538-3881/abc01b},
  
	url = {https://doi.org/10.3847\%2F1538-3881\%2Fabc01b},
  
	year = 2020,
	month = {nov},
  
	publisher = {American Astronomical Society},
  
	volume = {160},
  
	number = {6},
  
	pages = {267},
  
	author = {M. Smith and C. B. D'Andrea and M. Sullivan and A. Möller and R. C. Nichol and R. C. Thomas and A. G. Kim and M. Sako and F. J. Castander and A. V. Filippenko and R. J. Foley and L. Galbany and S. Gonz{\'{a}
}lez-Gait{\'{a}}n and E. Kasai and R. P. Kirshner and C. Lidman and D. Scolnic and D. Brout and T. M. Davis and R. R. Gupta and S. R. Hinton and R. Kessler and J. Lasker and E. Macaulay and R. C. Wolf and B. Zhang and J. Asorey and A. Avelino and B. A. Bassett and J. Calcino and D. Carollo and R. Casas and P. Challis and M. Childress and A. Clocchiatti and S. Crawford and C. Frohmaier and K. Glazebrook and D. A. Goldstein and M. L. Graham and J. K. Hoormann and K. Kuehn and G. F. Lewis and K. S. Mandel and E. Morganson and D. Muthukrishna and P. Nugent and Y.-C. Pan and M. Pursiainen and R. Sharp and N. E. Sommer and E. Swann and B. P. Thomas and B. E. Tucker and S. A. Uddin and P. Wiseman and W. Zheng and T. M. C. Abbott and J. Annis and S. Avila and K. Bechtol and G. M. Bernstein and E. Bertin and D. Brooks and D. L. Burke and A. Carnero Rosell and M. Carrasco Kind and J. Carretero and C. E. Cunha and L. N. da Costa and C. Davis and J. De Vicente and H. T. Diehl and T. F. Eifler and J. Estrada and J. Frieman and J. Garc{\'{\i}}a-Bellido and E. Gaztanaga and D. W. Gerdes and D. Gruen and R. A. Gruendl and J. Gschwend and G. Gutierrez and W. G. Hartley and D. L. Hollowood and K. Honscheid and B. Hoyle and D. J. James and M. W. G. Johnson and M. D. Johnson and N. Kuropatkin and T. S. Li and M. Lima and M. A. G. Maia and M. March and J. L. Marshall and P. Martini and F. Menanteau and C. J. Miller and R. Miquel and E. Neilsen and R. L. C. Ogando and A. A. Plazas and A. K. Romer and E. Sanchez and V. Scarpine and M. Schubnell and S. Serrano and I. Sevilla-Noarbe and M. Soares-Santos and F. Sobreira and E. Suchyta and G. Tarle and D. L. Tucker and W. Wester},
  
	title = {First Cosmology Results using Supernovae Ia from the Dark Energy Survey: Survey Overview, Performance, and Supernova Spectroscopy},
  
	journal = {\aj}
}

@article{Kenworthy_2021,
	doi = {10.3847/1538-4357/ac30d8},
	url = {https://doi.org/10.3847/1538-4357/ac30d8},
	year = 2021,
	month = {dec},
	publisher = {American Astronomical Society},
	volume = {923},
	number = {2},
	pages = {265},
	author = {W. D. Kenworthy and D. O. Jones and M. Dai and R. Kessler and D. Scolnic and D. Brout and M. R. Siebert and J. D. R. Pierel and K. G. Dettman and G. Dimitriadis and R. J. Foley and S. W. Jha and Y.-C. Pan and A. Riess and S. Rodney and C. Rojas-Bravo},
	title = {{SALT}3: An Improved Type Ia Supernova Model for Measuring Cosmic Distances},
	journal = {\apj}
}

@article{kunz07,
  title = {Bayesian estimation applied to multiple species},
  author = {Kunz, Martin and Bassett, Bruce A. and Hlozek, Ren\'ee A.},
  journal = {\prd},
  volume = {75},
  issue = {10},
  pages = {103508},
  numpages = {12},
  year = {2007},
  month = {May},
  publisher = {American Physical Society},
  doi = {10.1103/PhysRevD.75.103508},
  url = {https://link.aps.org/doi/10.1103/PhysRevD.75.103508}
}

@article{Hlozek_2012,
doi = {10.1088/0004-637X/752/2/79},
url = {https://dx.doi.org/10.1088/0004-637X/752/2/79},
year = {2012},
month = {may},
publisher = {The American Astronomical Society},
volume = {752},
number = {2},
pages = {79},
author = {Renée Hlozek and Martin Kunz and Bruce Bassett and Mat Smith and James Newling and Melvin Varughese and Rick Kessler and Joseph P. Bernstein and Heather Campbell and Ben Dilday and Bridget Falck and Joshua Frieman and Steve Kuhlmann and Hubert Lampeitl and John Marriner and Robert C. Nichol and Adam G. Riess and Masao Sako and Donald P. Schneider},
title = {PHOTOMETRIC SUPERNOVA COSMOLOGY WITH BEAMS AND SDSS-II},
journal = {\apj}
}

@INPROCEEDINGS{Kunz2013,
       author = {{Kunz}, Martin and {Hlozek}, Ren{\'e}e and {Bassett}, Bruce A. and {Smith}, Mathew and {Newling}, James and {Varughese}, Melvin},
        title = "{BEAMS: Separating the Wheat from the Chaff in Supernova Analysis}",
    booktitle = {Astrostatistical Challenges for the New Astronomy},
         year = 2013,
       editor = {{Hilbe}, Joseph M.},
        month = jan,
        pages = {1013},
          doi = {10.1007/978-1-4614-3508-2_4},
archivePrefix = {arXiv},
       eprint = {1210.7762},
 primaryClass = {astro-ph.IM},
       adsurl = {https://ui.adsabs.harvard.edu/abs/2013acna.conf...63K}
}

@ARTICLE{2020MNRAS.491.4277M,
       author = {{M{\"o}ller}, A. and {de Boissi{\`e}re}, T.},
        title = "{SuperNNova: an open-source framework for Bayesian, neural network-based supernova classification}",
      journal = {\mnras},
         year = 2020,
        month = jan,
       volume = {491},
       number = {3},
        pages = {4277-4293},
          doi = {10.1093/mnras/stz3312},
archivePrefix = {arXiv},
       eprint = {1901.06384},
 primaryClass = {astro-ph.IM},
       adsurl = {https://ui.adsabs.harvard.edu/abs/2020MNRAS.491.4277M}
}

@ARTICLE{2023MNRAS.tmp..345T,
       author = {{Taylor}, G. and {Jones}, D.~O. and {Popovic}, B. and {Vincenzi}, M. and {Kessler}, R. and {Scolnic}, D. and {Dai}, M. and {Kenworthy}, W.~D. and {Pierel}, J.~D.~R.},
        title = "{SALT2 versus SALT3: Updated model surfaces and their impacts on type Ia supernova cosmology}",
      journal = {\mnras},
         year = 2023,
        month = jan,
          doi = {10.1093/mnras/stad320},
archivePrefix = {arXiv},
       eprint = {2301.10644},
 primaryClass = {astro-ph.CO},
      volume = {520},
       number = {4},
       adsurl = {https://ui.adsabs.harvard.edu/abs/2023MNRAS.tmp..345T}
}

@article{Peterson_2022,
doi = {10.3847/1538-4357/ac4698},
url = {https://dx.doi.org/10.3847/1538-4357/ac4698},
year = {2022},
month = {oct},
publisher = {The American Astronomical Society},
volume = {938},
number = {2},
pages = {112},
author = {Erik R. Peterson and W. D’Arcy Kenworthy and Daniel Scolnic and Adam G. Riess and Dillon Brout and Anthony Carr and Hélène Courtois and Tamara Davis and Arianna Dwomoh and David O. Jones and Brodie Popovic and Benjamin M. Rose and Khaled Said},
title = {The Pantheon+ Analysis: Evaluating Peculiar Velocity Corrections in Cosmological Analyses with Nearby Type Ia Supernovae},
journal = {\apj}
}

@article{Qu_hostMismatch,
  title={The Dark Energy Survey Supernova Program: cosmological biases from host galaxy mismatch of type Ia supernovae},
  author={Qu, H and Sako, M and Vincenzi, M and S{\'a}nchez, C and Brout, D and Kessler, R and Chen, R and Davis, T and Galbany, L and Kelsey, L and others},
  journal={\apj},
  volume={964},
  number={2},
  pages={134},
  year={2024},
  publisher={IOP Publishing}
}

@article{Carr_2022,
	doi = {10.1017/pasa.2022.41},
  
	url = {https://doi.org/10.1017%2Fpasa.2022.41},
  
	year = 2022,
	publisher = {Cambridge University Press ({CUP})},
  
	volume = {39},
  
	author = {Anthony Carr and Tamara M. Davis and Dan Scolnic and Khaled Said and Dillon Brout and Erik R. Peterson and Richard Kessler},
  
	title = {The Pantheon$\mathplus$ analysis: Improving the redshifts and peculiar velocities of Type Ia supernovae used in cosmological analyses},
  
	journal={\pasa}
}

@ARTICLE{2015MNRAS.450..317C,
       author = {{Carrick}, Jonathan and {Turnbull}, Stephen J. and {Lavaux}, Guilhem and {Hudson}, Michael J.},
        title = "{Cosmological parameters from the comparison of peculiar velocities with predictions from the 2M++ density field}",
      journal = {\mnras},
         year = 2015,
        month = jun,
       volume = {450},
       number = {1},
        pages = {317-332},
          doi = {10.1093/mnras/stv547},
archivePrefix = {arXiv},
       eprint = {1504.04627},
 primaryClass = {astro-ph.CO},
       adsurl = {https://ui.adsabs.harvard.edu/abs/2015MNRAS.450..317C}
}

@article{kessler09,
doi = {10.1086/605984},
url = {https://dx.doi.org/10.1086/605984},
year = {2009},
month = {aug},
publisher = {University of Chicago Press},
volume = {121},
number = {883},
pages = {1028},
author = {Richard Kessler and Joseph P. Bernstein and David Cinabro and Benjamin Dilday and Joshua A. Frieman and Saurabh Jha and Stephen Kuhlmann and Gajus Miknaitis and Masao Sako and Matt Taylor and Jake Vanderplas},
title = {SNANA: A Public Software Package for Supernova Analysis},
journal = {\pasp}
}

@ARTICLE{handley19,
       author = {{Handley}, Will and {Lemos}, Pablo},
        title = "{Quantifying dimensionality: Bayesian cosmological model complexities}",
      journal = {\prd},
         year = 2019,
        month = jul,
       volume = {100},
       number = {2},
          eid = {023512},
        pages = {023512},
          doi = {10.1103/PhysRevD.100.023512},
archivePrefix = {arXiv},
       eprint = {1903.06682},
 primaryClass = {astro-ph.CO},
       adsurl = {https://ui.adsabs.harvard.edu/abs/2019PhRvD.100b3512H}
}

@ARTICLE{vincenzi24,
       author = {{Vincenzi}, M. and {Brout}, D. and {Armstrong}, P. and {Popovic}, B. and {Taylor}, G. and {Acevedo}, M. and {Camilleri}, R. and {Chen}, R. and {Davis}, T.~M. and {Lee}, J. and {Lidman}, C. and {Hinton}, S.~R. and {Kelsey}, L. and {Kessler}, R. and {M{\"o}ller}, A. and {Qu}, H. and {Sako}, M. and {Sanchez}, B. and {Scolnic}, D. and {Smith}, M. and {Sullivan}, M. and {Wiseman}, P. and {Asorey}, J. and {Bassett}, B.~A. and {Carollo}, D. and {Carr}, A. and {Foley}, R.~J. and {Frohmaier}, C. and {Galbany}, L. and {Glazebrook}, K. and {Graur}, O. and {Kovacs}, E. and {Kuehn}, K. and {Malik}, U. and {Nichol}, R.~C. and {Rose}, B. and {Tucker}, B.~E. and {Toy}, M. and {Tucker}, D.~L. and {Yuan}, F. and {Abbott}, T.~M.~C. and {Aguena}, M. and {Alves}, O. and {Allam}, S.~S. and {Andrade-Oliveira}, F. and {Annis}, J. and {Bacon}, D. and {Bechtol}, K. and {Bernstein}, G.~M. and {Brooks}, D. and {Burke}, D.~L. and {Carnero Rosell}, A. and {Carretero}, J. and {Castander}, F.~J. and {Conselice}, C. and {da Costa}, L.~N. and {Pereira}, M.~E.~S. and {Desai}, S. and {Diehl}, H.~T. and {Doel}, P. and {Ferrero}, I. and {Flaugher}, B. and {Friedel}, D. and {Frieman}, J. and {Garc{\'\i}a-Bellido}, J. and {Gatti}, M. and {Giannini}, G. and {Gruen}, D. and {Gruendl}, R.~A. and {Hollowood}, D.~L. and {Honscheid}, K. and {Huterer}, D. and {James}, D.~J. and {Kuropatkin}, N. and {Lahav}, O. and {Lee}, S. and {Lin}, H. and {Marshall}, J.~L. and {Mena-Fern{\'a}ndez}, J. and {Menanteau}, F. and {Miquel}, R. and {Palmese}, A. and {Pieres}, A. and {Plazas Malag{\'o}n}, A.~A. and {Porredon}, A. and {Romer}, A.~K. and {Roodman}, A. and {Sanchez}, E. and {Sanchez Cid}, D. and {Schubnell}, M. and {Sevilla-Noarbe}, I. and {Suchyta}, E. and {Swanson}, M.~E.~C. and {Tarle}, G. and {To}, C. and {Walker}, A.~R. and {Weaverdyck}, N. and {Yamamoto}, M.},
        title = "{The Dark Energy Survey Supernova Program: Cosmological Analysis and Systematic Uncertainties}",
      journal = {\apj},
         year = 2024,
        month = nov,
       volume = {975},
       number = {1},
          eid = {86},
        pages = {86},
          doi = {10.3847/1538-4357/ad5e6c},
archivePrefix = {arXiv},
       eprint = {2401.02945},
 primaryClass = {astro-ph.CO},
       adsurl = {https://ui.adsabs.harvard.edu/abs/2024ApJ...975...86V}
}

@ARTICLE{DESDATA24,
       author = {{S{\'a}nchez}, B.~O. and {Brout}, D. and {Vincenzi}, M. and {Sako}, M. and {Herner}, K. and {Kessler}, R. and {Davis}, T.~M. and {Scolnic}, D. and {Acevedo}, M. and {Lee}, J. and {M{\"o}ller}, A. and {Qu}, H. and {Kelsey}, L. and {Wiseman}, P. and {Armstrong}, P. and {Rose}, B. and {Camilleri}, R. and {Chen}, R. and {Galbany}, L. and {Kovacs}, E. and {Lidman}, C. and {Popovic}, B. and {Smith}, M. and {Shah}, P. and {Sullivan}, M. and {Toy}, M. and {Abbott}, T.~M.~C. and {Aguena}, M. and {Allam}, S. and {Alves}, O. and {Annis}, J. and {Asorey}, J. and {Avila}, S. and {Bacon}, D. and {Brooks}, D. and {Burke}, D.~L. and {Carnero Rosell}, A. and {Carollo}, D. and {Carretero}, J. and {da Costa}, L.~N. and {Castander}, F.~J. and {Desai}, S. and {Diehl}, H.~T. and {Duarte}, J. and {Everett}, S. and {Ferrero}, I. and {Flaugher}, B. and {Frieman}, J. and {Garc{\'\i}a-Bellido}, J. and {Gatti}, M. and {Gaztanaga}, E. and {Giannini}, G. and {Glazebrook}, K. and {Gonz{\'a}lez-Gait{\'a}n}, S. and {Gruendl}, R.~A. and {Gutierrez}, G. and {Hinton}, S.~R. and {Hollowood}, D.~L. and {Honscheid}, K. and {James}, D.~J. and {Kuehn}, K. and {Lahav}, O. and {Lee}, S. and {Lewis}, G.~F. and {Lin}, H. and {Marshall}, J.~L. and {Mena-Fern{\'a}ndez}, J. and {Miquel}, R. and {Myles}, J. and {Nichol}, R.~C. and {Ogando}, R.~L.~C. and {Palmese}, A. and {Pereira}, M.~E.~S. and {Pieres}, A. and {Plazas Malag{\'o}n}, A.~A. and {Porredon}, A. and {Romer}, A.~K. and {Sanchez}, E. and {Sanchez Cid}, D. and {Sevilla-Noarbe}, I. and {Suchyta}, E. and {Swanson}, M.~E.~C. and {Tarle}, G. and {Tucker}, B.~E. and {Tucker}, D.~L. and {Vikram}, V. and {Walker}, A.~R. and {Weaverdyck}, N.},
        title = "{The Dark Energy Survey Supernova Program: Light Curves and 5 Yr Data Release}",
      journal = {\apj},
         year = 2024,
        month = nov,
       volume = {975},
       number = {1},
          eid = {5},
        pages = {5},
          doi = {10.3847/1538-4357/ad739a},
archivePrefix = {arXiv},
       eprint = {2406.05046},
 primaryClass = {astro-ph.CO},
       adsurl = {https://ui.adsabs.harvard.edu/abs/2024ApJ...975....5S}
}

@ARTICLE{Trotta08,
       author = {{Trotta}, Roberto},
        title = "{Bayes in the sky: Bayesian inference and model selection in cosmology}",
      journal = {Contemporary Phys.},
         year = 2008,
        month = mar,
       volume = {49},
       number = {2},
        pages = {71-104},
          doi = {10.1080/00107510802066753},
archivePrefix = {arXiv},
       eprint = {0803.4089},
 primaryClass = {astro-ph},
       adsurl = {https://ui.adsabs.harvard.edu/abs/2008ConPh..49...71T}
}

@article{10.1093/mnras/stac1691,
    author = {Möller, A and Smith, M and Sako, M and Sullivan, M and Vincenzi, M and Wiseman, P and Armstrong, P and Asorey, J and Brout, D and Carollo, D and Davis, T M and Frohmaier, C and Galbany, L and Glazebrook, K and Kelsey, L and Kessler, R and Lewis, G F and Lidman, C and Malik, U and Nichol, R C and Scolnic, D and Tucker, B E and Abbott, T M C and Aguena, M and Allam, S and Annis, J and Bertin, E and Bocquet, S and Brooks, D and Burke, D L and Carnero Rosell, A and Carrasco Kind, M and Carretero, J and Castander, F J and Conselice, C and Costanzi, M and Crocce, M and da Costa, L N and De Vicente, J and Desai, S and Diehl, H T and Doel, P and Everett, S and Ferrero, I and Finley, D A and Flaugher, B and Friedel, D and Frieman, J and García-Bellido, J and Gerdes, D W and Gruen, D and Gruendl, R A and Gschwend, J and Gutierrez, G and Herner, K and Hinton, S R and Hollowood, D L and Honscheid, K and James, D J and Kuehn, K and Kuropatkin, N and Lahav, O and March, M and Marshall, J L and Menanteau, F and Miquel, R and Morgan, R and Palmese, A and Paz-Chinchón, F and Pieres, A and Plazas Malagón, A A and Romer, A K and Roodman, A and Sanchez, E and Scarpine, V and Schubnell, M and Serrano, S and Sevilla-Noarbe, I and Suchyta, E and Tarle, G and Thomas, D and To, C and Varga, T N},
    title = "{The dark energy survey 5-yr photometrically identified type Ia supernovae}",
    journal = {\mnras},
    volume = {514},
    number = {4},
    pages = {5159-5177},
    year = {2022},
    month = {06},
    issn = {0035-8711},
    doi = {10.1093/mnras/stac1691},
    url = {https://doi.org/10.1093/mnras/stac1691},
    eprint = {https://academic.oup.com/mnras/article-pdf/514/4/5159/44589908/stac1691.pdf},
}

@ARTICLE{desbao24,
       author = {{DES Collaboration} and {Abbott}, T.~M.~C. and {Adamow}, M. and {Aguena}, M. and {Allam}, S. and {Alves}, O. and {Amon}, A. and {Andrade-Oliveira}, F. and {Asorey}, J. and {Avila}, S. and {Bacon}, D. and {Bechtol}, K. and {Bernstein}, G.~M. and {Bertin}, E. and {Blazek}, J. and {Bocquet}, S. and {Brooks}, D. and {Burke}, D.~L. and {Camacho}, H. and {Carnero Rosell}, A. and {Carollo}, D. and {Carr}, A. and {Carretero}, J. and {Castander}, F.~J. and {Cawthon}, R. and {Chan}, K.~C. and {Chang}, C. and {Conselice}, C. and {Costanzi}, M. and {Crocce}, M. and {da Costa}, L.~N. and {Pereira}, M.~E.~S. and {Davis}, T.~M. and {De Vicente}, J. and {Deiosso}, N. and {Desai}, S. and {Diehl}, H.~T. and {Dodelson}, S. and {Doux}, C. and {Drlica-Wagner}, A. and {Elvin-Poole}, J. and {Everett}, S. and {Ferrero}, I. and {Fert{\'e}}, A. and {Flaugher}, B. and {Fosalba}, P. and {Frieman}, J. and {Garc{\'\i}a-Bellido}, J. and {Gaztanaga}, E. and {Giannini}, G. and {Glazebrook}, K. and {Gruendl}, R.~A. and {Gutierrez}, G. and {Hartley}, W.~G. and {Hinton}, S.~R. and {Hollowood}, D.~L. and {Honscheid}, K. and {Huterer}, D. and {James}, D.~J. and {Kent}, S. and {Kuehn}, K. and {Lahav}, O. and {Lee}, S. and {Lewis}, G.~F. and {Lidman}, C. and {Lima}, M. and {Lin}, H. and {Malik}, U. and {Maraston}, C. and {Marshall}, J.~L. and {Martini}, P. and {Mena-Fern{\'a}ndez}, J. and {Menanteau}, F. and {Miquel}, R. and {Mohr}, J.~J. and {Myles}, J. and {M{\"o}ller}, A. and {Nichol}, R.~C. and {Ogando}, R.~L.~C. and {Palmese}, A. and {Percival}, W.~J. and {Pieres}, A. and {Plazas Malag{\'o}n}, A.~A. and {Porredon}, A. and {Prat}, J. and {Rodr{\'\i}guez-Monroy}, M. and {Romer}, A.~K. and {Roodman}, A. and {Rosenfeld}, R. and {Ross}, A.~J. and {Rykoff}, E.~S. and {Sako}, M. and {Samuroff}, S. and {S{\'a}nchez}, C. and {Sanchez}, E. and {Sanchez Cid}, D. and {Santiago}, B. and {Schubnell}, M. and {Sevilla-Noarbe}, I. and {Sheldon}, E. and {Smith}, M. and {Suchyta}, E. and {Swanson}, M.~E.~C. and {Tarle}, G. and {Thomas}, D. and {To}, C. and {Toribio San Cipriano}, L. and {Troxel}, M.~A. and {Tucker}, B.~E. and {Tucker}, D.~L. and {Walker}, A.~R. and {Weaverdyck}, N. and {Weller}, J. and {Wiseman}, P. and {Yanny}, B. and {DES Collaboration}},
        title = "{Dark Energy Survey: A 2.1\% measurement of the angular baryonic acoustic oscillation scale at redshift zeff=0.85 from the final dataset}",
      journal = {\prd},
         year = 2024,
        month = sep,
       volume = {110},
       number = {6},
          eid = {063515},
        pages = {063515},
          doi = {10.1103/PhysRevD.110.063515},
archivePrefix = {arXiv},
       eprint = {2402.10696},
 primaryClass = {astro-ph.CO},
       adsurl = {https://ui.adsabs.harvard.edu/abs/2024PhRvD.110f3515A}
}

@ARTICLE{desicollaboration2024desi,
       author = {{DESI Collaboration} and {Adame}, A.~G. and {Aguilar}, J. and {Ahlen}, S. and {Alam}, S. and {Alexander}, D.~M. and {Alvarez}, M. and {Alves}, O. and {Anand}, A. and {Andrade}, U. and {Armengaud}, E. and {Avila}, S. and {Aviles}, A. and {Awan}, H. and {Bahr-Kalus}, B. and {Bailey}, S. and {Baltay}, C. and {Bault}, A. and {Behera}, J. and {BenZvi}, S. and {Bera}, A. and {Beutler}, F. and {Bianchi}, D. and {Blake}, C. and {Blum}, R. and {Brieden}, S. and {Brodzeller}, A. and {Brooks}, D. and {Buckley-Geer}, E. and {Burtin}, E. and {Calderon}, R. and {Canning}, R. and {Carnero Rosell}, A. and {Cereskaite}, R. and {Cervantes-Cota}, J.~L. and {Chabanier}, S. and {Chaussidon}, E. and {Chaves-Montero}, J. and {Chen}, S. and {Chen}, X. and {Claybaugh}, T. and {Cole}, S. and {Cuceu}, A. and {Davis}, T.~M. and {Dawson}, K. and {de la Macorra}, A. and {de Mattia}, A. and {Deiosso}, N. and {Dey}, A. and {Dey}, B. and {Ding}, Z. and {Doel}, P. and {Edelstein}, J. and {Eftekharzadeh}, S. and {Eisenstein}, D.~J. and {Elliott}, A. and {Fagrelius}, P. and {Fanning}, K. and {Ferraro}, S. and {Ereza}, J. and {Findlay}, N. and {Flaugher}, B. and {Font-Ribera}, A. and {Forero-S{\'a}nchez}, D. and {Forero-Romero}, J.~E. and {Frenk}, C.~S. and {Garcia-Quintero}, C. and {Gazta{\~n}aga}, E. and {Gil-Mar{\'\i}n}, H. and {Gontcho a Gontcho}, S. and {Gonzalez-Morales}, A.~X. and {Gonzalez-Perez}, V. and {Gordon}, C. and {Green}, D. and {Gruen}, D. and {Gsponer}, R. and {Gutierrez}, G. and {Guy}, J. and {Hadzhiyska}, B. and {Hahn}, C. and {Hanif}, M.~M.~S. and {Herrera-Alcantar}, H.~K. and {Honscheid}, K. and {Howlett}, C. and {Huterer}, D. and {Ir{\v{s}}i{\v{c}}}, V. and {Ishak}, M. and {Juneau}, S. and {Kara{\c{c}}ayl{\i}}, N.~G. and {Kehoe}, R. and {Kent}, S. and {Kirkby}, D. and {Kremin}, A. and {Krolewski}, A. and {Lai}, Y. and {Lan}, T.-W. and {Landriau}, M. and {Lang}, D. and {Lasker}, J. and {Le Goff}, J.~M. and {Le Guillou}, L. and {Leauthaud}, A. and {Levi}, M.~E. and {Li}, T.~S. and {Linder}, E. and {Lodha}, K. and {Magneville}, C. and {Manera}, M. and {Margala}, D. and {Martini}, P. and {Maus}, M. and {McDonald}, P. and {Medina-Varela}, L. and {Meisner}, A. and {Mena-Fern{\'a}ndez}, J. and {Miquel}, R. and {Moon}, J. and {Moore}, S. and {Moustakas}, J. and {Mueller}, E. and {Mu{\~n}oz-Guti{\'e}rrez}, A. and {Myers}, A.~D. and {Nadathur}, S. and {Napolitano}, L. and {Neveux}, R. and {Newman}, J.~A. and {Nguyen}, N.~M. and {Nie}, J. and {Niz}, G. and {Noriega}, H.~E. and {Padmanabhan}, N. and {Paillas}, E. and {Palanque-Delabrouille}, N. and {Pan}, J. and {Penmetsa}, S. and {Percival}, W.~J. and {Pieri}, M.~M. and {Pinon}, M. and {Poppett}, C. and {Porredon}, A. and {Prada}, F. and {P{\'e}rez-Fern{\'a}ndez}, A. and {P{\'e}rez-R{\`a}fols}, I. and {Rabinowitz}, D. and {Raichoor}, A. and {Ram{\'\i}rez-P{\'e}rez}, C. and {Ramirez-Solano}, S. and {Rashkovetskyi}, M. and {Ravoux}, C. and {Rezaie}, M. and {Rich}, J. and {Rocher}, A. and {Rockosi}, C. and {Roe}, N.~A. and {Rosado-Marin}, A. and {Ross}, A.~J. and {Rossi}, G. and {Ruggeri}, R. and {Ruhlmann-Kleider}, V. and {Samushia}, L. and {Sanchez}, E. and {Saulder}, C. and {Schlafly}, E.~F. and {Schlegel}, D. and {Schubnell}, M. and {Seo}, H. and {Shafieloo}, A. and {Sharples}, R. and {Silber}, J. and {Slosar}, A. and {Smith}, A. and {Sprayberry}, D. and {Tan}, T. and {Tarl{\'e}}, G. and {Taylor}, P. and {Trusov}, S. and {Ure{\~n}a-L{\'o}pez}, L.~A. and {Vaisakh}, R. and {Valcin}, D. and {Valdes}, F. and {Vargas-Maga{\~n}a}, M. and {Verde}, L. and {Walther}, M. and {Wang}, B. and {Wang}, M.~S. and {Weaver}, B.~A. and {Weaverdyck}, N. and {Wechsler}, R.~H. and {Weinberg}, D.~H. and {White}, M. and {Yu}, J. and {Yu}, Y. and {Yuan}, S. and {Y{\`e}che}, C. and {Zaborowski}, E.~A. and {Zarrouk}, P. and {Zhang}, H. and {Zhao}, C. and {Zhao}, R. and {Zhou}, R. and {Zhuang}, T.},
        title = "{DESI 2024 VI: cosmological constraints from the measurements of baryon acoustic oscillations}",
      journal = {\jcap},
         year = 2025,
        month = feb,
       volume = {2025},
       number = {2},
          eid = {021},
        pages = {021},
          doi = {10.1088/1475-7516/2025/02/021},
archivePrefix = {arXiv},
       eprint = {2404.03002},
 primaryClass = {astro-ph.CO},
       adsurl = {https://ui.adsabs.harvard.edu/abs/2025JCAP...02..021A}
}

@article{camilleri24,
    author = {Camilleri, R and Davis, T M and Vincenzi, M and Shah, P and Frieman, J and Kessler, R and Armstrong, P and Brout, D and Carr, A and Chen, R and Galbany, L and Glazebrook, K and Hinton, S R and Lee, J and Lidman, C and Möller, A and Popovic, B and Qu, H and Sako, M and Scolnic, D and Smith, M and Sullivan, M and Sánchez, B O and Taylor, G and Toy, M and Wiseman, P and Abbott, T M C and Aguena, M and Allam, S and Alves, O and Annis, J and Avila, S and Bacon, D and Bertin, E and Bocquet, S and Brooks, D and Burke, D L and Rosell, A Carnero and Carretero, J and Castander, F J and da Costa, L N and Pereira, M E S and Desai, S and Diehl, H T and Doel, P and Doux, C and Everett, S and Ferrero, I and Flaugher, B and Fosalba, P and García-Bellido, J and Gatti, M and Gaztanaga, E and Giannini, G and Gruen, D and Hollowood, D L and Honscheid, K and James, D J and Kuehn, K and Lahav, O and Lee, S and Lewis, G F and Marshall, J L and Mena-Fernández, J and Miquel, R and Muir, J and Myles, J and Ogando, R L C and Pieres, A and Malagón, A A Plazas and Porredon, A and Rodriguez-Monroy, M and Sanchez, E and Cid, D Sanchez and Schubnell, M and Sevilla-Noarbe, I and Suchyta, E and Swanson, M E C and Tarle, G and Walker, A R and Weaverdyck, N and (DES Collaboration)},
    title = {The dark energy survey supernova program: investigating beyond-ΛCDM},
    journal = {\mnras},
    volume = {533},
    number = {3},
    pages = {2615-2639},
    year = {2024},
    month = {09},
    issn = {0035-8711},
    doi = {10.1093/mnras/stae1988},
    url = {https://doi.org/10.1093/mnras/stae1988},
    eprint = {https://academic.oup.com/mnras/article-pdf/533/3/2615/58966760/stae1988.pdf},
}

@article{Linder_2003,
   title={Baryon oscillations as a cosmological probe},
   volume={68},
   ISSN={1089-4918},
   url={http://dx.doi.org/10.1103/PhysRevD.68.083504},
   DOI={10.1103/physrevd.68.083504},
   number={8},
   journal={\prd},
   publisher={American Physical Society (APS)},
   author={Linder, Eric V.},
   year={2003},
   month=oct }

@ARTICLE{2009ApJ...700..331H,
       author = {{Hicken}, Malcolm and {Challis}, Peter and {Jha}, Saurabh and {Kirshner}, Robert P. and {Matheson}, Tom and {Modjaz}, Maryam and {Rest}, Armin and {Wood-Vasey}, W. Michael and {Bakos}, Gaspar and {Barton}, Elizabeth J. and {Berlind}, Perry and {Bragg}, Ann and {Brice{\~n}o}, Cesar and {Brown}, Warren R. and {Caldwell}, Nelson and {Calkins}, Mike and {Cho}, Richard and {Ciupik}, Larry and {Contreras}, Maria and {Dendy}, Kristi-Concannon and {Dosaj}, Anil and {Durham}, Nick and {Eriksen}, Kris and {Esquerdo}, Gil and {Everett}, Mark and {Falco}, Emilio and {Fernandez}, Jose and {Gaba}, Alejandro and {Garnavich}, Peter and {Graves}, Genevieve and {Green}, Paul and {Groner}, Ted and {Hergenrother}, Carl and {Holman}, Matthew J. and {Hradecky}, Vit and {Huchra}, John and {Hutchison}, Bob and {Jerius}, Diab and {Jordan}, Andres and {Kilgard}, Roy and {Krauss}, Miriam and {Luhman}, Kevin and {Macri}, Lucas and {Marrone}, Daniel and {McDowell}, Jonathan and {McIntosh}, Daniel and {McNamara}, Brian and {Megeath}, Tom and {Mochejska}, Barbara and {Munoz}, Diego and {Muzerolle}, James and {Naranjo}, Orlando and {Narayan}, Gautham and {Pahre}, Michael and {Peters}, Wayne and {Peterson}, Dawn and {Rines}, Ken and {Ripman}, Ben and {Roussanova}, Anna and {Schild}, Rudolph and {Sicilia-Aguilar}, Aurora and {Sokoloski}, Jennifer and {Smalley}, Kyle and {Smith}, Andy and {Spahr}, Tim and {Stanek}, K.~Z. and {Barmby}, Pauline and {Blondin}, St{\'e}phane and {Stubbs}, Christopher W. and {Szentgyorgyi}, Andrew and {Torres}, Manuel A.~P. and {Vaz}, Amili and {Vikhlinin}, Alexey and {Wang}, Zhong and {Westover}, Mike and {Woods}, Deborah and {Zhao}, Ping},
        title = "{CfA3: 185 Type Ia Supernova Light Curves from the CfA}",
      journal = {\apj},
         year = 2009,
        month = jul,
       volume = {700},
       number = {1},
        pages = {331-357},
          doi = {10.1088/0004-637X/700/1/331},
archivePrefix = {arXiv},
       eprint = {0901.4787},
 primaryClass = {astro-ph.CO},
       adsurl = {https://ui.adsabs.harvard.edu/abs/2009ApJ...700..331H}
}

@ARTICLE{2012ApJS..200...12H,
       author = {{Hicken}, Malcolm and {Challis}, Peter and {Kirshner}, Robert P. and {Rest}, Armin and {Cramer}, Claire E. and {Wood-Vasey}, W. Michael and {Bakos}, Gaspar and {Berlind}, Perry and {Brown}, Warren R. and {Caldwell}, Nelson and {Calkins}, Mike and {Currie}, Thayne and {de Kleer}, Kathy and {Esquerdo}, Gil and {Everett}, Mark and {Falco}, Emilio and {Fernandez}, Jose and {Friedman}, Andrew S. and {Groner}, Ted and {Hartman}, Joel and {Holman}, Matthew J. and {Hutchins}, Robert and {Keys}, Sonia and {Kipping}, David and {Latham}, Dave and {Marion}, George H. and {Narayan}, Gautham and {Pahre}, Michael and {Pal}, Andras and {Peters}, Wayne and {Perumpilly}, Gopakumar and {Ripman}, Ben and {Sipocz}, Brigitta and {Szentgyorgyi}, Andrew and {Tang}, Sumin and {Torres}, Manuel A.~P. and {Vaz}, Amali and {Wolk}, Scott and {Zezas}, Andreas},
        title = "{CfA4: Light Curves for 94 Type Ia Supernovae}",
      journal = {\apjs},
         year = 2012,
        month = jun,
       volume = {200},
       number = {2},
          eid = {12},
        pages = {12},
          doi = {10.1088/0067-0049/200/2/12},
archivePrefix = {arXiv},
       eprint = {1205.4493},
 primaryClass = {astro-ph.CO},
       adsurl = {https://ui.adsabs.harvard.edu/abs/2012ApJS..200...12H}
}

@ARTICLE{2017AJ....154..211K,
       author = {{Krisciunas}, Kevin and {Contreras}, Carlos and {Burns}, Christopher R. and {Phillips}, M.~M. and {Stritzinger}, Maximilian D. and {Morrell}, Nidia and {Hamuy}, Mario and {Anais}, Jorge and {Boldt}, Luis and {Busta}, Luis and {Campillay}, Abdo and {Castell{\'o}n}, Sergio and {Folatelli}, Gast{\'o}n and {Freedman}, Wendy L. and {Gonz{\'a}lez}, Consuelo and {Hsiao}, Eric Y. and {Krzeminski}, Wojtek and {Persson}, Sven Eric and {Roth}, Miguel and {Salgado}, Francisco and {Ser{\'o}n}, Jacqueline and {Suntzeff}, Nicholas B. and {Torres}, Sim{\'o}n and {Filippenko}, Alexei V. and {Li}, Weidong and {Madore}, Barry F. and {DePoy}, D.~L. and {Marshall}, Jennifer L. and {Rheault}, Jean-Philippe and {Villanueva}, Steven},
        title = "{The Carnegie Supernova Project. I. Third Photometry Data Release of Low-redshift Type Ia Supernovae and Other White Dwarf Explosions}",
      journal = {\aj},
         year = 2017,
        month = nov,
       volume = {154},
       number = {5},
          eid = {211},
        pages = {211},
          doi = {10.3847/1538-3881/aa8df0},
archivePrefix = {arXiv},
       eprint = {1709.05146},
 primaryClass = {astro-ph.IM},
       adsurl = {https://ui.adsabs.harvard.edu/abs/2017AJ....154..211K}
}

@article{Scolnic_2018,
doi = {10.3847/1538-4357/aab9bb},
url = {https://dx.doi.org/10.3847/1538-4357/aab9bb},
year = {2018},
month = {may},
publisher = {The American Astronomical Society},
volume = {859},
number = {2},
pages = {101},
author = {Scolnic, D. M. and Jones, D. O. and Rest, A. and Pan, Y. C. and Chornock, R. and Foley, R. J. and Huber, M. E. and Kessler, R. and Narayan, G. and Riess, A. G. and Rodney, S. and Berger, E. and Brout, D. J. and Challis, P. J. and Drout, M. and Finkbeiner, D. and Lunnan, R. and Kirshner, R. P. and Sanders, N. E. and Schlafly, E. and Smartt, S. and Stubbs, C. W. and Tonry, J. and Wood-Vasey, W. M. and Foley, M. and Hand, J. and Johnson, E. and Burgett, W. S. and Chambers, K. C. and Draper, P. W. and Hodapp, K. W. and Kaiser, N. and Kudritzki, R. P. and Magnier, E. A. and Metcalfe, N. and Bresolin, F. and Gall, E. and Kotak, R. and McCrum, M. and Smith, K. W.},
title = {The Complete Light-curve Sample of Spectroscopically Confirmed SNe Ia from Pan-STARRS1 and Cosmological Constraints from the Combined Pantheon Sample},
journal = {\apj}
}

@article{Sako_2011,
doi = {10.1088/0004-637X/738/2/162},
url = {https://dx.doi.org/10.1088/0004-637X/738/2/162},
year = {2011},
month = {aug},
publisher = {The American Astronomical Society},
volume = {738},
number = {2},
pages = {162},
author = {Sako, Masao and Bassett, Bruce and Connolly, Brian and Dilday, Benjamin and Cambell, Heather and Frieman, Joshua A. and Gladney, Larry and Kessler, Richard and Lampeitl, Hubert and Marriner, John and Miquel, Ramon and Nichol, Robert C. and Schneider, Donald P. and Smith, Mathew and Sollerman, Jesper},
title = {PHOTOMETRIC TYPE Ia SUPERNOVA CANDIDATES FROM THE THREE-YEAR SDSS-II SN SURVEY DATA},
journal = {\apj}
}

@article{Ganeshalingam_2010,
doi = {10.1088/0067-0049/190/2/418},
url = {https://dx.doi.org/10.1088/0067-0049/190/2/418},
year = {2010},
month = {sep},
publisher = {The American Astronomical Society},
volume = {190},
number = {2},
pages = {418},
author = {Ganeshalingam, Mohan and Li, Weidong and Filippenko, Alexei V. and Anderson, Carmen and Foster, Griffin and Gates, Elinor L. and Griffith, Christopher V. and Grigsby, Bryant J. and Joubert, Niels and Leja, Joel and Lowe, Thomas B. and Macomber, Brent and Pritchard, Tyler and Thrasher, Patrick and Winslow, Dustin},
title = {RESULTS OF THE LICK OBSERVATORY SUPERNOVA SEARCH FOLLOW-UP PHOTOMETRY PROGRAM: BVRI LIGHT CURVES OF 165 TYPE Ia SUPERNOVAE},
journal = {\apjs}
}

@article{Gilliland_1999,
doi = {10.1086/307549},
url = {https://dx.doi.org/10.1086/307549},
year = {1999},
month = {aug},
publisher = {},
volume = {521},
number = {1},
pages = {30},
author = {Gilliland, Ronald L. and Nugent, Peter E. and Phillips, M. M.},
title = {High-Redshift Supernovae in the Hubble Deep Field*},
journal = {\apj}
}

@article{Chen_2022,
doi = {10.3847/1538-4365/ac50b7},
url = {https://dx.doi.org/10.3847/1538-4365/ac50b7},
year = {2022},
month = {mar},
publisher = {The American Astronomical Society},
volume = {259},
number = {2},
pages = {53},
author = {Chen, Ping and Dong, Subo and Kochanek, C. S. and Stanek, K. Z. and Post, R. S. and Stritzinger, M. D. and Prieto, J. L. and Filippenko, Alexei V. and Kollmeier, Juna A. and Elias-Rosa, N. and Katz, Boaz and Tomasella, Lina and Bose, S. and Ashall, Chris and Benetti, S. and Bersier, D. and Brimacombe, Joseph and Brink, Thomas G. and Brown, P. and Buckley, David A. H. and Cappellaro, Enrico and Christie, Grant W. and Fraser, Morgan and Gromadzki, Mariusz and Holoien, Thomas W.-S. and Hu, Shaoming and Kankare, Erkki and Koff, Robert and Lundqvist, P. and Mattila, S. and Milne, P. A. and Morrell, Nidia and Muñoz, J. A. and Mutel, Robert and Natusch, Tim and Nicolas, Joel and Pastorello, A. and Prentice, Simon and Roth, Tyler and Shappee, B. J. and Stone, Geoffrey and Thompson, Todd A. and Villanueva, Steven and Zheng, WeiKang},
title = {The First Data Release of CNIa0.02—A Complete Nearby (Redshift &lt;0.02) Sample of Type Ia Supernova Light Curves*},
journal = {\apjs}
}

@article{Riess_1999,
doi = {10.1086/300738},
url = {https://dx.doi.org/10.1086/300738},
year = {1999},
month = {feb},
publisher = {},
volume = {117},
number = {2},
pages = {707},
author = {Riess, Adam G. and Kirshner, Robert P. and Schmidt, Brian P. and Jha, Saurabh and Challis, Peter and Garnavich, Peter M. and Esin, Ann A. and Carpenter, Chris and Grashius, Randy and Schild, Rudolph E. and Berlind, Perry L. and Huchra, John P. and Prosser, Charles F. and Falco, Emilio E. and Benson, Priscilla J. and Briceño, César and Brown, Warren R. and Caldwell, Nelson and Dell'Antonio, Ian P. and Filippenko, Alexei V. and Goodman, Alyssa A. and Grogin, Norman A. and Groner, Ted and Hughes, John P. and Green, Paul J. and Jansen, Rolf A. and Kleyna, Jan T. and Luu, Jane X. and Macri, Lucas M. and McLeod, Brian A. and McLeod, Kim K. and McNamara, Brian R. and McLean, Brian and Milone, Alejandra A. E. and Mohr, Joseph J. and Moraru, Dan and Peng, Chien and Peters, Jim and Prestwich, Andrea H. and Stanek, Krzysztof Z. and Szentgyorgyi, Andy and Zhao, Ping},
title = {BVRI Light Curves for 22 Type Ia Supernovae},
journal = {\aj}
}

@article{Jha_2006,
doi = {10.1086/497989},
url = {https://dx.doi.org/10.1086/497989},
year = {2006},
month = {jan},
publisher = {},
volume = {131},
number = {1},
pages = {527},
author = {Jha, Saurabh and Kirshner, Robert P. and Challis, Peter and Garnavich, Peter M. and Matheson, Thomas and Soderberg, Alicia M. and Graves, Genevieve J. M. and Hicken, Malcolm and Alves, João F. and Arce, Héctor G. and Balog, Zoltan and Barmby, Pauline and Barton, Elizabeth J. and Berlind, Perry and Bragg, Ann E. and Briceño, César and Brown, Warren R. and Buckley, James H. and Caldwell, Nelson and Calkins, Michael L. and Carter, Barbara J. and Concannon, Kristi Dendy and Donnelly, R. Hank and Eriksen, Kristoffer A. and Fabricant, Daniel G. and Falco, Emilio E. and Fiore, Fabrizio and Garcia, Michael R. and Gómez, Mercedes and Grogin, Norman A. and Groner, Ted and Groot, Paul J. and Haisch, Jr., Karl E. and Hartmann, Lee and Hergenrother, Carl W. and Holman, Matthew J. and Huchra, John P. and Jayawardhana, Ray and Jerius, Diab and Kannappan, Sheila J. and Kim, Dong-Woo and Kleyna, Jan T. and Kochanek, Christopher S. and Koranyi, Daniel M. and Krockenberger, Martin and Lada, Charles J. and Luhman, Kevin L. and Luu, Jane X. and Macri, Lucas M. and Mader, Jeff A. and Mahdavi, Andisheh and Marengo, Massimo and Marsden, Brian G. and McLeod, Brian A. and McNamara, Brian R. and Megeath, S. Thomas and Moraru, Dan and Mossman, Amy E. and Muench, August A. and Muñoz, Jose A. and Muzerolle, James and Naranjo, Orlando and Nelson-Patel, Kristin and Pahre, Michael A. and Patten, Brian M. and Peters, James and Peters, Wayne and Raymond, John C. and Rines, Kenneth and Schild, Rudolph E. and Sobczak, Gregory J. and Spahr, Timothy B. and Stauffer, John R. and Stefanik, Robert P. and Szentgyorgyi, Andrew H. and Tollestrup, Eric V. and Väisänen, Petri and Vikhlinin, Alexey and Wang, Zhong and Willner, S. P. and Wolk, Scott J. and Zajac, Joseph M. and Zhao, Ping and Stanek, Krzysztof Z.},
title = {UBVRI Light Curves of 44 Type Ia Supernovae},
journal = {\aj}
}

@article{stahl_2019,
    author = {Stahl, Benjamin E and Zheng, WeiKang and de Jaeger, Thomas and Filippenko, Alexei V and Bigley, Andrew and Blanchard, Kyle and Blanchard, Peter K and Brink, Thomas G and Cargill, Samantha K and Casper, Chadwick and Channa, Sanyum and Choi, Byung Yun and Choksi, Nick and Chu, Jason and Clubb, Kelsey I and Cohen, Daniel P and Ellison, Michael and Falcon, Edward and Fazeli, Pegah and Fuller, Kiera and Ganeshalingam, Mohan and Gates, Elinor L and Gould, Carolina and Halevi, Goni and Hayakawa, Kevin T and Hestenes, Julia and Jeffers, Benjamin T and Joubert, Niels and Kandrashoff, Michael T and Kim, Minkyu and Kim, Haejung and Kislak, Michelle E and Kleiser, Io and Kong, Jason J and de Kouchkovsky, Maxime and Krishnan, Daniel and Kumar, Sahana and Leja, Joel and Leonard, Erin J and Li, Gary Z and Li, Weidong and Lu, Philip and Mason, Michelle N and Molloy, Jeffrey and Pina, Kenia and Rex, Jacob and Ross, Timothy W and Stegman, Samantha and Tang, Kevin and Thrasher, Patrick and Wang, Xianggao and Wilkins, Andrew and Yuk, Heechan and Yunus, Sameen and Zhang, Keto},
    title = {Lick Observatory Supernova Search follow-up program: photometry data release of 93 Type Ia supernovae},
    journal = {\mnras},
    volume = {490},
    number = {3},
    pages = {3882-3907},
    year = {2019},
    month = {10},
    issn = {0035-8711},
    doi = {10.1093/mnras/stz2742},
    url = {https://doi.org/10.1093/mnras/stz2742},
    eprint = {https://academic.oup.com/mnras/article-pdf/490/3/3882/30370150/stz2742.pdf},
}

@Article{Brown2014,
author={Brown, Peter J.
and Breeveld, Alice A.
and Holland, Stephen
and Kuin, Paul
and Pritchard, Tyler},
title={SOUSA: the Swift Optical/Ultraviolet Supernova Archive},
journal={Astrophysics and Space Science},
year={2014},
month={Nov},
day={01},
volume={354},
number={1},
pages={89-96},
issn={1572-946X},
doi={10.1007/s10509-014-2059-8},
url={https://doi.org/10.1007/s10509-014-2059-8}
}

@article{Riess_2001,
doi = {10.1086/322348},
url = {https://dx.doi.org/10.1086/322348},
year = {2001},
month = {oct},
publisher = {},
volume = {560},
number = {1},
pages = {49},
author = {Riess, Adam G. and Nugent, Peter E. and Gilliland, Ronald L. and Schmidt, Brian P. and Tonry, John and Dickinson, Mark and Thompson, Rodger I. and Budavári, Tamás and Casertano, Stefano and Evans, Aaron S. and Filippenko, Alexei V. and Livio, Mario and Sanders, David B. and Shapley, Alice E. and Spinrad, Hyron and Steidel, Charles C. and Stern, Daniel and Surace, Jason and Veilleux, Sylvain},
title = {The Farthest Known Supernova: Support for an Accelerating Universe and a Glimpse of the Epoch of Deceleration*},
journal = {\apj}
}

@article{Riess_2004,
doi = {10.1086/383612},
url = {https://dx.doi.org/10.1086/383612},
year = {2004},
month = {jun},
publisher = {},
volume = {607},
number = {2},
pages = {665},
author = {Riess, Adam G. and Strolger, Louis-Gregory and Tonry, John and Casertano, Stefano and Ferguson, Henry C. and Mobasher, Bahram and Challis, Peter and Filippenko, Alexei V. and Jha, Saurabh and Li, Weidong and Chornock, Ryan and Kirshner, Robert P. and Leibundgut, Bruno and Dickinson, Mark and Livio, Mario and Giavalisco, Mauro and Steidel, Charles C. and Benítez, Txitxo and Tsvetanov, Zlatan},
title = {Type Ia Supernova Discoveries at z &gt; 1 from the Hubble Space Telescope: Evidence for Past Deceleration and Constraints on Dark Energy Evolution*},
journal = {\apj}
}

@article{Riess_2018,
doi = {10.3847/1538-4357/aaa5a9},
url = {https://dx.doi.org/10.3847/1538-4357/aaa5a9},
year = {2018},
month = {jan},
publisher = {The American Astronomical Society},
volume = {853},
number = {2},
pages = {126},
author = {Riess, Adam G. and Rodney, Steven A. and Scolnic, Daniel M. and Shafer, Daniel L. and Strolger, Louis-Gregory and Ferguson, Henry C. and Postman, Marc and Graur, Or and Maoz, Dan and Jha, Saurabh W. and Mobasher, Bahram and Casertano, Stefano and Hayden, Brian and Molino, Alberto and Hjorth, Jens and Garnavich, Peter M. and Jones, David O. and Kirshner, Robert P. and Koekemoer, Anton M. and Grogin, Norman A. and Brammer, Gabriel and Hemmati, Shoubaneh and Dickinson, Mark and Challis, Peter M. and Wolff, Schuyler and Clubb, Kelsey I. and Filippenko, Alexei V. and Nayyeri, Hooshang and Vivian, U and Koo, David C. and Faber, Sandra M. and Kocevski, Dale and Bradley, Larry and Coe, Dan},
title = {Type Ia Supernova Distances at Redshift &gt;1.5 from the Hubble Space Telescope Multi-cycle Treasury Programs: The Early Expansion Rate},
journal = {\apj}
}

@article{Suzuki_2012,
doi = {10.1088/0004-637X/746/1/85},
url = {https://dx.doi.org/10.1088/0004-637X/746/1/85},
year = {2012},
month = {jan},
publisher = {The American Astronomical Society},
volume = {746},
number = {1},
pages = {85},
author = {Suzuki, N. and Rubin, D. and Lidman, C. and Aldering, G. and Amanullah, R. and Barbary, K. and Barrientos, L. F. and Botyanszki, J. and Brodwin, M. and Connolly, N. and Dawson, K. S. and Dey, A. and Doi, M. and Donahue, M. and Deustua, S. and Eisenhardt, P. and Ellingson, E. and Faccioli, L. and Fadeyev, V. and Fakhouri, H. K. and Fruchter, A. S. and Gilbank, D. G. and Gladders, M. D. and Goldhaber, G. and Gonzalez, A. H. and Goobar, A. and Gude, A. and Hattori, T. and Hoekstra, H. and Hsiao, E. and Huang, X. and Ihara, Y. and Jee, M. J. and Johnston, D. and Kashikawa, N. and Koester, B. and Konishi, K. and Kowalski, M. and Linder, E. V. and Lubin, L. and Melbourne, J. and Meyers, J. and Morokuma, T. and Munshi, F. and Mullis, C. and Oda, T. and Panagia, N. and Perlmutter, S. and Postman, M. and Pritchard, T. and Rhodes, J. and Ripoche, P. and Rosati, P. and Schlegel, D. J. and Spadafora, A. and Stanford, S. A. and Stanishev, V. and Stern, D. and Strovink, M. and Takanashi, N. and Tokita, K. and Wagner, M. and Wang, L. and Yasuda, N. and Yee, H. K. C. and (The Supernova Cosmology Project)},
title = {THE HUBBLE SPACE TELESCOPE CLUSTER SUPERNOVA SURVEY. V. IMPROVING THE DARK-ENERGY CONSTRAINTS ABOVE z &gt; 1 AND BUILDING AN EARLY-TYPE-HOSTED SUPERNOVA SAMPLE*},
journal = {\apj}
}

@ARTICLE{tsvetkov2010sn2008fvtypeia,
       author = {{Tsvetkov}, D.~Y. and {Elenin}, L.},
        title = "{SN 2008fv: the Third Type Ia Supernova in NGC 3147}",
      journal = {Peremennye Zvezdy},
         year = 2010,
        month = mar,
       volume = {30},
       number = {2},
        pages = {2},
          doi = {10.48550/arXiv.1003.2558},
archivePrefix = {arXiv},
       eprint = {1003.2558},
 primaryClass = {astro-ph.SR},
       adsurl = {https://ui.adsabs.harvard.edu/abs/2010PZ.....30....2T}
}

@article{Kawabata_2020,
doi = {10.3847/1538-4357/ab8236},
url = {https://dx.doi.org/10.3847/1538-4357/ab8236},
year = {2020},
month = {apr},
publisher = {The American Astronomical Society},
volume = {893},
number = {2},
pages = {143},
author = {Kawabata, Miho and Maeda, Keiichi and Yamanaka, Masayuki and Nakaoka, Tatsuya and Kawabata, Koji S. and Adachi, Ryo and Akitaya, Hiroshi and Burgaz, Umut and Hanayama, Hidekazu and Horiuchi, Takashi and Hosokawa, Ryohei and Iida, Kota and Imazato, Fumiya and Isogai, Keisuke and Jiang, Ji-an and Katoh, Noriyuki and Kimura, Hiroki and Kino, Masaru and Kuroda, Daisuke and Maehara, Hiroyuki and Matsubayashi, Kazuya and Morihana, Kumiko and Murata, Katsuhiro L. and Nagao, Takashi and Niwano, Masafumi and Nogami, Daisaku and Oeda, Motoki and Ono, Tatsuharu and Onozato, Hiroki and Otsuka, Masaaki and Saito, Tomoki and Sasada, Mahito and Shiraishi, Kazuki and Sugiyama, Haruki and Taguchi, Kenta and Takahashi, Jun and Takagi, Kengo and Takagi, Seiko and Takayama, Masaki and Tozuka, Miyako and Sekiguchi, Kazuhiro},
title = {SN 2019ein: New Insights into the Similarities and Diversity among High-velocity Type Ia Supernovae},
journal = {\apj}
}

@article{Zhang_2010,
doi = {10.1086/649851},
url = {https://dx.doi.org/10.1086/649851},
year = {2009},
month = {dec},
publisher = {University of Chicago Press},
volume = {122},
number = {887},
pages = {1},
author = {Zhang, T. and Wang, X. and Li, W. and Filippenko, A. V. and Wang, L. and Zhou, X. and Brown, P. J. and Silverman, J. M. and Steele, T. N. and Ganeshalingam, M. and Li, J. and Deng, J. and Li, T. and Qiu, Y. and Zhai, M. and Shang, R.},
title = {Optical Observations of the Rapidly Expanding Type Ia Supernova 2007gi},
journal = {\pasp}
}

@article{Stritzinger_2010,
doi = {10.1088/0004-6256/140/6/2036},
url = {https://dx.doi.org/10.1088/0004-6256/140/6/2036},
year = {2010},
month = {nov},
publisher = {The American Astronomical Society},
volume = {140},
number = {6},
pages = {2036},
author = {Stritzinger, Maximilian and Burns, Christopher R. and Phillips, Mark M. and Folatelli, Gastón and Krisciunas, Kevin and Kattner, ShiAnne and Persson, Sven E. and Boldt, Luis and Campillay, Abdo and Contreras, Carlos and Krzeminski, Wojtek and Morrell, Nidia and Salgado, Francisco and Freedman, Wendy L. and Hamuy, Mario and Madore, Barry F. and Roth, Miguel and Suntzeff, Nicholas B.},
title = {THE DISTANCE TO NGC 1316 (FORNAX A) FROM OBSERVATIONS OF FOUR TYPE Ia SUPERNOVAE*},
journal = {\aj}
}

@article{Milne_2010,
doi = {10.1088/0004-637X/721/2/1627},
url = {https://dx.doi.org/10.1088/0004-637X/721/2/1627},
year = {2010},
month = {sep},
publisher = {The American Astronomical Society},
volume = {721},
number = {2},
pages = {1627},
author = {Milne, Peter A. and Brown, Peter J. and Roming, Peter W. A. and Holland, Stephen T. and Immler, Stefan and Filippenko, Alexei V. and Ganeshalingam, Mohan and Li, Weidong and Stritzinger, Maximilian and Phillips, Mark M. and Hicken, Malcolm and Kirshner, Robert P. and Challis, Peter J. and Mazzali, Paolo and Schmidt, Brian P. and Bufano, Filomena and Gehrels, Neil and Vanden Berk, Daniel},
title = {NEAR-ULTRAVIOLET PROPERTIES OF A LARGE SAMPLE OF TYPE Ia SUPERNOVAE AS OBSERVED WITH THE Swift UVOT},
journal = {\apj}
}

@article{Burns_2020,
doi = {10.3847/1538-4357/ab8e3e},
url = {https://dx.doi.org/10.3847/1538-4357/ab8e3e},
year = {2020},
month = {jun},
publisher = {The American Astronomical Society},
volume = {895},
number = {2},
pages = {118},
author = {Burns, Christopher R. and Ashall, Chris and Contreras, Carlos and Brown, Peter and Stritzinger, Maximilian and Phillips, M. M. and Flores, Ricardo and Suntzeff, Nicholas B. and Hsiao, Eric Y. and Uddin, Syed and Simon, Joshua D. and Krisciunas, Kevin and Campillay, Abdo and Foley, Ryan J. and Freedman, Wendy L. and Galbany, Lluís and González, Consuelo and Hoeflich, Peter and Holmbo, S. and Kilpatrick, Charles D. and Kirshner, Robert P. and Morrell, Nidia and Muñoz-Elgueta, Nahir and Piro, Anthony L. and Rojas-Bravo, César and Sand, David and Vargas-González, Jaime and Ulloa, Natalie and Vilchez, Jorge Anais},
title = {SN 2013aa and SN 2017cbv: Two Sibling Type Ia Supernovae in the Spiral Galaxy NGC 5643},
journal = {\apj}
}

@article{Burns_2018,
doi = {10.3847/1538-4357/aae51c},
url = {https://dx.doi.org/10.3847/1538-4357/aae51c},
year = {2018},
month = {dec},
publisher = {The American Astronomical Society},
volume = {869},
number = {1},
pages = {56},
author = {Burns, Christopher R. and Parent, Emilie and Phillips, M. M. and Stritzinger, Maximilian and Krisciunas, Kevin and Suntzeff, Nicholas B. and Hsiao, Eric Y. and Contreras, Carlos and Anais, Jorge and Boldt, Luis and Busta, Luis and Campillay, Abdo and Castellón, Sergio and Folatelli, Gastón and Freedman, Wendy L. and González, Consuelo and Hamuy, Mario and Heoflich, Peter and Krzeminski, Wojtek and Madore, Barry F. and Morrell, Nidia and Persson, S. E. and Roth, Miguel and Salgado, Francisco and Serón, Jacqueline and Torres, Simón},
title = {The Carnegie Supernova Project: Absolute Calibration and the Hubble Constant},
journal = {\apj}
}

@article{Gall_2018,
	author = {{Gall, C.} and {Stritzinger, M. D.} and {Ashall, C.} and {Baron, E.} and {Burns, C. R.} and {Hoeflich, P.} and {Hsiao, E. Y.} and {Mazzali, P. A.} and {Phillips, M. M.} and {Filippenko, A. V.} and {Anderson, J. P.} and {Benetti, S.} and {Brown, P. J.} and {Campillay, A.} and {Challis, P.} and {Contreras, C.} and {Elias de la Rosa, N.} and {Folatelli, G.} and {Foley, R. J.} and {Fraser, M.} and {Holmbo, S.} and {Marion, G. H.} and {Morrell, N.} and {Pan, Y.-C.} and {Pignata, G.} and {Suntzeff, N. B.} and {Taddia, F.} and {Torres Robledo, S.} and {Valenti, S.}},
	title = {Two transitional type Ia supernovae located in the Fornax cluster member NGC 1404: SN 2007on and SN 2011iv★},
	DOI= "10.1051/0004-6361/201730886",
	url= "https://doi.org/10.1051/0004-6361/201730886",
	journal = {A&A},
	year = 2018,
	volume = 611,
	pages = "A58",
}

@ARTICLE{vincenzi25,
       author = {{Vincenzi}, M. and {Kessler}, R. and {Shah}, P. and {Lee}, J. and {Davis}, T.~M. and {Scolnic}, D. and {Armstrong}, P. and {Brout}, D. and {Camilleri}, R. and {Chen}, R. and {Galbany}, L. and {Lidman}, C. and {M{\"o}ller}, A. and {Popovic}, B. and {Rose}, B. and {Sako}, M. and {S{\'a}nchez}, B.~O. and {Smith}, M. and {Sullivan}, M. and {Wiseman}, P. and {Abbott}, T.~M.~C. and {Aguena}, M. and {Allam}, S. and {Andrade-Oliveira}, F. and {Bocquet}, S. and {Brooks}, D. and {Carnero Rosell}, A. and {Carretero}, J. and {da Costa}, L.~N. and {Pereira}, M.~E.~S. and {Diehl}, H.~T. and {Doel}, P. and {Everett}, S. and {Flaugher}, B. and {Frieman}, J. and {Garc{\'\i}a-Bellido}, J. and {Gaztanaga}, E. and {Gruen}, D. and {Gruendl}, R.~A. and {Gutierrez}, G. and {Hinton}, S.~R. and {Hollowood}, D.~L. and {Honscheid}, K. and {James}, D.~J. and {Kuehn}, K. and {Lahav}, O. and {Lee}, S. and {Marshall}, J.~L. and {Mena-Fern{\'a}ndez}, J. and {Miquel}, R. and {Muir}, J. and {Myles}, J. and {Palmese}, A. and {Plazas Malag{\'o}n}, A.~A. and {Porredon}, A. and {Samuroff}, S. and {Sanchez}, E. and {Sanchez Cid}, D. and {Sevilla-Noarbe}, I. and {Suchyta}, E. and {Tarle}, G. and {To}, C. and {Tucker}, D.~L. and {Vikram}, V. and {Walker}, A.~R. and {Weaverdyck}, N. and {Weller}, J.},
        title = "{Comparing the DES-SN5YR and Pantheon+ SN cosmology analyses: investigation based on 'evolving dark energy or supernovae systematics'?}",
      journal = {\mnras},
         year = 2025,
        month = aug,
       volume = {541},
       number = {3},
        pages = {2585-2593},
          doi = {10.1093/mnras/staf943},
archivePrefix = {arXiv},
       eprint = {2501.06664},
 primaryClass = {astro-ph.CO},
       adsurl = {https://ui.adsabs.harvard.edu/abs/2025MNRAS.541.2585V}
}

@article{Wiseman20,
    author = {Wiseman, P and Smith, M and Childress, M and Kelsey, L and Möller, A and Gupta, R R and Swann, E and Angus, C R and Brout, D and Davis, T M and Foley, R J and Frohmaier, C and Galbany, L and Gutiérrez, C P and Inserra, C and Kessler, R and Lewis, G F and Lidman, C and Macaulay, E and Nichol, R C and Pursiainen, M and Sako, M and Scolnic, D and Sommer, N E and Sullivan, M and Tucker, B E and Abbott, T M C and Aguena, M and Allam, S and Avila, S and Bertin, E and Brooks, D and Buckley-Geer, E and Burke, D L and Carnero Rosell, A and Carollo, D and Carrasco Kind, M and da Costa, L N and De Vicente, J and Desai, S and Diehl, H T and Doel, P and Eifler, T F and Everett, S and Fosalba, P and Frieman, J and García-Bellido, J and Gaztanaga, E and Gerdes, D W and Gill, M S S and Glazebrook, K and Gruendl, R A and Gschwend, J and Hartley, W G and Hinton, S R and Hollowood, D L and Honscheid, K and James, D J and Kuehn, K and Kuropatkin, N and Lima, M and Maia, M A G and March, M and Martini, P and Melchior, P and Menanteau, F and Miquel, R and Ogando, R L C and Paz-Chinchón, F and Plazas, A A and Romer, A K and Roodman, A and Sanchez, E and Scarpine, V and Serrano, S and Suchyta, E and Swanson, M E C and Tarle, G and Thomas, D and Tucker, D L and Varga, T N and Walker, A R and Wilkinson, R D and (DES Collaboration)},
    title = {Supernova host galaxies in the dark energy survey: I. Deep coadds, photometry, and stellar masses},
    journal = {\mnras},
    volume = {495},
    number = {4},
    pages = {4040-4060},
    year = {2020},
    month = {05},
    issn = {0035-8711},
    doi = {10.1093/mnras/staa1302},
    url = {https://doi.org/10.1093/mnras/staa1302},
    eprint = {https://academic.oup.com/mnras/article-pdf/495/4/4040/33371290/staa1302.pdf},
}

@article{guy05,
	author = {Guy, J. and {Astier, P.} and {Nobili, S.} and {Regnault, N.} and {Pain, R.}},
	title = {SALT: a spectral adaptive light curve template  for type Ia supernovae},
	DOI= "10.1051/0004-6361:20053025",
	url= "https://doi.org/10.1051/0004-6361:20053025",
	journal = {A&A},
	year = 2005,
	volume = 443,
	number = 3,
	pages = "781-791",
}

@article{Brout_2022scal,
doi = {10.3847/1538-4357/ac8bcc},
url = {https://dx.doi.org/10.3847/1538-4357/ac8bcc},
year = {2022},
month = {oct},
publisher = {The American Astronomical Society},
volume = {938},
number = {2},
pages = {111},
author = {Brout, Dillon and Taylor, Georgie and Scolnic, Dan and Wood, Charlotte M. and Rose, Benjamin M. and Vincenzi, Maria and Dwomoh, Arianna and Lidman, Christopher and Riess, Adam and Ali, Noor and Qu, Helen and Dai, Mi},
title = {The Pantheon+ Analysis: SuperCal-fragilistic Cross Calibration, Retrained SALT2 Light-curve Model, and Calibration Systematic Uncertainty},
journal = {\apj}
}

@article{Brout_2021dust,
doi = {10.3847/1538-4357/abd69b},
url = {https://dx.doi.org/10.3847/1538-4357/abd69b},
year = {2021},
month = {mar},
publisher = {The American Astronomical Society},
volume = {909},
number = {1},
pages = {26},
author = {Brout, Dillon and Scolnic, Daniel},
title = {It’s Dust: Solving the Mysteries of the Intrinsic Scatter and Host-galaxy Dependence of Standardized Type Ia Supernova Brightnesses},
journal = {\apj}
}

@ARTICLE{chevallier01,
       author = {{Chevallier}, Michel and {Polarski}, David},
        title = "{Accelerating Universes with Scaling Dark Matter}",
      journal = {IJMP D},
         year = 2001,
        month = jan,
       volume = {10},
       number = {2},
        pages = {213-223},
          doi = {10.1142/S0218271801000822},
archivePrefix = {arXiv},
       eprint = {gr-qc/0009008},
 primaryClass = {gr-qc},
       adsurl = {https://ui.adsabs.harvard.edu/abs/2001IJMPD..10..213C}
}

@ARTICLE{efstathiou25,
       author = {{Efstathiou}, George},
        title = "{Evolving dark energy or supernovae systematics?}",
      journal = {\mnras},
         year = 2025,
        month = apr,
       volume = {538},
       number = {2},
        pages = {875-882},
          doi = {10.1093/mnras/staf301},
archivePrefix = {arXiv},
       eprint = {2408.07175},
 primaryClass = {astro-ph.CO},
       adsurl = {https://ui.adsabs.harvard.edu/abs/2025MNRAS.538..875E}
}

@ARTICLE{huang25,
       author = {{Huang}, Lu and {Cai}, Rong-Gen and {Wang}, Shao-Jiang},
        title = "{The DESI DR1/DR2 evidence for dynamical dark energy is biased by low-redshift supernovae}",
      journal = {Science China Physics, Mechanics, and Astronomy},
         year = 2025,
        month = aug,
       volume = {68},
       number = {10},
          eid = {100413},
        pages = {100413},
          doi = {10.1007/s11433-025-2754-5},
archivePrefix = {arXiv},
       eprint = {2502.04212},
 primaryClass = {astro-ph.CO},
       adsurl = {https://ui.adsabs.harvard.edu/abs/2025SCPMA..6800413H}
}

@ARTICLE{Notari24,
       author = {{Notari}, Alessio and {Redi}, Michele and {Tesi}, Andrea},
        title = "{BAO vs. SN evidence for evolving dark energy}",
      journal = {\jcap},
         year = 2025,
        month = apr,
       volume = {2025},
       number = {4},
          eid = {048},
        pages = {048},
          doi = {10.1088/1475-7516/2025/04/048},
archivePrefix = {arXiv},
       eprint = {2411.11685},
 primaryClass = {astro-ph.CO},
       adsurl = {https://ui.adsabs.harvard.edu/abs/2025JCAP...04..048N}
}

@ARTICLE{UNITY_2015,
       author = {{Rubin}, D. and {Aldering}, G. and {Barbary}, K. and {Boone}, K. and {Chappell}, G. and {Currie}, M. and {Deustua}, S. and {Fagrelius}, P. and {Fruchter}, A. and {Hayden}, B. and {Lidman}, C. and {Nordin}, J. and {Perlmutter}, S. and {Saunders}, C. and {Sofiatti}, C. and {Supernova Cosmology Project}, The},
        title = "{UNITY: Confronting Supernova Cosmology's Statistical and Systematic Uncertainties in a Unified Bayesian Framework}",
      journal = {\apj},
         year = 2015,
        month = nov,
       volume = {813},
       number = {2},
          eid = {137},
        pages = {137},
          doi = {10.1088/0004-637X/813/2/137},
archivePrefix = {arXiv},
       eprint = {1507.01602},
 primaryClass = {astro-ph.CO},
       adsurl = {https://ui.adsabs.harvard.edu/abs/2015ApJ...813..137R}
}

@ARTICLE{desidr2,
       author = {{DESI Collaboration} and {Abdul Karim}, M. and {Aguilar}, J. and {Ahlen}, S. and {Alam}, S. and {Allen}, L. and {Allende Prieto}, C. and {Alves}, O. and {Anand}, A. and {Andrade}, U. and {Armengaud}, E. and {Aviles}, A. and {Bailey}, S. and {Baltay}, C. and {Bansal}, P. and {Bault}, A. and {Behera}, J. and {BenZvi}, S. and {Bianchi}, D. and {Blake}, C. and {Brieden}, S. and {Brodzeller}, A. and {Brooks}, D. and {Buckley-Geer}, E. and {Burtin}, E. and {Calderon}, R. and {Canning}, R. and {Rosell}, A. Carnero and {Carrilho}, P. and {Casas}, L. and {Castander}, F.~J. and {Charles}, M. and {Chaussidon}, E. and {Chaves-Montero}, J. and {Chebat}, D. and {Chen}, X. and {Claybaugh}, T. and {Cole}, S. and {Cooper}, A.~P. and {Cuceu}, A. and {Dawson}, K.~S. and {de la Macorra}, A. and {de Mattia}, A. and {Deiosso}, N. and {Della Costa}, J. and {Demina}, R. and {Dey}, A. and {Dey}, B. and {Ding}, Z. and {Doel}, P. and {Edelstein}, J. and {Eisenstein}, D.~J. and {Elbers}, W. and {Fagrelius}, P. and {Fanning}, K. and {Fern{\'a}ndez-Garc{\'\i}a}, E. and {Ferraro}, S. and {Font-Ribera}, A. and {Forero-Romero}, J.~E. and {Frenk}, C.~S. and {Garcia-Quintero}, C. and {Garrison}, L.~H. and {Gazta{\~n}aga}, E. and {Gil-Mar{\'\i}n}, H. and {Gontcho A Gontcho}, S. and {Gonzalez}, D. and {Gonzalez-Morales}, A.~X. and {Gordon}, C. and {Green}, D. and {Gutierrez}, G. and {Guy}, J. and {Hadzhiyska}, B. and {Hahn}, C. and {He}, S. and {Herbold}, M. and {Herrera-Alcantar}, H.~K. and {Ho}, M.-F. and {Honscheid}, K. and {Howlett}, C. and {Huterer}, D. and {Ishak}, M. and {Juneau}, S. and {Kamble}, N.~V. and {Kara{\c{c}}ayl{\i}}, N.~G. and {Kehoe}, R. and {Kent}, S. and {Kim}, A.~G. and {Kirkby}, D. and {Kisner}, T. and {Koposov}, S.~E. and {Kremin}, A. and {Krolewski}, A. and {Lahav}, O. and {Lamman}, C. and {Landriau}, M. and {Lang}, D. and {Lasker}, J. and {Le Goff}, J.~M. and {Le Guillou}, L. and {Leauthaud}, A. and {Levi}, M.~E. and {Li}, Q. and {Li}, T.~S. and {Lodha}, K. and {Lokken}, M. and {Lozano-Rodr{\'\i}guez}, F. and {Magneville}, C. and {Manera}, M. and {Martini}, P. and {Matthewson}, W.~L. and {Meisner}, A. and {Mena-Fern{\'a}ndez}, J. and {Menegas}, A. and {Mergulh{\~a}o}, T. and {Miquel}, R. and {Moustakas}, J. and {Mu{\~n}oz-Guti{\'e}rrez}, A. and {Mu{\~n}oz-Santos}, D. and {Myers}, A.~D. and {Nadathur}, S. and {Naidoo}, K. and {Napolitano}, L. and {Newman}, J.~A. and {Niz}, G. and {Noriega}, H.~E. and {Paillas}, E. and {Palanque-Delabrouille}, N. and {Pan}, J. and {Peacock}, J.~A. and {Pellejero Ibanez}, M. and {Percival}, W.~J. and {P{\'e}rez-Fern{\'a}ndez}, A. and {P{\'e}rez-R{\`a}fols}, I. and {Pieri}, M.~M. and {Poppett}, C. and {Prada}, F. and {Rabinowitz}, D. and {Raichoor}, A. and {Ram{\'\i}rez-P{\'e}rez}, C. and {Rashkovetskyi}, M. and {Ravoux}, C. and {Rich}, J. and {Rocher}, A. and {Rockosi}, C. and {Rohlf}, J. and {Rom{\'a}n-Herrera}, J.~O. and {Ross}, A.~J. and {Rossi}, G. and {Ruggeri}, R. and {Ruhlmann-Kleider}, V. and {Samushia}, L. and {Sanchez}, E. and {Sanders}, N. and {Schlegel}, D. and {Schubnell}, M. and {Seo}, H. and {Shafieloo}, A. and {Sharples}, R. and {Silber}, J. and {Sinigaglia}, F. and {Sprayberry}, D. and {Tan}, T. and {Tarl{\'e}}, G. and {Taylor}, P. and {Turner}, W. and {Ure{\~n}a-L{\'o}pez}, L.~A. and {Vaisakh}, R. and {Valdes}, F. and {Valogiannis}, G. and {Vargas-Maga{\~n}a}, M. and {Verde}, L. and {Walther}, M. and {Weaver}, B.~A. and {Weinberg}, D.~H. and {White}, M. and {Wolfson}, M. and {Y{\`e}che}, C. and {Yu}, J. and {Zaborowski}, E.~A. and {Zarrouk}, P. and {Zhai}, Z. and {Zhang}, H. and {Zhao}, C. and {Zhao}, G.~B. and {Zhou}, R. and {Zou}, H. and {DESI Collaboration}},
        title = "{DESI DR2 results. II. Measurements of baryon acoustic oscillations and cosmological constraints}",
      journal = {\prd},
         year = 2025,
        month = oct,
       volume = {112},
       number = {8},
          eid = {083515},
        pages = {083515},
          doi = {10.1103/tr6y-kpc6},
archivePrefix = {arXiv},
       eprint = {2503.14738},
 primaryClass = {astro-ph.CO},
       adsurl = {https://ui.adsabs.harvard.edu/abs/2025PhRvD.112h3515A}
}

@article{ZUNTZ201545,
title = {CosmoSIS: Modular cosmological parameter estimation},
journal = {Astronomy and Computing},
volume = {12},
pages = {45-59},
year = {2015},
issn = {2213-1337},
doi = {https://doi.org/10.1016/j.ascom.2015.05.005},
url = {https://www.sciencedirect.com/science/article/pii/S2213133715000591},
author = {J. Zuntz and M. Paterno and E. Jennings and D. Rudd and A. Manzotti and S. Dodelson and S. Bridle and S. Sehrish and J. Kowalkowski}
}

@ARTICLE{p25a,
       author = {{Popovic}, B. and {Kenworthy}, W.~D. and {Ginolin}, M. and {Goobar}, A. and {Shah}, P. and {Boyd}, B.~M. and {Do}, A. and {Brout}, D. and {Scolnic}, D. and {Vincenzi}, M. and {Dhawan}, S. and {Jones}, D.~O. and {Smith}, M. and {Rigault}, M. and {Racine}, B. and {Hayes}, E.~E. and {Chen}, R. and {Wiseman}, P. and {Galbany}, L. and {Grayling}, M. and {LaCroix}, L. and {Barjou-Delayre}, C. and {Kuhn}, D. and {Lemon}, C.},
        title = "{A Reassessment of the Pantheon+ and DES 5YR Calibration Uncertainties: Dovekie}",
      journal = {arXiv e-prints},
         year = 2025,
        month = jun,
          eid = {arXiv:2506.05471},
        pages = {arXiv:2506.05471},
          doi = {10.48550/arXiv.2506.05471},
archivePrefix = {arXiv},
       eprint = {2506.05471},
 primaryClass = {astro-ph.CO},
       adsurl = {https://ui.adsabs.harvard.edu/abs/2025arXiv250605471P}
}

@ARTICLE{Boyd25,
       author = {{Boyd}, Benjamin M. and {Narayan}, Gautham and {Mandel}, Kaisey S. and {Grayling}, Matthew and {Saha}, Abhijit and {Axelrod}, Tim and {Matheson}, Thomas and {Olszewski}, Edward W. and {Calamida}, Annalisa and {Do}, Aaron and {Bohlin}, Ralph C. and {Holberg}, Jay B. and {Hubeny}, Ivan and {Deustua}, Susana and {Rest}, Armin and {Stubbs}, Christopher W. and {Berres}, Aidan and {Li}, Mai and {Mackenty}, John W. and {Sabbi}, Elena},
        title = "{DAmodel: hierarchical Bayesian modelling of DA white dwarfs for spectrophotometric calibration}",
      journal = {\mnras},
         year = 2025,
        month = jun,
       volume = {540},
       number = {1},
        pages = {385-415},
          doi = {10.1093/mnras/staf629},
archivePrefix = {arXiv},
       eprint = {2412.08809},
 primaryClass = {astro-ph.IM},
       adsurl = {https://ui.adsabs.harvard.edu/abs/2025MNRAS.540..385B}
}

@ARTICLE{axelrod2023,
       author = {{Axelrod}, Tim and {Saha}, Abhijit and {Matheson}, Thomas and {Olszewski}, Edward W. and {Bohlin}, Ralph C. and {Calamida}, Annalisa and {Claver}, Jenna and {Deustua}, Susana and {Holberg}, Jay B. and {Hubeny}, Ivan and {Mackenty}, John W. and {Malanchev}, Konstantin and {Narayan}, Gautham and {Points}, Sean and {Rest}, Armin and {Sabbi}, Elena and {Stubbs}, Christopher W.},
        title = "{All-sky Faint DA White Dwarf Spectrophotometric Standards for Astrophysical Observatories: The Complete Sample}",
      journal = {\apj},
         year = 2023,
        month = jul,
       volume = {951},
       number = {1},
          eid = {78},
        pages = {78},
          doi = {10.3847/1538-4357/acd333},
archivePrefix = {arXiv},
       eprint = {2305.07563},
 primaryClass = {astro-ph.SR},
       adsurl = {https://ui.adsabs.harvard.edu/abs/2023ApJ...951...78A}
}

@ARTICLE{narayan2019,
       author = {{Narayan}, Gautham and {Matheson}, Thomas and {Saha}, Abhijit and {Axelrod}, Tim and {Calamida}, Annalisa and {Olszewski}, Edward and {Claver}, Jenna and {Mandel}, Kaisey S. and {Bohlin}, Ralph C. and {Holberg}, Jay B. and {Deustua}, Susana and {Rest}, Armin and {Stubbs}, Christopher W. and {Shanahan}, Clare E. and {Vaz}, Amali L. and {Zenteno}, Alfredo and {Strampelli}, Giovanni and {Hubeny}, Ivan and {Points}, Sean and {Sabbi}, Elena and {Mackenty}, John},
        title = "{Subpercent Photometry: Faint DA White Dwarf Spectrophotometric Standards for Astrophysical Observatories}",
      journal = {\apjs},
         year = 2019,
        month = apr,
       volume = {241},
       number = {2},
          eid = {20},
        pages = {20},
          doi = {10.3847/1538-4365/ab0557},
archivePrefix = {arXiv},
       eprint = {1811.12534},
 primaryClass = {astro-ph.IM},
       adsurl = {https://ui.adsabs.harvard.edu/abs/2019ApJS..241...20N}
}

@ARTICLE{GAIAONE,
       author = {{Gaia Collaboration} and {Prusti}, T. and {de Bruijne}, J.~H.~J. and {Brown}, A.~G.~A. and {Vallenari}, A. and {Babusiaux}, C. and {Bailer-Jones}, C.~A.~L. and {Bastian}, U. and {Biermann}, M. and {Evans}, D.~W. and {Eyer}, L. and {Jansen}, F. and {Jordi}, C. and {Klioner}, S.~A. and {Lammers}, U. and {Lindegren}, L. and {Luri}, X. and {Mignard}, F. and {Milligan}, D.~J. and {Panem}, C. and {Poinsignon}, V. and {Pourbaix}, D. and {Randich}, S. and {Sarri}, G. and {Sartoretti}, P. and {Siddiqui}, H.~I. and {Soubiran}, C. and {Valette}, V. and {van Leeuwen}, F. and {Walton}, N.~A. and {Aerts}, C. and {Arenou}, F. and {Cropper}, M. and {Drimmel}, R. and {H{\o}g}, E. and {Katz}, D. and {Lattanzi}, M.~G. and {O'Mullane}, W. and {Grebel}, E.~K. and {Holland}, A.~D. and {Huc}, C. and {Passot}, X. and {Bramante}, L. and {Cacciari}, C. and {Casta{\~n}eda}, J. and {Chaoul}, L. and {Cheek}, N. and {De Angeli}, F. and {Fabricius}, C. and {Guerra}, R. and {Hern{\'a}ndez}, J. and {Jean-Antoine-Piccolo}, A. and {Masana}, E. and {Messineo}, R. and {Mowlavi}, N. and {Nienartowicz}, K. and {Ord{\'o}{\~n}ez-Blanco}, D. and {Panuzzo}, P. and {Portell}, J. and {Richards}, P.~J. and {Riello}, M. and {Seabroke}, G.~M. and {Tanga}, P. and {Th{\'e}venin}, F. and {Torra}, J. and {Els}, S.~G. and {Gracia-Abril}, G. and {Comoretto}, G. and {Garcia-Reinaldos}, M. and {Lock}, T. and {Mercier}, E. and {Altmann}, M. and {Andrae}, R. and {Astraatmadja}, T.~L. and {Bellas-Velidis}, I. and {Benson}, K. and {Berthier}, J. and {Blomme}, R. and {Busso}, G. and {Carry}, B. and {Cellino}, A. and {Clementini}, G. and {Cowell}, S. and {Creevey}, O. and {Cuypers}, J. and {Davidson}, M. and {De Ridder}, J. and {de Torres}, A. and {Delchambre}, L. and {Dell'Oro}, A. and {Ducourant}, C. and {Fr{\'e}mat}, Y. and {Garc{\'\i}a-Torres}, M. and {Gosset}, E. and {Halbwachs}, J. -L. and {Hambly}, N.~C. and {Harrison}, D.~L. and {Hauser}, M. and {Hestroffer}, D. and {Hodgkin}, S.~T. and {Huckle}, H.~E. and {Hutton}, A. and {Jasniewicz}, G. and {Jordan}, S. and {Kontizas}, M. and {Korn}, A.~J. and {Lanzafame}, A.~C. and {Manteiga}, M. and {Moitinho}, A. and {Muinonen}, K. and {Osinde}, J. and {Pancino}, E. and {Pauwels}, T. and {Petit}, J. -M. and {Recio-Blanco}, A. and {Robin}, A.~C. and {Sarro}, L.~M. and {Siopis}, C. and {Smith}, M. and {Smith}, K.~W. and {Sozzetti}, A. and {Thuillot}, W. and {van Reeven}, W. and {Viala}, Y. and {Abbas}, U. and {Abreu Aramburu}, A. and {Accart}, S. and {Aguado}, J.~J. and {Allan}, P.~M. and {Allasia}, W. and {Altavilla}, G. and {{\'A}lvarez}, M.~A. and {Alves}, J. and {Anderson}, R.~I. and {Andrei}, A.~H. and {Anglada Varela}, E. and {Antiche}, E. and {Antoja}, T. and {Ant{\'o}n}, S. and {Arcay}, B. and {Atzei}, A. and {Ayache}, L. and {Bach}, N. and {Baker}, S.~G. and {Balaguer-N{\'u}{\~n}ez}, L. and {Barache}, C. and {Barata}, C. and {Barbier}, A. and {Barblan}, F. and {Baroni}, M. and {Barrado y Navascu{\'e}s}, D. and {Barros}, M. and {Barstow}, M.~A. and {Becciani}, U. and {Bellazzini}, M. and {Bellei}, G. and {Bello Garc{\'\i}a}, A. and {Belokurov}, V. and {Bendjoya}, P. and {Berihuete}, A. and {Bianchi}, L. and {Bienaym{\'e}}, O. and {Billebaud}, F. and {Blagorodnova}, N. and {Blanco-Cuaresma}, S. and {Boch}, T. and {Bombrun}, A. and {Borrachero}, R. and {Bouquillon}, S. and {Bourda}, G. and {Bouy}, H. and {Bragaglia}, A. and {Breddels}, M.~A. and {Brouillet}, N. and {Br{\"u}semeister}, T. and {Bucciarelli}, B. and {Budnik}, F. and {Burgess}, P. and {Burgon}, R. and {Burlacu}, A. and {Busonero}, D. and {Buzzi}, R. and {Caffau}, E. and {Cambras}, J. and {Campbell}, H. and {Cancelliere}, R. and {Cantat-Gaudin}, T. and {Carlucci}, T. and {Carrasco}, J.~M. and {Castellani}, M. and {Charlot}, P. and {Charnas}, J. and {Charvet}, P. and {Chassat}, F. and {Chiavassa}, A. and {Clotet}, M. and {Cocozza}, G. and {Collins}, R.~S. and {Collins}, P. and {Costigan}, G. and {Crifo}, F. and {Cross}, N.~J.~G. and {Crosta}, M. and {Crowley}, C. and {Dafonte}, C. and {Damerdji}, Y. and {Dapergolas}, A. and {David}, P. and {David}, M. and {De Cat}, P. and {de Felice}, F. and {de Laverny}, P. and {De Luise}, F. and {De March}, R. and {de Martino}, D. and {de Souza}, R. and {Debosscher}, J. and {del Pozo}, E. and {Delbo}, M. and {Delgado}, A. and {Delgado}, H.~E. and {di Marco}, F. and {Di Matteo}, P. and {Diakite}, S. and {Distefano}, E. and {Dolding}, C. and {Dos Anjos}, S. and {Drazinos}, P. and {Dur{\'a}n}, J. and {Dzigan}, Y. and {Ecale}, E. and {Edvardsson}, B. and {Enke}, H. and {Erdmann}, M. and {Escolar}, D. and {Espina}, M. and {Evans}, N.~W. and {Eynard Bontemps}, G. and {Fabre}, C. and {Fabrizio}, M. and {Faigler}, S. and {Falc{\~a}o}, A.~J. and {Farr{\`a}s Casas}, M. and {Faye}, F. and {Federici}, L. and {Fedorets}, G. and {Fern{\'a}ndez-Hern{\'a}ndez}, J. and {Fernique}, P. and {Fienga}, A. and {Figueras}, F. and {Filippi}, F. and {Findeisen}, K. and {Fonti}, A. and {Fouesneau}, M. and {Fraile}, E. and {Fraser}, M. and {Fuchs}, J. and {Furnell}, R. and {Gai}, M. and {Galleti}, S. and {Galluccio}, L. and {Garabato}, D. and {Garc{\'\i}a-Sedano}, F. and {Gar{\'e}}, P. and {Garofalo}, A. and {Garralda}, N. and {Gavras}, P. and {Gerssen}, J. and {Geyer}, R. and {Gilmore}, G. and {Girona}, S. and {Giuffrida}, G. and {Gomes}, M. and {Gonz{\'a}lez-Marcos}, A. and {Gonz{\'a}lez-N{\'u}{\~n}ez}, J. and {Gonz{\'a}lez-Vidal}, J.~J. and {Granvik}, M. and {Guerrier}, A. and {Guillout}, P. and {Guiraud}, J. and {G{\'u}rpide}, A. and {Guti{\'e}rrez-S{\'a}nchez}, R. and {Guy}, L.~P. and {Haigron}, R. and {Hatzidimitriou}, D. and {Haywood}, M. and {Heiter}, U. and {Helmi}, A. and {Hobbs}, D. and {Hofmann}, W. and {Holl}, B. and {Holland}, G. and {Hunt}, J.~A.~S. and {Hypki}, A. and {Icardi}, V. and {Irwin}, M. and {Jevardat de Fombelle}, G. and {Jofr{\'e}}, P. and {Jonker}, P.~G. and {Jorissen}, A. and {Julbe}, F. and {Karampelas}, A. and {Kochoska}, A. and {Kohley}, R. and {Kolenberg}, K. and {Kontizas}, E. and {Koposov}, S.~E. and {Kordopatis}, G. and {Koubsky}, P. and {Kowalczyk}, A. and {Krone-Martins}, A. and {Kudryashova}, M. and {Kull}, I. and {Bachchan}, R.~K. and {Lacoste-Seris}, F. and {Lanza}, A.~F. and {Lavigne}, J. -B. and {Le Poncin-Lafitte}, C. and {Lebreton}, Y. and {Lebzelter}, T. and {Leccia}, S. and {Leclerc}, N. and {Lecoeur-Taibi}, I. and {Lemaitre}, V. and {Lenhardt}, H. and {Leroux}, F. and {Liao}, S. and {Licata}, E. and {Lindstr{\o}m}, H.~E.~P. and {Lister}, T.~A. and {Livanou}, E. and {Lobel}, A. and {L{\"o}ffler}, W. and {L{\'o}pez}, M. and {Lopez-Lozano}, A. and {Lorenz}, D. and {Loureiro}, T. and {MacDonald}, I. and {Magalh{\~a}es Fernandes}, T. and {Managau}, S. and {Mann}, R.~G. and {Mantelet}, G. and {Marchal}, O. and {Marchant}, J.~M. and {Marconi}, M. and {Marie}, J. and {Marinoni}, S. and {Marrese}, P.~M. and {Marschalk{\'o}}, G. and {Marshall}, D.~J. and {Mart{\'\i}n-Fleitas}, J.~M. and {Martino}, M. and {Mary}, N. and {Matijevi{\v{c}}}, G. and {Mazeh}, T. and {McMillan}, P.~J. and {Messina}, S. and {Mestre}, A. and {Michalik}, D. and {Millar}, N.~R. and {Miranda}, B.~M.~H. and {Molina}, D. and {Molinaro}, R. and {Molinaro}, M. and {Moln{\'a}r}, L. and {Moniez}, M. and {Montegriffo}, P. and {Monteiro}, D. and {Mor}, R. and {Mora}, A. and {Morbidelli}, R. and {Morel}, T. and {Morgenthaler}, S. and {Morley}, T. and {Morris}, D. and {Mulone}, A.~F. and {Muraveva}, T. and {Musella}, I. and {Narbonne}, J. and {Nelemans}, G. and {Nicastro}, L. and {Noval}, L. and {Ord{\'e}novic}, C. and {Ordieres-Mer{\'e}}, J. and {Osborne}, P. and {Pagani}, C. and {Pagano}, I. and {Pailler}, F. and {Palacin}, H. and {Palaversa}, L. and {Parsons}, P. and {Paulsen}, T. and {Pecoraro}, M. and {Pedrosa}, R. and {Pentik{\"a}inen}, H. and {Pereira}, J. and {Pichon}, B. and {Piersimoni}, A.~M. and {Pineau}, F. -X. and {Plachy}, E. and {Plum}, G. and {Poujoulet}, E. and {Pr{\v{s}}a}, A. and {Pulone}, L. and {Ragaini}, S. and {Rago}, S. and {Rambaux}, N. and {Ramos-Lerate}, M. and {Ranalli}, P. and {Rauw}, G. and {Read}, A. and {Regibo}, S. and {Renk}, F. and {Reyl{\'e}}, C. and {Ribeiro}, R.~A. and {Rimoldini}, L. and {Ripepi}, V. and {Riva}, A. and {Rixon}, G. and {Roelens}, M. and {Romero-G{\'o}mez}, M. and {Rowell}, N. and {Royer}, F. and {Rudolph}, A. and {Ruiz-Dern}, L. and {Sadowski}, G. and {Sagrist{\`a} Sell{\'e}s}, T. and {Sahlmann}, J. and {Salgado}, J. and {Salguero}, E. and {Sarasso}, M. and {Savietto}, H. and {Schnorhk}, A. and {Schultheis}, M. and {Sciacca}, E. and {Segol}, M. and {Segovia}, J.~C. and {Segransan}, D. and {Serpell}, E. and {Shih}, I. -C. and {Smareglia}, R. and {Smart}, R.~L. and {Smith}, C. and {Solano}, E. and {Solitro}, F. and {Sordo}, R. and {Soria Nieto}, S. and {Souchay}, J. and {Spagna}, A. and {Spoto}, F. and {Stampa}, U. and {Steele}, I.~A. and {Steidelm{\"u}ller}, H. and {Stephenson}, C.~A. and {Stoev}, H. and {Suess}, F.~F. and {S{\"u}veges}, M. and {Surdej}, J. and {Szabados}, L. and {Szegedi-Elek}, E. and {Tapiador}, D. and {Taris}, F. and {Tauran}, G. and {Taylor}, M.~B. and {Teixeira}, R. and {Terrett}, D. and {Tingley}, B. and {Trager}, S.~C. and {Turon}, C. and {Ulla}, A. and {Utrilla}, E. and {Valentini}, G. and {van Elteren}, A. and {Van Hemelryck}, E. and {van Leeuwen}, M. and {Varadi}, M. and {Vecchiato}, A. and {Veljanoski}, J. and {Via}, T. and {Vicente}, D. and {Vogt}, S. and {Voss}, H. and {Votruba}, V. and {Voutsinas}, S. and {Walmsley}, G. and {Weiler}, M. and {Weingrill}, K. and {Werner}, D. and {Wevers}, T. and {Whitehead}, G. and {Wyrzykowski}, {\L}. and {Yoldas}, A. and {{\v{Z}}erjal}, M. and {Zucker}, S. and {Zurbach}, C. and {Zwitter}, T. and {Alecu}, A. and {Allen}, M. and {Allende Prieto}, C. and {Amorim}, A. and {Anglada-Escud{\'e}}, G. and {Arsenijevic}, V. and {Azaz}, S. and {Balm}, P. and {Beck}, M. and {Bernstein}, H. -H. and {Bigot}, L. and {Bijaoui}, A. and {Blasco}, C. and {Bonfigli}, M. and {Bono}, G. and {Boudreault}, S. and {Bressan}, A. and {Brown}, S. and {Brunet}, P. -M. and {Bunclark}, P. and {Buonanno}, R. and {Butkevich}, A.~G. and {Carret}, C. and {Carrion}, C. and {Chemin}, L. and {Ch{\'e}reau}, F. and {Corcione}, L. and {Darmigny}, E. and {de Boer}, K.~S. and {de Teodoro}, P. and {de Zeeuw}, P.~T. and {Delle Luche}, C. and {Domingues}, C.~D. and {Dubath}, P. and {Fodor}, F. and {Fr{\'e}zouls}, B. and {Fries}, A. and {Fustes}, D. and {Fyfe}, D. and {Gallardo}, E. and {Gallegos}, J. and {Gardiol}, D. and {Gebran}, M. and {Gomboc}, A. and {G{\'o}mez}, A. and {Grux}, E. and {Gueguen}, A. and {Heyrovsky}, A. and {Hoar}, J. and {Iannicola}, G. and {Isasi Parache}, Y. and {Janotto}, A. -M. and {Joliet}, E. and {Jonckheere}, A. and {Keil}, R. and {Kim}, D. -W. and {Klagyivik}, P. and {Klar}, J. and {Knude}, J. and {Kochukhov}, O. and {Kolka}, I. and {Kos}, J. and {Kutka}, A. and {Lainey}, V. and {LeBouquin}, D. and {Liu}, C. and {Loreggia}, D. and {Makarov}, V.~V. and {Marseille}, M.~G. and {Martayan}, C. and {Martinez-Rubi}, O. and {Massart}, B. and {Meynadier}, F. and {Mignot}, S. and {Munari}, U. and {Nguyen}, A. -T. and {Nordlander}, T. and {Ocvirk}, P. and {O'Flaherty}, K.~S. and {Olias Sanz}, A. and {Ortiz}, P. and {Osorio}, J. and {Oszkiewicz}, D. and {Ouzounis}, A. and {Palmer}, M. and {Park}, P. and {Pasquato}, E. and {Peltzer}, C. and {Peralta}, J. and {P{\'e}turaud}, F. and {Pieniluoma}, T. and {Pigozzi}, E. and {Poels}, J. and {Prat}, G. and {Prod'homme}, T. and {Raison}, F. and {Rebordao}, J.~M. and {Risquez}, D. and {Rocca-Volmerange}, B. and {Rosen}, S. and {Ruiz-Fuertes}, M.~I. and {Russo}, F. and {Sembay}, S. and {Serraller Vizcaino}, I. and {Short}, A. and {Siebert}, A. and {Silva}, H. and {Sinachopoulos}, D. and {Slezak}, E. and {Soffel}, M. and {Sosnowska}, D. and {Strai{\v{z}}ys}, V. and {ter Linden}, M. and {Terrell}, D. and {Theil}, S. and {Tiede}, C. and {Troisi}, L. and {Tsalmantza}, P. and {Tur}, D. and {Vaccari}, M. and {Vachier}, F. and {Valles}, P. and {Van Hamme}, W. and {Veltz}, L. and {Virtanen}, J. and {Wallut}, J. -M. and {Wichmann}, R. and {Wilkinson}, M.~I. and {Ziaeepour}, H. and {Zschocke}, S.},
        title = "{The Gaia mission}",
      journal = {\aap},
         year = 2016,
        month = nov,
       volume = {595},
          eid = {A1},
        pages = {A1},
          doi = {10.1051/0004-6361/201629272},
archivePrefix = {arXiv},
       eprint = {1609.04153},
 primaryClass = {astro-ph.IM},
       adsurl = {https://ui.adsabs.harvard.edu/abs/2016A&A...595A...1G}
}

@ARTICLE{GAIATWO,
       author = {{Gaia Collaboration} and {Vallenari}, A. and {Brown}, A.~G.~A. and {Prusti}, T. and {de Bruijne}, J.~H.~J. and {Arenou}, F. and {Babusiaux}, C. and {Biermann}, M. and {Creevey}, O.~L. and {Ducourant}, C. and {Evans}, D.~W. and {Eyer}, L. and {Guerra}, R. and {Hutton}, A. and {Jordi}, C. and {Klioner}, S.~A. and {Lammers}, U.~L. and {Lindegren}, L. and {Luri}, X. and {Mignard}, F. and {Panem}, C. and {Pourbaix}, D. and {Randich}, S. and {Sartoretti}, P. and {Soubiran}, C. and {Tanga}, P. and {Walton}, N.~A. and {Bailer-Jones}, C.~A.~L. and {Bastian}, U. and {Drimmel}, R. and {Jansen}, F. and {Katz}, D. and {Lattanzi}, M.~G. and {van Leeuwen}, F. and {Bakker}, J. and {Cacciari}, C. and {Casta{\~n}eda}, J. and {De Angeli}, F. and {Fabricius}, C. and {Fouesneau}, M. and {Fr{\'e}mat}, Y. and {Galluccio}, L. and {Guerrier}, A. and {Heiter}, U. and {Masana}, E. and {Messineo}, R. and {Mowlavi}, N. and {Nicolas}, C. and {Nienartowicz}, K. and {Pailler}, F. and {Panuzzo}, P. and {Riclet}, F. and {Roux}, W. and {Seabroke}, G.~M. and {Sordo}, R. and {Th{\'e}venin}, F. and {Gracia-Abril}, G. and {Portell}, J. and {Teyssier}, D. and {Altmann}, M. and {Andrae}, R. and {Audard}, M. and {Bellas-Velidis}, I. and {Benson}, K. and {Berthier}, J. and {Blomme}, R. and {Burgess}, P.~W. and {Busonero}, D. and {Busso}, G. and {C{\'a}novas}, H. and {Carry}, B. and {Cellino}, A. and {Cheek}, N. and {Clementini}, G. and {Damerdji}, Y. and {Davidson}, M. and {de Teodoro}, P. and {Nu{\~n}ez Campos}, M. and {Delchambre}, L. and {Dell'Oro}, A. and {Esquej}, P. and {Fern{\'a}ndez-Hern{\'a}ndez}, J. and {Fraile}, E. and {Garabato}, D. and {Garc{\'\i}a-Lario}, P. and {Gosset}, E. and {Haigron}, R. and {Halbwachs}, J. -L. and {Hambly}, N.~C. and {Harrison}, D.~L. and {Hern{\'a}ndez}, J. and {Hestroffer}, D. and {Hodgkin}, S.~T. and {Holl}, B. and {Jan{\ss}en}, K. and {Jevardat de Fombelle}, G. and {Jordan}, S. and {Krone-Martins}, A. and {Lanzafame}, A.~C. and {L{\"o}ffler}, W. and {Marchal}, O. and {Marrese}, P.~M. and {Moitinho}, A. and {Muinonen}, K. and {Osborne}, P. and {Pancino}, E. and {Pauwels}, T. and {Recio-Blanco}, A. and {Reyl{\'e}}, C. and {Riello}, M. and {Rimoldini}, L. and {Roegiers}, T. and {Rybizki}, J. and {Sarro}, L.~M. and {Siopis}, C. and {Smith}, M. and {Sozzetti}, A. and {Utrilla}, E. and {van Leeuwen}, M. and {Abbas}, U. and {{\'A}brah{\'a}m}, P. and {Abreu Aramburu}, A. and {Aerts}, C. and {Aguado}, J.~J. and {Ajaj}, M. and {Aldea-Montero}, F. and {Altavilla}, G. and {{\'A}lvarez}, M.~A. and {Alves}, J. and {Anders}, F. and {Anderson}, R.~I. and {Anglada Varela}, E. and {Antoja}, T. and {Baines}, D. and {Baker}, S.~G. and {Balaguer-N{\'u}{\~n}ez}, L. and {Balbinot}, E. and {Balog}, Z. and {Barache}, C. and {Barbato}, D. and {Barros}, M. and {Barstow}, M.~A. and {Bartolom{\'e}}, S. and {Bassilana}, J. -L. and {Bauchet}, N. and {Becciani}, U. and {Bellazzini}, M. and {Berihuete}, A. and {Bernet}, M. and {Bertone}, S. and {Bianchi}, L. and {Binnenfeld}, A. and {Blanco-Cuaresma}, S. and {Blazere}, A. and {Boch}, T. and {Bombrun}, A. and {Bossini}, D. and {Bouquillon}, S. and {Bragaglia}, A. and {Bramante}, L. and {Breedt}, E. and {Bressan}, A. and {Brouillet}, N. and {Brugaletta}, E. and {Bucciarelli}, B. and {Burlacu}, A. and {Butkevich}, A.~G. and {Buzzi}, R. and {Caffau}, E. and {Cancelliere}, R. and {Cantat-Gaudin}, T. and {Carballo}, R. and {Carlucci}, T. and {Carnerero}, M.~I. and {Carrasco}, J.~M. and {Casamiquela}, L. and {Castellani}, M. and {Castro-Ginard}, A. and {Chaoul}, L. and {Charlot}, P. and {Chemin}, L. and {Chiaramida}, V. and {Chiavassa}, A. and {Chornay}, N. and {Comoretto}, G. and {Contursi}, G. and {Cooper}, W.~J. and {Cornez}, T. and {Cowell}, S. and {Crifo}, F. and {Cropper}, M. and {Crosta}, M. and {Crowley}, C. and {Dafonte}, C. and {Dapergolas}, A. and {David}, M. and {David}, P. and {de Laverny}, P. and {De Luise}, F. and {De March}, R. and {De Ridder}, J. and {de Souza}, R. and {de Torres}, A. and {del Peloso}, E.~F. and {del Pozo}, E. and {Delbo}, M. and {Delgado}, A. and {Delisle}, J. -B. and {Demouchy}, C. and {Dharmawardena}, T.~E. and {Di Matteo}, P. and {Diakite}, S. and {Diener}, C. and {Distefano}, E. and {Dolding}, C. and {Edvardsson}, B. and {Enke}, H. and {Fabre}, C. and {Fabrizio}, M. and {Faigler}, S. and {Fedorets}, G. and {Fernique}, P. and {Fienga}, A. and {Figueras}, F. and {Fournier}, Y. and {Fouron}, C. and {Fragkoudi}, F. and {Gai}, M. and {Garcia-Gutierrez}, A. and {Garcia-Reinaldos}, M. and {Garc{\'\i}a-Torres}, M. and {Garofalo}, A. and {Gavel}, A. and {Gavras}, P. and {Gerlach}, E. and {Geyer}, R. and {Giacobbe}, P. and {Gilmore}, G. and {Girona}, S. and {Giuffrida}, G. and {Gomel}, R. and {Gomez}, A. and {Gonz{\'a}lez-N{\'u}{\~n}ez}, J. and {Gonz{\'a}lez-Santamar{\'\i}a}, I. and {Gonz{\'a}lez-Vidal}, J.~J. and {Granvik}, M. and {Guillout}, P. and {Guiraud}, J. and {Guti{\'e}rrez-S{\'a}nchez}, R. and {Guy}, L.~P. and {Hatzidimitriou}, D. and {Hauser}, M. and {Haywood}, M. and {Helmer}, A. and {Helmi}, A. and {Sarmiento}, M.~H. and {Hidalgo}, S.~L. and {Hilger}, T. and {H{\l}adczuk}, N. and {Hobbs}, D. and {Holland}, G. and {Huckle}, H.~E. and {Jardine}, K. and {Jasniewicz}, G. and {Jean-Antoine Piccolo}, A. and {Jim{\'e}nez-Arranz}, {\'O}. and {Jorissen}, A. and {Juaristi Campillo}, J. and {Julbe}, F. and {Karbevska}, L. and {Kervella}, P. and {Khanna}, S. and {Kontizas}, M. and {Kordopatis}, G. and {Korn}, A.~J. and {K{\'o}sp{\'a}l}, {\'A}. and {Kostrzewa-Rutkowska}, Z. and {Kruszy{\'n}ska}, K. and {Kun}, M. and {Laizeau}, P. and {Lambert}, S. and {Lanza}, A.~F. and {Lasne}, Y. and {Le Campion}, J. -F. and {Lebreton}, Y. and {Lebzelter}, T. and {Leccia}, S. and {Leclerc}, N. and {Lecoeur-Taibi}, I. and {Liao}, S. and {Licata}, E.~L. and {Lindstr{\o}m}, H.~E.~P. and {Lister}, T.~A. and {Livanou}, E. and {Lobel}, A. and {Lorca}, A. and {Loup}, C. and {Madrero Pardo}, P. and {Magdaleno Romeo}, A. and {Managau}, S. and {Mann}, R.~G. and {Manteiga}, M. and {Marchant}, J.~M. and {Marconi}, M. and {Marcos}, J. and {Marcos Santos}, M.~M.~S. and {Mar{\'\i}n Pina}, D. and {Marinoni}, S. and {Marocco}, F. and {Marshall}, D.~J. and {Martin Polo}, L. and {Mart{\'\i}n-Fleitas}, J.~M. and {Marton}, G. and {Mary}, N. and {Masip}, A. and {Massari}, D. and {Mastrobuono-Battisti}, A. and {Mazeh}, T. and {McMillan}, P.~J. and {Messina}, S. and {Michalik}, D. and {Millar}, N.~R. and {Mints}, A. and {Molina}, D. and {Molinaro}, R. and {Moln{\'a}r}, L. and {Monari}, G. and {Mongui{\'o}}, M. and {Montegriffo}, P. and {Montero}, A. and {Mor}, R. and {Mora}, A. and {Morbidelli}, R. and {Morel}, T. and {Morris}, D. and {Muraveva}, T. and {Murphy}, C.~P. and {Musella}, I. and {Nagy}, Z. and {Noval}, L. and {Oca{\~n}a}, F. and {Ogden}, A. and {Ordenovic}, C. and {Osinde}, J.~O. and {Pagani}, C. and {Pagano}, I. and {Palaversa}, L. and {Palicio}, P.~A. and {Pallas-Quintela}, L. and {Panahi}, A. and {Payne-Wardenaar}, S. and {Pe{\~n}alosa Esteller}, X. and {Penttil{\"a}}, A. and {Pichon}, B. and {Piersimoni}, A.~M. and {Pineau}, F. -X. and {Plachy}, E. and {Plum}, G. and {Poggio}, E. and {Pr{\v{s}}a}, A. and {Pulone}, L. and {Racero}, E. and {Ragaini}, S. and {Rainer}, M. and {Raiteri}, C.~M. and {Rambaux}, N. and {Ramos}, P. and {Ramos-Lerate}, M. and {Re Fiorentin}, P. and {Regibo}, S. and {Richards}, P.~J. and {Rios Diaz}, C. and {Ripepi}, V. and {Riva}, A. and {Rix}, H. -W. and {Rixon}, G. and {Robichon}, N. and {Robin}, A.~C. and {Robin}, C. and {Roelens}, M. and {Rogues}, H.~R.~O. and {Rohrbasser}, L. and {Romero-G{\'o}mez}, M. and {Rowell}, N. and {Royer}, F. and {Ruz Mieres}, D. and {Rybicki}, K.~A. and {Sadowski}, G. and {S{\'a}ez N{\'u}{\~n}ez}, A. and {Sagrist{\`a} Sell{\'e}s}, A. and {Sahlmann}, J. and {Salguero}, E. and {Samaras}, N. and {Sanchez Gimenez}, V. and {Sanna}, N. and {Santove{\~n}a}, R. and {Sarasso}, M. and {Schultheis}, M. and {Sciacca}, E. and {Segol}, M. and {Segovia}, J.~C. and {S{\'e}gransan}, D. and {Semeux}, D. and {Shahaf}, S. and {Siddiqui}, H.~I. and {Siebert}, A. and {Siltala}, L. and {Silvelo}, A. and {Slezak}, E. and {Slezak}, I. and {Smart}, R.~L. and {Snaith}, O.~N. and {Solano}, E. and {Solitro}, F. and {Souami}, D. and {Souchay}, J. and {Spagna}, A. and {Spina}, L. and {Spoto}, F. and {Steele}, I.~A. and {Steidelm{\"u}ller}, H. and {Stephenson}, C.~A. and {S{\"u}veges}, M. and {Surdej}, J. and {Szabados}, L. and {Szegedi-Elek}, E. and {Taris}, F. and {Taylor}, M.~B. and {Teixeira}, R. and {Tolomei}, L. and {Tonello}, N. and {Torra}, F. and {Torra}, J. and {Torralba Elipe}, G. and {Trabucchi}, M. and {Tsounis}, A.~T. and {Turon}, C. and {Ulla}, A. and {Unger}, N. and {Vaillant}, M.~V. and {van Dillen}, E. and {van Reeven}, W. and {Vanel}, O. and {Vecchiato}, A. and {Viala}, Y. and {Vicente}, D. and {Voutsinas}, S. and {Weiler}, M. and {Wevers}, T. and {Wyrzykowski}, {\L}. and {Yoldas}, A. and {Yvard}, P. and {Zhao}, H. and {Zorec}, J. and {Zucker}, S. and {Zwitter}, T.},
        title = "{Gaia Data Release 3. Summary of the content and survey properties}",
      journal = {\aap},
         year = 2023,
        month = jun,
       volume = {674},
          eid = {A1},
        pages = {A1},
          doi = {10.1051/0004-6361/202243940},
archivePrefix = {arXiv},
       eprint = {2208.00211},
 primaryClass = {astro-ph.GA},
       adsurl = {https://ui.adsabs.harvard.edu/abs/2023A&A...674A...1G}
}

@ARTICLE{GAIATHREE,
       author = {{Babusiaux}, C. and {Fabricius}, C. and {Khanna}, S. and {Muraveva}, T. and {Reyl{\'e}}, C. and {Spoto}, F. and {Vallenari}, A. and {Luri}, X. and {Arenou}, F. and {{\'A}lvarez}, M.~A. and {Anders}, F. and {Antoja}, T. and {Balbinot}, E. and {Barache}, C. and {Bauchet}, N. and {Bossini}, D. and {Busonero}, D. and {Cantat-Gaudin}, T. and {Carrasco}, J.~M. and {Dafonte}, C. and {Diakit{\'e}}, S. and {Figueras}, F. and {Garcia-Gutierrez}, A. and {Garofalo}, A. and {Helmi}, A. and {Jim{\'e}nez-Arranz}, {\'O}. and {Jordi}, C. and {Kervella}, P. and {Kostrzewa-Rutkowska}, Z. and {Leclerc}, N. and {Licata}, E. and {Manteiga}, M. and {Masip}, A. and {Mongui{\'o}}, M. and {Ramos}, P. and {Robichon}, N. and {Robin}, A.~C. and {Romero-G{\'o}mez}, M. and {S{\'a}ez}, A. and {Santove{\~n}a}, R. and {Spina}, L. and {Torralba Elipe}, G. and {Weiler}, M.},
        title = "{Gaia Data Release 3. Catalogue validation}",
      journal = {\aap},
         year = 2023,
        month = jun,
       volume = {674},
          eid = {A32},
        pages = {A32},
          doi = {10.1051/0004-6361/202243790},
archivePrefix = {arXiv},
       eprint = {2206.05989},
 primaryClass = {astro-ph.SR},
       adsurl = {https://ui.adsabs.harvard.edu/abs/2023A&A...674A..32B}
}

@ARTICLE{Fitzpatrick99,
   author = {{Fitzpatrick}, E.~L.},
    title = "{Correcting for the Effects of Interstellar Extinction}",
  journal = {\pasp},
   eprint = {arXiv:astro-ph/9809387},
     year = 1999,
    month = jan,
   volume = 111,
    pages = {63-75},
      doi = {10.1086/316293},
   adsurl = {http://adsabs.harvard.edu/abs/1999PASP..111...63F}
}

@ARTICLE{Cardelli89,
       author = {{Cardelli}, Jason A. and {Clayton}, Geoffrey C. and {Mathis}, John S.},
        title = "{The Relationship between Infrared, Optical, and Ultraviolet Extinction}",
      journal = {\apj},
         year = 1989,
        month = oct,
       volume = {345},
        pages = {245},
          doi = {10.1086/167900},
       adsurl = {https://ui.adsabs.harvard.edu/abs/1989ApJ...345..245C}
}

@ARTICLE{Wiseman22,
       author = {{Wiseman}, P. and {Vincenzi}, M. and {Sullivan}, M. and {Kelsey}, L. and {Popovic}, B. and {Rose}, B. and {Brout}, D. and {Davis}, T.~M. and {Frohmaier}, C. and {Galbany}, L. and {Lidman}, C. and {M{\"o}ller}, A. and {Scolnic}, D. and {Smith}, M. and {Aguena}, M. and {Allam}, S. and {Andrade-Oliveira}, F. and {Annis}, J. and {Bertin}, E. and {Bocquet}, S. and {Brooks}, D. and {Burke}, D.~L. and {Carnero Rosell}, A. and {Carrasco Kind}, M. and {Carretero}, J. and {Castander}, F.~J. and {Costanzi}, M. and {Pereira}, M.~E.~S. and {Desai}, S. and {Diehl}, H.~T. and {Doel}, P. and {Everett}, S. and {Ferrero}, I. and {Friedel}, D. and {Frieman}, J. and {Garc{\'\i}a-Bellido}, J. and {Gatti}, M. and {Gaztanaga}, E. and {Gruen}, D. and {Gschwend}, J. and {Gutierrez}, G. and {Hinton}, S.~R. and {Hollowood}, D.~L. and {Honscheid}, K. and {James}, D.~J. and {March}, M. and {Menanteau}, F. and {Miquel}, R. and {Morgan}, R. and {Palmese}, A. and {Paz-Chinch{\'o}n}, F. and {Pieres}, A. and {Plazas Malag{\'o}n}, A.~A. and {Romer}, A.~K. and {Sanchez}, E. and {Scarpine}, V. and {Sevilla-Noarbe}, I. and {Soares-Santos}, M. and {Suchyta}, E. and {Tarle}, G. and {To}, C. and {Varga}, T.~N. and {DES Collaboration}},
        title = "{A galaxy-driven model of type Ia supernova luminosity variations}",
      journal = {\mnras},
         year = 2022,
        month = sep,
       volume = {515},
       number = {3},
        pages = {4587-4605},
          doi = {10.1093/mnras/stac1984},
archivePrefix = {arXiv},
       eprint = {2207.05583},
 primaryClass = {astro-ph.GA},
       adsurl = {https://ui.adsabs.harvard.edu/abs/2022MNRAS.515.4587W}
}

@ARTICLE{ActDr6maps,
       author = {{Naess}, Sigurd and {Guan}, Yilun and {Duivenvoorden}, Adriaan J. and {Hasselfield}, Matthew and {Wang}, Yuhan and {Abril-Cabezas}, Irene and {Addison}, Graeme E. and {Ade}, Peter A.~R. and {Aiola}, Simone and {Alford}, Tommy and {Alonso}, David and {Amiri}, Mandana and {An}, Rui and {Atkins}, Zachary and {Austermann}, Jason E. and {Barbavara}, Eleonora and {Battaglia}, Nicholas and {Battistelli}, Elia Stefano and {Beall}, James A. and {Bean}, Rachel and {Beheshti}, Ali and {Beringue}, Benjamin and {Bhandarkar}, Tanay and {Biermann}, Emily and {Bolliet}, Boris and {Bond}, J. Richard and {Calabrese}, Erminia and {Capalbo}, Valentina and {Carrero}, Felipe and {Chen}, Stephen and {Chesmore}, Grace and {Cho}, Hsiao-mei and {Choi}, Steve K. and {Clark}, Susan E. and {Rosado}, Rodrigo Cordova and {Cothard}, Nicholas F. and {Coughlin}, Kevin and {Coulton}, William and {Crichton}, Devin and {Crowley}, Kevin T. and {Devlin}, Mark J. and {Dicker}, Simon and {Duell}, Cody J. and {Duff}, Shannon M. and {Dunkley}, Jo and {Dunner}, Rolando and {Embil Villagra}, Carmen and {Fankhanel}, Max and {Farren}, Gerrit S. and {Ferraro}, Simone and {Foster}, Allen and {Freundt}, Rodrigo and {Fuzia}, Brittany and {Gallardo}, Patricio A. and {Garrido}, Xavier and {Giardiello}, Serena and {Gill}, Ajay and {Givans}, Jahmour and {Gluscevic}, Vera and {Golec}, Joseph E. and {Gong}, Yulin and {Halpern}, Mark and {Harrison}, Ian and {Healy}, Erin and {Henderson}, Shawn and {Hensley}, Brandon and {Herv{\'\i}as-Caimapo}, Carlos and {Hill}, J. Colin and {Hilton}, Gene C. and {Hilton}, Matt and {Hincks}, Adam D. and {Hlo{\v{z}}ek}, Ren{\'e}e and {Ho}, Shuay-Pwu Patty and {Hood}, John and {Hornecker}, Erika and {Huber}, Zachary B. and {Hubmayr}, Johannes and {Huffenberger}, Kevin M. and {Hughes}, John P. and {Ikape}, Margaret and {Irwin}, Kent and {Isopi}, Giovanni and {Jense}, Hidde T. and {Joshi}, Neha and {Keller}, Ben and {Kim}, Joshua and {Knowles}, Kenda and {Koopman}, Brian J. and {Kosowsky}, Arthur and {Kramer}, Darby and {Kusiak}, Aleksandra and {La Posta}, Adrien and {Lagu{\"e}}, Alex and {Lakey}, Victoria and {Lee}, Eunseong and {Li}, Yaqiong and {Li}, Zack and {Limon}, Michele and {Lokken}, Martine and {Louis}, Thibaut and {Lungu}, Marius and {MacCrann}, Niall and {MacInnis}, Amanda and {Madhavacheril}, Mathew S. and {Maldonado}, Diego and {Maldonado}, Felipe and {Mallaby-Kay}, Maya and {Marques}, Gabriela A. and {van Marrewijk}, Joshiwa and {McCarthy}, Fiona and {McMahon}, Jeff and {Mehta}, Yogesh and {Menanteau}, Felipe and {Moodley}, Kavilan and {Morris}, Thomas W. and {Mroczkowski}, Tony and {Namikawa}, Toshiya and {Nati}, Federico and {Nerval}, Simran K. and {Newburgh}, Laura and {Nicola}, Andrina and {Niemack}, Michael D. and {Nolta}, Michael R. and {Orlowski-Scherer}, John and {Page}, Lyman A. and {Pandey}, Shivam and {Partridge}, Bruce and {Perez Sarmiento}, Karen and {Prince}, Heather and {Puddu}, Roberto and {Qu}, Frank J. and {Ragavan}, Damien C. and {Ried Guachalla}, Bernardita and {Rogers}, Keir K. and {Rojas}, Felipe and {Sakuma}, Tai and {Schaan}, Emmanuel and {Schmitt}, Benjamin L. and {Sehgal}, Neelima and {Shaikh}, Shabbir and {Sherwin}, Blake D. and {Sierra}, Carlos and {Sievers}, Jon and {Sif{\'o}n}, Crist{\'o}bal and {Simon}, Sara and {Sonka}, Rita and {London}, Alexander Spencer and {Spergel}, David N. and {Staggs}, Suzanne T. and {Storer}, Emilie and {Surrao}, Kristen and {Switzer}, Eric R. and {Tampier}, Niklas and {Thornton}, Robert and {Trac}, Hy and {Tucker}, Carole and {Ullom}, Joel and {Vale}, Leila R. and {Van Engelen}, Alexander and {Van Lanen}, Jeff and {Vargas}, Cristian and {Vavagiakis}, Eve M. and {Wagoner}, Kasey and {Wenzl}, Lukas and {Wollack}, Edward J. and {Zheng}, Kaiwen and {The Atacama Cosmology Telescope collaboration}},
        title = "{The Atacama Cosmology Telescope: DR6 maps}",
      journal = {\jcap},
         year = 2025,
        month = nov,
       volume = {2025},
       number = {11},
          eid = {061},
        pages = {061},
          doi = {10.1088/1475-7516/2025/11/061},
archivePrefix = {arXiv},
       eprint = {2503.14451},
 primaryClass = {astro-ph.CO},
       adsurl = {https://ui.adsabs.harvard.edu/abs/2025JCAP...11..061N}
}

@ARTICLE{ActDr6like,
       author = {{Louis}, Thibaut and {La Posta}, Adrien and {Atkins}, Zachary and {Jense}, Hidde T. and {Abril-Cabezas}, Irene and {Addison}, Graeme E. and {Ade}, Peter A.~R. and {Aiola}, Simone and {Alford}, Tommy and {Alonso}, David and {Amiri}, Mandana and {An}, Rui and {Austermann}, Jason E. and {Barbavara}, Eleonora and {Battaglia}, Nicholas and {Battistelli}, Elia Stefano and {Beall}, James A. and {Bean}, Rachel and {Beheshti}, Ali and {Beringue}, Benjamin and {Bhandarkar}, Tanay and {Biermann}, Emily and {Bolliet}, Boris and {Bond}, J. Richard and {Calabrese}, Erminia and {Capalbo}, Valentina and {Carrero}, Felipe and {Chen}, Shi-Fan and {Chesmore}, Grace and {Cho}, Hsiao-mei and {Choi}, Steve K. and {Clark}, Susan E. and {Cothard}, Nicholas F. and {Coughlin}, Kevin and {Coulton}, William and {Crichton}, Devin and {Crowley}, Kevin T. and {Darwish}, Omar and {Devlin}, Mark J. and {Dicker}, Simon and {Duell}, Cody J. and {Duff}, Shannon M. and {Duivenvoorden}, Adriaan J. and {Dunkley}, Jo and {Dunner}, Rolando and {Embil Villagra}, Carmen and {Fankhanel}, Max and {Farren}, Gerrit S. and {Ferraro}, Simone and {Foster}, Allen and {Freundt}, Rodrigo and {Fuzia}, Brittany and {Gallardo}, Patricio A. and {Garrido}, Xavier and {Gerbino}, Martina and {Giardiello}, Serena and {Gill}, Ajay and {Givans}, Jahmour and {Gluscevic}, Vera and {Goldstein}, Samuel and {Golec}, Joseph E. and {Gong}, Yulin and {Guan}, Yilun and {Halpern}, Mark and {Harrison}, Ian and {Hasselfield}, Matthew and {Healy}, Erin and {Henderson}, Shawn and {Hensley}, Brandon and {Herv{\'\i}as-Caimapo}, Carlos and {Hill}, J. Colin and {Hilton}, Gene C. and {Hilton}, Matt and {Hincks}, Adam D. and {Hlo{\v{z}}ek}, Ren{\'e}e and {Ho}, Shuay-Pwu Patty and {Hood}, John and {Hornecker}, Erika and {Huber}, Zachary B. and {Hubmayr}, Johannes and {Huffenberger}, Kevin M. and {Hughes}, John P. and {Ikape}, Margaret and {Irwin}, Kent and {Isopi}, Giovanni and {Joshi}, Neha and {Keller}, Ben and {Kim}, Joshua and {Knowles}, Kenda and {Koopman}, Brian J. and {Kosowsky}, Arthur and {Kramer}, Darby and {Kusiak}, Aleksandra and {Lagu{\"e}}, Alex and {Lakey}, Victoria and {Lee}, Eunseong and {Li}, Yaqiong and {Li}, Zack and {Limon}, Michele and {Lokken}, Martine and {Lungu}, Marius and {MacCrann}, Niall and {MacInnis}, Amanda and {Madhavacheril}, Mathew S. and {Maldonado}, Diego and {Maldonado}, Felipe and {Mallaby-Kay}, Maya and {Marques}, Gabriela A. and {van Marrewijk}, Joshiwa and {McCarthy}, Fiona and {McMahon}, Jeff and {Mehta}, Yogesh and {Menanteau}, Felipe and {Moodley}, Kavilan and {Morris}, Thomas W. and {Mroczkowski}, Tony and {Naess}, Sigurd and {Namikawa}, Toshiya and {Nati}, Federico and {Nerval}, Simran K. and {Newburgh}, Laura and {Nicola}, Andrina and {Niemack}, Michael D. and {Nolta}, Michael R. and {Orlowski-Scherer}, John and {Pagano}, Luca and {Page}, Lyman A. and {Pandey}, Shivam and {Partridge}, Bruce and {Perez Sarmiento}, Karen and {Prince}, Heather and {Puddu}, Roberto and {Qu}, Frank J. and {Ragavan}, Damien C. and {Ried Guachalla}, Bernardita and {Rogers}, Keir K. and {Rojas}, Felipe and {Sakuma}, Tai and {Schaan}, Emmanuel and {Schmitt}, Benjamin L. and {Sehgal}, Neelima and {Shaikh}, Shabbir and {Sherwin}, Blake D. and {Sierra}, Carlos and {Sievers}, Jon and {Sif{\'o}n}, Crist{\'o}bal and {Simon}, Sara and {Sonka}, Rita and {Spergel}, David N. and {Staggs}, Suzanne T. and {Storer}, Emilie and {Surrao}, Kristen and {Switzer}, Eric R. and {Tampier}, Niklas and {Thornton}, Robert and {Trac}, Hy and {Tucker}, Carole and {Ullom}, Joel and {Vale}, Leila R. and {Van Engelen}, Alexander and {Van Lanen}, Jeff and {Vargas}, Cristian and {Vavagiakis}, Eve M. and {Wagoner}, Kasey and {Wang}, Yuhan and {Wenzl}, Lukas and {Wollack}, Edward J. and {Zheng}, Kaiwen and {The Atacama Cosmology Telescope collaboration}},
        title = "{The Atacama Cosmology Telescope: DR6 power spectra, likelihoods and {\ensuremath{\Lambda}}CDM parameters}",
      journal = {\jcap},
         year = 2025,
        month = nov,
       volume = {2025},
       number = {11},
          eid = {062},
        pages = {062},
          doi = {10.1088/1475-7516/2025/11/062},
archivePrefix = {arXiv},
       eprint = {2503.14452},
 primaryClass = {astro-ph.CO},
       adsurl = {https://ui.adsabs.harvard.edu/abs/2025JCAP...11..062L}
}

@ARTICLE{PrinceDunkley19,
       author = {{Prince}, Heather and {Dunkley}, Jo},
        title = "{Data compression in cosmology: A compressed likelihood for Planck data}",
      journal = {\prd},
         year = 2019,
        month = oct,
       volume = {100},
       number = {8},
          eid = {083502},
        pages = {083502},
          doi = {10.1103/PhysRevD.100.083502},
archivePrefix = {arXiv},
       eprint = {1909.05869},
 primaryClass = {astro-ph.CO},
       adsurl = {https://ui.adsabs.harvard.edu/abs/2019PhRvD.100h3502P}
}

@ARTICLE{Camphuis2025,
       author = {{Camphuis}, E. and {Quan}, W. and {Balkenhol}, L. and {Khalife}, A.~R. and {Ge}, F. and {Guidi}, F. and {Huang}, N. and {Lynch}, G.~P. and {Omori}, Y. and {Trendafilova}, C. and {Anderson}, A.~J. and {Ansarinejad}, B. and {Archipley}, M. and {Barry}, P.~S. and {Benabed}, K. and {Bender}, A.~N. and {Benson}, B.~A. and {Bianchini}, F. and {Bleem}, L.~E. and {Bouchet}, F.~R. and {Bryant}, L. and {Campitiello}, M.~G. and {Carlstrom}, J.~E. and {Chang}, C.~L. and {Chaubal}, P. and {Chichura}, P.~M. and {Chokshi}, A. and {Chou}, T.-L. and {Coerver}, A. and {Crawford}, T.~M. and {Daley}, C. and {de Haan}, T. and {Dibert}, K.~R. and {Dobbs}, M.~A. and {Doohan}, M. and {Doussot}, A. and {Dutcher}, D. and {Everett}, W. and {Feng}, C. and {Ferguson}, K.~R. and {Fichman}, K. and {Foster}, A. and {Galli}, S. and {Gambrel}, A.~E. and {Gardner}, R.~W. and {Goeckner-Wald}, N. and {Gualtieri}, R. and {Guns}, S. and {Halverson}, N.~W. and {Hivon}, E. and {Holder}, G.~P. and {Holzapfel}, W.~L. and {Hood}, J.~C. and {Hryciuk}, A. and {K{\'e}ruzor{\'e}}, F. and {Knox}, L. and {Korman}, M. and {Kornoelje}, K. and {Kuo}, C.-L. and {Levy}, K. and {Lowitz}, A.~E. and {Lu}, C. and {Maniyar}, A. and {Martsen}, E.~S. and {Menanteau}, F. and {Millea}, M. and {Montgomery}, J. and {Nakato}, Y. and {Natoli}, T. and {Noble}, G.~I. and {Ouellette}, A. and {Pan}, Z. and {Paschos}, P. and {Phadke}, K.~A. and {Pollak}, A.~W. and {Prabhu}, K. and {Raghunathan}, S. and {Rahimi}, M. and {Rahlin}, A. and {Reichardt}, C.~L. and {Rouble}, M. and {Ruhl}, J.~E. and {Schiappucci}, E. and {Simpson}, A. and {Sobrin}, J.~A. and {Stark}, A.~A. and {Stephen}, J. and {Tandoi}, C. and {Thorne}, B. and {Umilta}, C. and {Vieira}, J.~D. and {Vitrier}, A. and {Wan}, Y. and {Whitehorn}, N. and {Wu}, W.~L.~K. and {Young}, M.~R. and {Zebrowski}, J.~A. and {SPT-3G Collaboration}},
        title = "{SPT-3G D1: CMB temperature and polarization power spectra and cosmology from 2019 and 2020 observations of the SPT-3G main field}",
      journal = {\prd},
         year = 2026,
        month = apr,
       volume = {113},
       number = {8},
          eid = {083504},
        pages = {083504},
          doi = {10.1103/7wt3-9v2y},
archivePrefix = {arXiv},
       eprint = {2506.20707},
 primaryClass = {astro-ph.CO},
       adsurl = {https://ui.adsabs.harvard.edu/abs/2026PhRvD.113h3504C}
}

@ARTICLE{Balkenhol2024,
       author = {{Balkenhol}, L. and {Trendafilova}, C. and {Benabed}, K. and {Galli}, S.},
        title = "{candl: cosmic microwave background analysis with a differentiable likelihood}",
      journal = {\aap},
         year = 2024,
        month = jun,
       volume = {686},
          eid = {A10},
        pages = {A10},
          doi = {10.1051/0004-6361/202449432},
archivePrefix = {arXiv},
       eprint = {2401.13433},
 primaryClass = {astro-ph.CO},
       adsurl = {https://ui.adsabs.harvard.edu/abs/2024A&A...686A..10B}
}

@ARTICLE{Madhavacheril2024,
       author = {{Madhavacheril}, Mathew S. and {Qu}, Frank J. and {Sherwin}, Blake D. and {MacCrann}, Niall and {Li}, Yaqiong and {Abril-Cabezas}, Irene and {Ade}, Peter A.~R. and {Aiola}, Simone and {Alford}, Tommy and {Amiri}, Mandana and {Amodeo}, Stefania and {An}, Rui and {Atkins}, Zachary and {Austermann}, Jason E. and {Battaglia}, Nicholas and {Battistelli}, Elia Stefano and {Beall}, James A. and {Bean}, Rachel and {Beringue}, Benjamin and {Bhandarkar}, Tanay and {Biermann}, Emily and {Bolliet}, Boris and {Bond}, J. Richard and {Cai}, Hongbo and {Calabrese}, Erminia and {Calafut}, Victoria and {Capalbo}, Valentina and {Carrero}, Felipe and {Challinor}, Anthony and {Chesmore}, Grace E. and {Cho}, Hsiao-mei and {Choi}, Steve K. and {Clark}, Susan E. and {C{\'o}rdova Rosado}, Rodrigo and {Cothard}, Nicholas F. and {Coughlin}, Kevin and {Coulton}, William and {Crowley}, Kevin T. and {Dalal}, Roohi and {Darwish}, Omar and {Devlin}, Mark J. and {Dicker}, Simon and {Doze}, Peter and {Duell}, Cody J. and {Duff}, Shannon M. and {Duivenvoorden}, Adriaan J. and {Dunkley}, Jo and {D{\"u}nner}, Rolando and {Fanfani}, Valentina and {Fankhanel}, Max and {Farren}, Gerrit and {Ferraro}, Simone and {Freundt}, Rodrigo and {Fuzia}, Brittany and {Gallardo}, Patricio A. and {Garrido}, Xavier and {Givans}, Jahmour and {Gluscevic}, Vera and {Golec}, Joseph E. and {Guan}, Yilun and {Hall}, Kirsten R. and {Halpern}, Mark and {Han}, Dongwon and {Harrison}, Ian and {Hasselfield}, Matthew and {Healy}, Erin and {Henderson}, Shawn and {Hensley}, Brandon and {Herv{\'\i}as-Caimapo}, Carlos and {Hill}, J. Colin and {Hilton}, Gene C. and {Hilton}, Matt and {Hincks}, Adam D. and {Hlo{\v{z}}ek}, Ren{\'e}e and {Ho}, Shuay-Pwu Patty and {Huber}, Zachary B. and {Hubmayr}, Johannes and {Huffenberger}, Kevin M. and {Hughes}, John P. and {Irwin}, Kent and {Isopi}, Giovanni and {Jense}, Hidde T. and {Keller}, Ben and {Kim}, Joshua and {Knowles}, Kenda and {Koopman}, Brian J. and {Kosowsky}, Arthur and {Kramer}, Darby and {Kusiak}, Aleksandra and {La Posta}, Adrien and {Lague}, Alex and {Lakey}, Victoria and {Lee}, Eunseong and {Li}, Zack and {Limon}, Michele and {Lokken}, Martine and {Louis}, Thibaut and {Lungu}, Marius and {MacInnis}, Amanda and {Maldonado}, Diego and {Maldonado}, Felipe and {Mallaby-Kay}, Maya and {Marques}, Gabriela A. and {McMahon}, Jeff and {Mehta}, Yogesh and {Menanteau}, Felipe and {Moodley}, Kavilan and {Morris}, Thomas W. and {Mroczkowski}, Tony and {Naess}, Sigurd and {Namikawa}, Toshiya and {Nati}, Federico and {Newburgh}, Laura and {Nicola}, Andrina and {Niemack}, Michael D. and {Nolta}, Michael R. and {Orlowski-Scherer}, John and {Page}, Lyman A. and {Pandey}, Shivam and {Partridge}, Bruce and {Prince}, Heather and {Puddu}, Roberto and {Radiconi}, Federico and {Robertson}, Naomi and {Rojas}, Felipe and {Sakuma}, Tai and {Salatino}, Maria and {Schaan}, Emmanuel and {Schmitt}, Benjamin L. and {Sehgal}, Neelima and {Shaikh}, Shabbir and {Sierra}, Carlos and {Sievers}, Jon and {Sif{\'o}n}, Crist{\'o}bal and {Simon}, Sara and {Sonka}, Rita and {Spergel}, David N. and {Staggs}, Suzanne T. and {Storer}, Emilie and {Switzer}, Eric R. and {Tampier}, Niklas and {Thornton}, Robert and {Trac}, Hy and {Treu}, Jesse and {Tucker}, Carole and {Ullom}, Joel and {Vale}, Leila R. and {Van Engelen}, Alexander and {Van Lanen}, Jeff and {van Marrewijk}, Joshiwa and {Vargas}, Cristian and {Vavagiakis}, Eve M. and {Wagoner}, Kasey and {Wang}, Yuhan and {Wenzl}, Lukas and {Wollack}, Edward J. and {Xu}, Zhilei and {Zago}, Fernando and {Zheng}, Kaiwen},
        title = "{The Atacama Cosmology Telescope: DR6 Gravitational Lensing Map and Cosmological Parameters}",
      journal = {\apj},
         year = 2024,
        month = feb,
       volume = {962},
       number = {2},
          eid = {113},
        pages = {113},
          doi = {10.3847/1538-4357/acff5f},
archivePrefix = {arXiv},
       eprint = {2304.05203},
 primaryClass = {astro-ph.CO},
       adsurl = {https://ui.adsabs.harvard.edu/abs/2024ApJ...962..113M}
}

@ARTICLE{Qu2024,
       author = {{Qu}, Frank J. and {Sherwin}, Blake D. and {Madhavacheril}, Mathew S. and {Han}, Dongwon and {Crowley}, Kevin T. and {Abril-Cabezas}, Irene and {Ade}, Peter A.~R. and {Aiola}, Simone and {Alford}, Tommy and {Amiri}, Mandana and {Amodeo}, Stefania and {An}, Rui and {Atkins}, Zachary and {Austermann}, Jason E. and {Battaglia}, Nicholas and {Battistelli}, Elia Stefano and {Beall}, James A. and {Bean}, Rachel and {Beringue}, Benjamin and {Bhandarkar}, Tanay and {Biermann}, Emily and {Bolliet}, Boris and {Bond}, J. Richard and {Cai}, Hongbo and {Calabrese}, Erminia and {Calafut}, Victoria and {Capalbo}, Valentina and {Carrero}, Felipe and {Carron}, Julien and {Challinor}, Anthony and {Chesmore}, Grace E. and {Cho}, Hsiao-mei and {Choi}, Steve K. and {Clark}, Susan E. and {C{\'o}rdova Rosado}, Rodrigo and {Cothard}, Nicholas F. and {Coughlin}, Kevin and {Coulton}, William and {Dalal}, Roohi and {Darwish}, Omar and {Devlin}, Mark J. and {Dicker}, Simon and {Doze}, Peter and {Duell}, Cody J. and {Duff}, Shannon M. and {Duivenvoorden}, Adriaan J. and {Dunkley}, Jo and {D{\"u}nner}, Rolando and {Fanfani}, Valentina and {Fankhanel}, Max and {Farren}, Gerrit and {Ferraro}, Simone and {Freundt}, Rodrigo and {Fuzia}, Brittany and {Gallardo}, Patricio A. and {Garrido}, Xavier and {Gluscevic}, Vera and {Golec}, Joseph E. and {Guan}, Yilun and {Halpern}, Mark and {Harrison}, Ian and {Hasselfield}, Matthew and {Healy}, Erin and {Henderson}, Shawn and {Hensley}, Brandon and {Herv{\'\i}as-Caimapo}, Carlos and {Hill}, J. Colin and {Hilton}, Gene C. and {Hilton}, Matt and {Hincks}, Adam D. and {Hlo{\v{z}}ek}, Ren{\'e}e and {Ho}, Shuay-Pwu Patty and {Huber}, Zachary B. and {Hubmayr}, Johannes and {Huffenberger}, Kevin M. and {Hughes}, John P. and {Irwin}, Kent and {Isopi}, Giovanni and {Jense}, Hidde T. and {Keller}, Ben and {Kim}, Joshua and {Knowles}, Kenda and {Koopman}, Brian J. and {Kosowsky}, Arthur and {Kramer}, Darby and {Kusiak}, Aleksandra and {La Posta}, Adrien and {Lague}, Alex and {Lakey}, Victoria and {Lee}, Eunseong and {Li}, Zack and {Li}, Yaqiong and {Limon}, Michele and {Lokken}, Martine and {Louis}, Thibaut and {Lungu}, Marius and {MacCrann}, Niall and {MacInnis}, Amanda and {Maldonado}, Diego and {Maldonado}, Felipe and {Mallaby-Kay}, Maya and {Marques}, Gabriela A. and {McMahon}, Jeff and {Mehta}, Yogesh and {Menanteau}, Felipe and {Moodley}, Kavilan and {Morris}, Thomas W. and {Mroczkowski}, Tony and {Naess}, Sigurd and {Namikawa}, Toshiya and {Nati}, Federico and {Newburgh}, Laura and {Nicola}, Andrina and {Niemack}, Michael D. and {Nolta}, Michael R. and {Orlowski-Scherer}, John and {Page}, Lyman A. and {Pandey}, Shivam and {Partridge}, Bruce and {Prince}, Heather and {Puddu}, Roberto and {Radiconi}, Federico and {Robertson}, Naomi and {Rojas}, Felipe and {Sakuma}, Tai and {Salatino}, Maria and {Schaan}, Emmanuel and {Schmitt}, Benjamin L. and {Sehgal}, Neelima and {Shaikh}, Shabbir and {Sierra}, Carlos and {Sievers}, Jon and {Sif{\'o}n}, Crist{\'o}bal and {Simon}, Sara and {Sonka}, Rita and {Spergel}, David N. and {Staggs}, Suzanne T. and {Storer}, Emilie and {Switzer}, Eric R. and {Tampier}, Niklas and {Thornton}, Robert and {Trac}, Hy and {Treu}, Jesse and {Tucker}, Carole and {Ullom}, Joel and {Vale}, Leila R. and {Van Engelen}, Alexander and {Van Lanen}, Jeff and {van Marrewijk}, Joshiwa and {Vargas}, Cristian and {Vavagiakis}, Eve M. and {Wagoner}, Kasey and {Wang}, Yuhan and {Wenzl}, Lukas and {Wollack}, Edward J. and {Xu}, Zhilei and {Zago}, Fernando and {Zheng}, Kaiwen},
        title = "{The Atacama Cosmology Telescope: A Measurement of the DR6 CMB Lensing Power Spectrum and Its Implications for Structure Growth}",
      journal = {\apj},
         year = 2024,
        month = feb,
       volume = {962},
       number = {2},
          eid = {112},
        pages = {112},
          doi = {10.3847/1538-4357/acfe06},
archivePrefix = {arXiv},
       eprint = {2304.05202},
 primaryClass = {astro-ph.CO},
       adsurl = {https://ui.adsabs.harvard.edu/abs/2024ApJ...962..112Q}
}

@ARTICLE{Carron2022,
       author = {{Carron}, Julien and {Mirmelstein}, Mark and {Lewis}, Antony},
        title = "{CMB lensing from Planck PR4 maps}",
      journal = {\jcap},
         year = 2022,
        month = sep,
       volume = {2022},
       number = {9},
          eid = {039},
        pages = {039},
          doi = {10.1088/1475-7516/2022/09/039},
archivePrefix = {arXiv},
       eprint = {2206.07773},
 primaryClass = {astro-ph.CO},
       adsurl = {https://ui.adsabs.harvard.edu/abs/2022JCAP...09..039C}
}

@ARTICLE{Pan2023,
       author = {{Pan}, Z. and {Bianchini}, F. and {Wu}, W.~L.~K. and {Ade}, P.~A.~R. and {Ahmed}, Z. and {Anderes}, E. and {Anderson}, A.~J. and {Ansarinejad}, B. and {Archipley}, M. and {Aylor}, K. and {Balkenhol}, L. and {Barry}, P.~S. and {Basu Thakur}, R. and {Benabed}, K. and {Bender}, A.~N. and {Benson}, B.~A. and {Bleem}, L.~E. and {Bouchet}, F.~R. and {Bryant}, L. and {Byrum}, K. and {Camphuis}, E. and {Carlstrom}, J.~E. and {Carter}, F.~W. and {Cecil}, T.~W. and {Chang}, C.~L. and {Chaubal}, P. and {Chen}, G. and {Chichura}, P.~M. and {Cho}, H. -M. and {Chou}, T. -L. and {Cliche}, J. -F. and {Coerver}, A. and {Crawford}, T.~M. and {Cukierman}, A. and {Daley}, C. and {de Haan}, T. and {Denison}, E.~V. and {Dibert}, K.~R. and {Ding}, J. and {Dobbs}, M.~A. and {Doussot}, A. and {Dutcher}, D. and {Everett}, W. and {Feng}, C. and {Ferguson}, K.~R. and {Fichman}, K. and {Foster}, A. and {Fu}, J. and {Galli}, S. and {Gambrel}, A.~E. and {Gardner}, R.~W. and {Ge}, F. and {Goeckner-Wald}, N. and {Gualtieri}, R. and {Guidi}, F. and {Guns}, S. and {Gupta}, N. and {Halverson}, N.~W. and {Harke-Hosemann}, A.~H. and {Harrington}, N.~L. and {Henning}, J.~W. and {Hilton}, G.~C. and {Hivon}, E. and {Holder}, G.~P. and {Holzapfel}, W.~L. and {Hood}, J.~C. and {Howe}, D. and {Huang}, N. and {Irwin}, K.~D. and {Jeong}, O. and {Jonas}, M. and {Jones}, A. and {K{\'e}ruzor{\'e}}, F. and {Khaire}, T.~S. and {Knox}, L. and {Kofman}, A.~M. and {Korman}, M. and {Kubik}, D.~L. and {Kuhlmann}, S. and {Kuo}, C. -L. and {Lee}, A.~T. and {Leitch}, E.~M. and {Levy}, K. and {Lowitz}, A.~E. and {Lu}, C. and {Maniyar}, A. and {Menanteau}, F. and {Meyer}, S.~S. and {Michalik}, D. and {Millea}, M. and {Montgomery}, J. and {Nadolski}, A. and {Nakato}, Y. and {Natoli}, T. and {Nguyen}, H. and {Noble}, G.~I. and {Novosad}, V. and {Omori}, Y. and {Padin}, S. and {Paschos}, P. and {Pearson}, J. and {Posada}, C.~M. and {Prabhu}, K. and {Quan}, W. and {Raghunathan}, S. and {Rahimi}, M. and {Rahlin}, A. and {Reichardt}, C.~L. and {Riebel}, D. and {Riedel}, B. and {Ruhl}, J.~E. and {Sayre}, J.~T. and {Schiappucci}, E. and {Shirokoff}, E. and {Smecher}, G. and {Sobrin}, J.~A. and {Stark}, A.~A. and {Stephen}, J. and {Story}, K.~T. and {Suzuki}, A. and {Takakura}, S. and {Tandoi}, C. and {Thompson}, K.~L. and {Thorne}, B. and {Trendafilova}, C. and {Tucker}, C. and {Umilta}, C. and {Vale}, L.~R. and {Vanderlinde}, K. and {Vieira}, J.~D. and {Wang}, G. and {Whitehorn}, N. and {Yefremenko}, V. and {Yoon}, K.~W. and {Young}, M.~R. and {Zebrowski}, J.~A.},
        title = "{Measurement of gravitational lensing of the cosmic microwave background using SPT-3G 2018 data}",
      journal = {\prd},
         year = 2023,
        month = dec,
       volume = {108},
       number = {12},
          eid = {122005},
        pages = {122005},
          doi = {10.1103/PhysRevD.108.122005},
archivePrefix = {arXiv},
       eprint = {2308.11608},
 primaryClass = {astro-ph.CO},
       adsurl = {https://ui.adsabs.harvard.edu/abs/2023PhRvD.108l2005P}
}

@INPROCEEDINGS{Smith04,
       author = {{Smith}, Greg A. and {Saunders}, Will and {Bridges}, Terry and {Churilov}, Vladimir and {Lankshear}, Allan and {Dawson}, John and {Correll}, David and {Waller}, Lew and {Haynes}, Roger and {Frost}, Gabriella},
        title = "{AAOmega: a multipurpose fiber-fed spectrograph for the AAT}",
    booktitle = {Ground-based Instrumentation for Astronomy},
         year = 2004,
       editor = {{Moorwood}, Alan F.~M. and {Iye}, Masanori},
       series = {Society of Photo-Optical Instrumentation Engineers (SPIE) Conference Series},
       volume = {5492},
        month = sep,
        pages = {410-420},
          doi = {10.1117/12.551013},
       adsurl = {https://ui.adsabs.harvard.edu/abs/2004SPIE.5492..410S}
}

\appendix
\renewcommand{\thefigure}{\thesection.\arabic{figure}}
\renewcommand{\thetable}{\thesection.\arabic{table}}

\section{Comparing systematic uncertainty budgets}\label{appendix1}
In Section~\ref{sec:systbudget} we present the systematic error budget on $Q_H(0.2)$ that allows us to present a single non-degenerate number summarising the constraints in the $w-\Omega_{\rm m}$ plane (similar to how $S_8$ summarises the constraints in the $\sigma_8-\Omega_{\rm m}$ plane for lensing analyses). Historically, the error budget has been presented for $w$; for completeness, we also provide the separate systematic error budgets on $\Omega_m$ and $w$, in Figures~\ref{fig:unc_budget_om} and~\ref{fig:unc_budget_w} respectively for the Unite analysis. 
\begin{figure}
    \centering \includegraphics[width=0.95\linewidth]{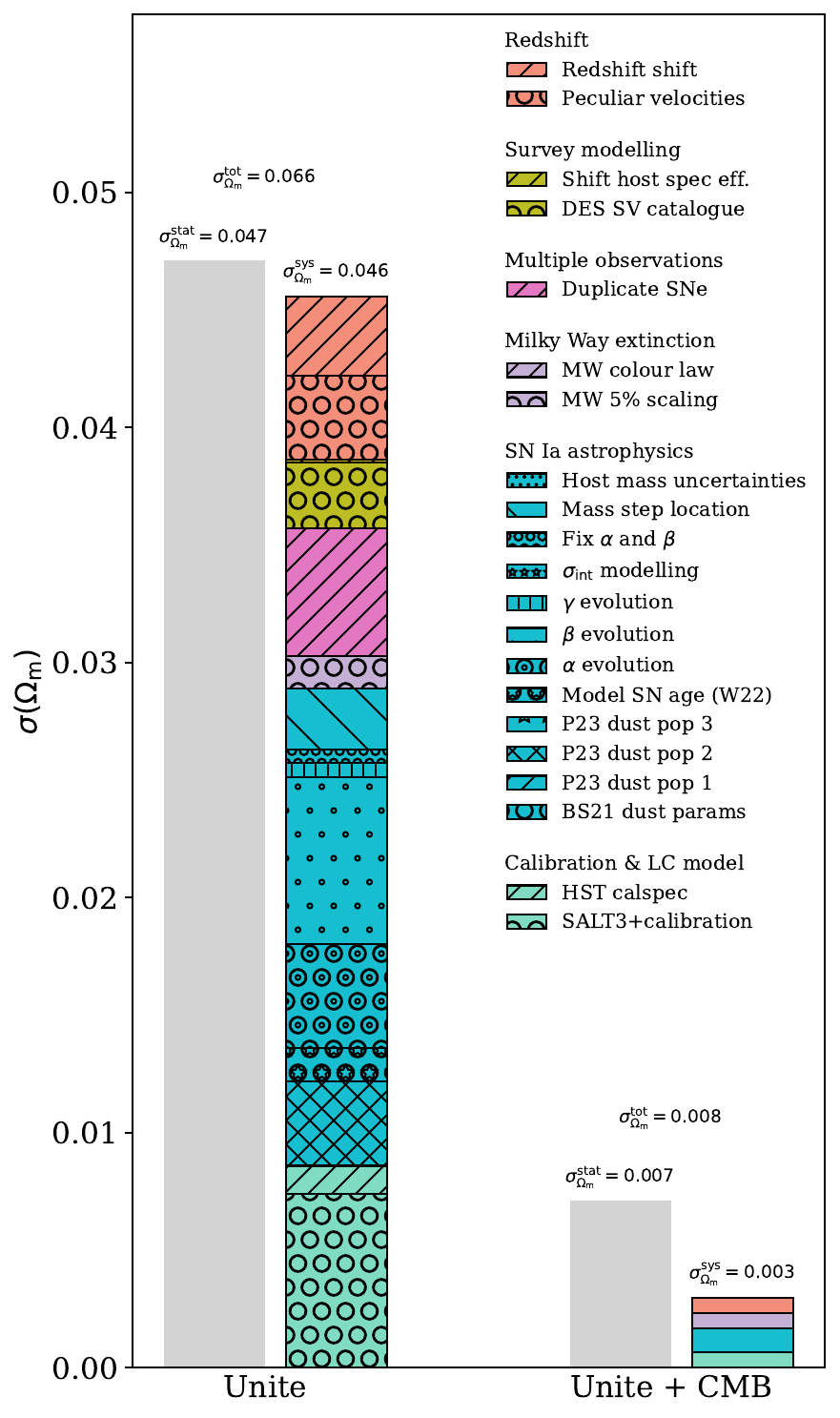}
    \caption{Systematic error budget on $\Omega_m$ with and without a CMB prior. The statistical uncertainty is also shown for comparison. The error budget that includes a CMB prior only shows the colour of the systematic categories for clarity. The various sources of systematic uncertainty in this analysis and labelling conventions are presented in Table~\ref{tab:systsources}.}
    \label{fig:unc_budget_om}
\end{figure}
\begin{figure}
    \centering \includegraphics[width=0.95\linewidth]{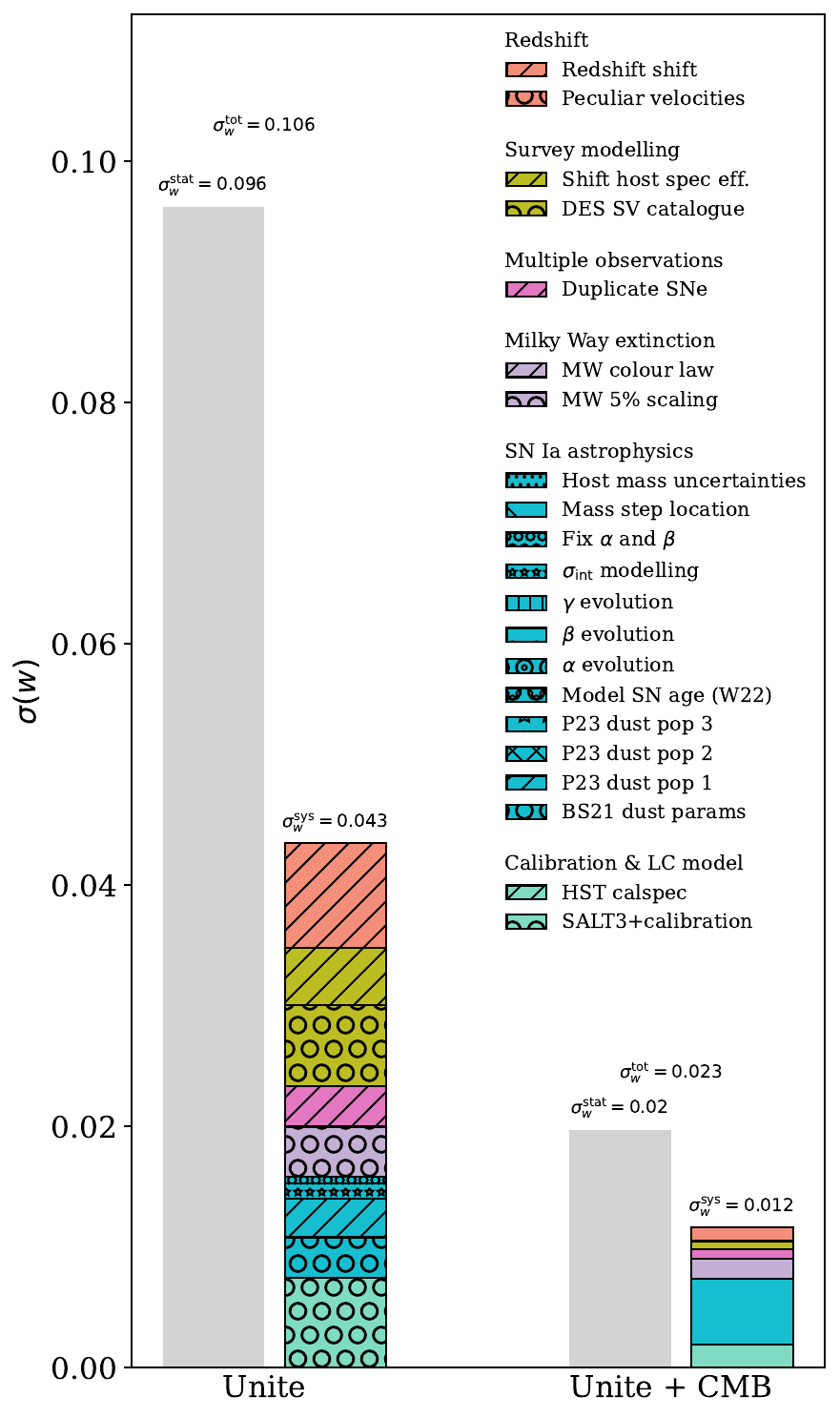}
    \caption{Systematic error budget on $w$ with and without a CMB prior. The statistical uncertainty is also shown for comparison. The error budget that includes a CMB prior only shows the colour of the systematic categories for clarity. The various sources of systematic uncertainty in this analysis and labelling conventions are presented in Table~\ref{tab:systsources}.}
    \label{fig:unc_budget_w}
\end{figure}

\section{Additional validation tests: Intrinsic scatter model}\label{appendix2}
The BBC framework relies on good agreement between the bias correction simulations and the observed data. Accordingly, we compare key simulated distributions to the data as part of our unblinding checks (see Section~\ref{sec:unblind}). In addition to the checks described above, we also examine the colour distribution, the mean Hubble residuals, and the RMS of the Hubble residuals for the Unite sample, which provide a useful consistency check on the intrinsic scatter model.

Towards the completion of this work, independent analyses on the Dust2dust fitting code suggested that the three sets of dust parameters used as systematic variations may not accurately capture the range of systematic uncertainty. Although we do not pursue this further in the present analysis, we perform additional validation tests to assess whether the nominal P23 model is robust to reasonable variations in the best fit dust parameters. Using parameters previously identified as having the most impact on cosmological constraints, we vary individual parameters and parameter combinations by $1\sigma$ and $2\sigma$ and assess their impact on the modelling of intrinsic scatter. We present our results in Figure~\ref{fig:d2d_tests}. The only discrepancy we find is between the simulations and data in the RMS of the residuals for $c>0.2$, such that our uncertainties may be overestimated. However, the model remains robust to the range of parameter shifts considered.
\begin{figure*}
    \centering \includegraphics[width=\linewidth]{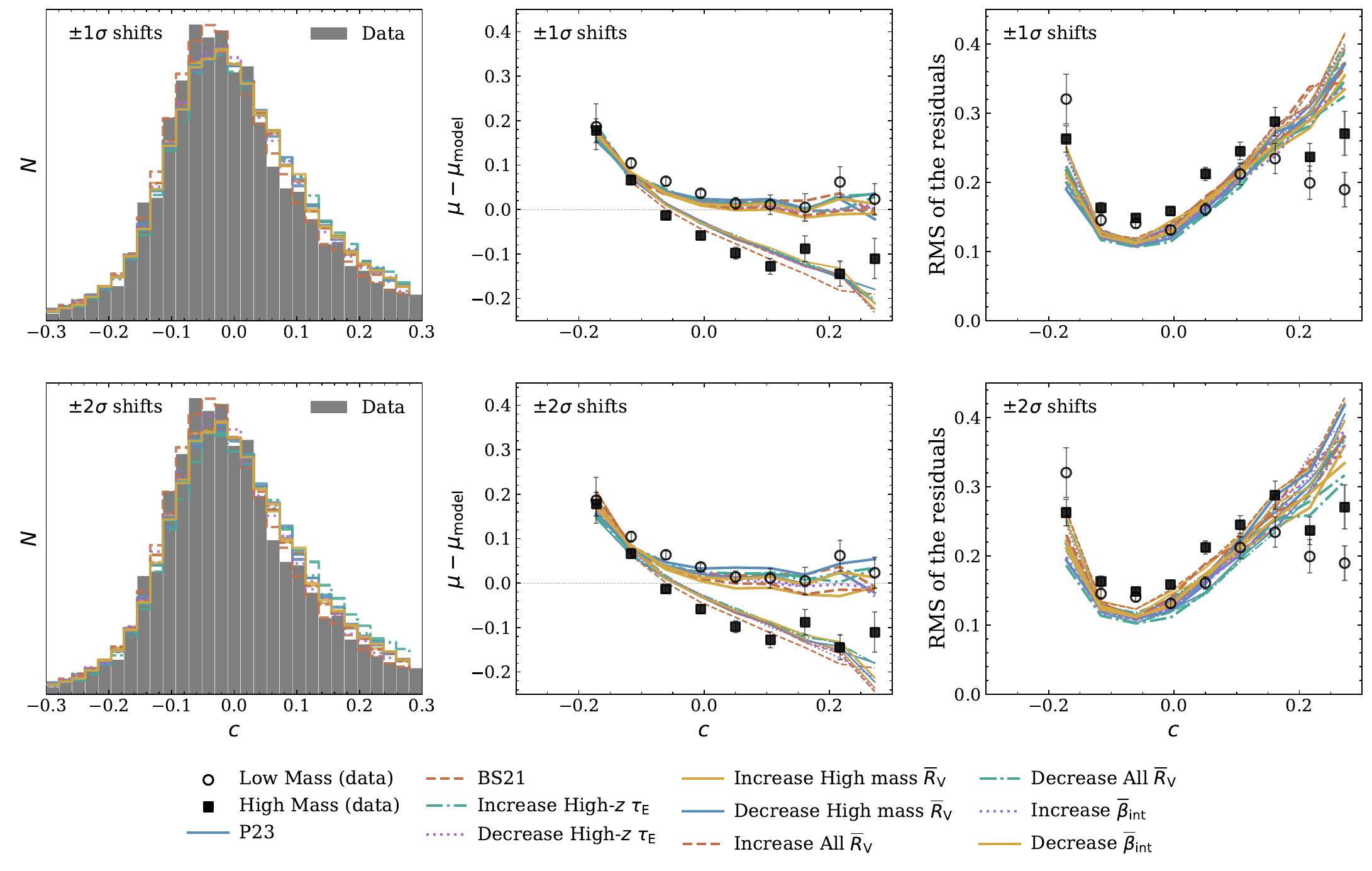}
    \caption{Comparison of the colour distribution (left), mean Hubble residuals (centre), and RMS of the Hubble residuals (right) for the Unite sample as a function of \texttt{SALT3} colour, $c$, between the nominal P23 model and variants obtained by applying parameter shifts. The individual parameters and parameter combinations that we shift are given in the figure legend. The upper panels show results for models with $1\sigma$ shifts applied, while the lower panels show results for models with $2\sigma$ shifts applied. The model is robust to these shifts.}
    \label{fig:d2d_tests}
\end{figure*}

\section{Impact of analysis variants on Cosmology}\label{appendix3}
\begin{table}
    \centering
    \caption{Summary of cosmological parameter constraints for four models fitted to the nominal Unite sample, Unite with no type Ia probability cut and redetermined bias corrections, Unite with a $z>0.01$ redshift cut applied to cosmology fits (the same cut used by \pplus\ during cosmology fitting to mitigate against peculiar velocity biases) and Unite without the lensing correction term defined in Section~\ref{sec:lens}. Reported values correspond to the medians of the marginalised posterior distributions, with 68.27\% credible intervals.}
    \renewcommand{\arraystretch}{1.5}
    \label{tab:appendix_results}
    \setlength{\tabcolsep}{5pt}
    \begin{tabular}{lcccc}
        \cline{1-5}
        & $\om$ & $\ok$ & $w_0$ & $w_a$ \\
        
        \cline{1-5}
        \multicolumn{5}{l}{\textbf{Unite} } \\ \cline{1-5}
        Flat-$\Lambda$CDM & $\mflcdmS$  &  -- & -- & -- \\
        $\Lambda$CDM & $\mlcdmS$  &  $\klcdmS$ & -- & -- \\
        Flat-$w$CDM & $\mfwcdmS$  &  -- & $\wfwcdmS$ & -- \\
        Flat-$w_0w_a$CDM & $\mfwacdmS$  &  -- & $\wfwacdmS$ & $\afwacdmS$ \\
        \cline{1-5}
        \multicolumn{5}{l}{\textbf{Unite: No $P_{\rm Ia}$ cut}} \\ \cline{1-5}
		Flat-$\Lambda$CDM & $0.310\pm0.011$  &  -- & -- & -- \\
		$\Lambda$CDM & $0.199^{+0.043}_{-0.041}$  &  $0.28^{+0.10}_{-0.11}$ & -- & -- \\
        Flat-$w$CDM & $0.181^{+0.053}_{-0.049}$  &  -- & $-0.741^{+0.066}_{-0.087}$ & -- \\
        Flat-$w_0w_a$CDM & $0.344^{+0.062}_{-0.114}$  &  -- & $-0.744^{+0.094}_{-0.091}$ & $-1.92^{+1.55}_{-2.02}$ \\
        \cline{1-5}
        \multicolumn{5}{l}{\textbf{Unite: $z>0.01$ redshift cut}} \\ \cline{1-5}
		Flat-$\Lambda$CDM & $0.311\pm0.011$  &  -- & -- & -- \\
		$\Lambda$CDM & $0.212^{+0.043}_{-0.045}$  & $0.26\pm0.11$ & -- & -- \\
        Flat-$w$CDM & $0.197\pm0.056$  &  -- & $-0.760^{+0.078}_{-0.098}$ & -- \\
        Flat-$w_0w_a$CDM & $0.311^{+0.084}_{-0.142}$  &  -- & $-0.774^{+0.098}_{-0.097}$ & $-1.07^{+1.17}_{-2.02}$ \\
        \cline{1-5}
        \multicolumn{5}{l}{\textbf{Unite: No $\Delta m_{\mathrm{lens}}$ term}} \\ \cline{1-5}
        Flat-$\Lambda$CDM & $0.310^{+0.012}_{-0.011}$  &  -- & -- & -- \\
        $\Lambda$CDM & $0.220^{+0.045}_{-0.044}$  & $0.23\pm0.11$ & -- & -- \\
        Flat-$w$CDM & $0.206^{+0.056}_{-0.058}$  &  -- & $-0.777^{+0.083}_{-0.101}$ & -- \\
        Flat-$w_0w_a$CDM & $0.336^{+0.077}_{-0.146}$  &  -- & $-0.790^{+0.111}_{-0.102}$ & $-1.39^{+1.40}_{-2.38}$ \\
        \cline{1-5}
    \end{tabular}
\end{table}

\subsection{Cosmology without the type Ia probability cut}\label{sec:appendix_piacut}

One difference between Unite and both DES-Dovekie and \des\ is that Unite keeps only those SNe for which the \texttt{SCONE} classifier assigns a probability of being type Ia greater than $0.8$, and then subsequently treats DES as a pure sample (see Section~\ref{sec:samplecuts} for details). Here, we redetermine bias corrections without this cut to see how this choice impacts our cosmological fits to the SN data alone. 
Without the type Ia probability cut, we obtain a cosmology sample of $3173$ SNe. We summarise our results in Table~\ref{tab:appendix_results} and compare posteriors in Figures~\ref{fig:flcdm_appendix},~\ref{fig:lcdm_appendix},~\ref{fig:fwcdm_appendix}, and~\ref{fig:fw0wacdm_appendix}. We find fully consistent cosmological parameter constraints across all four models. We find the largest shifts in our best fit time-varying dark energy parameters, although these lie along the degeneracy directions. 

The purpose of this cut was to eliminate SNe that the machine-learning classifiers confidently identified as non-type Ia, which would remove SNe with uncertainties $\gtrsim 100$ magnitudes. The inclusion of these likely non-type Ia SNe in the DES-SN5YR data release was a point of confusion for users that we want to alleviate. 

As discussed in Section~\ref{sec:samplecuts}, after applying a probability cut we found the BBC method resulted in small biases on our best fit cosmological parameters when fitting simulated data. This arises because the distribution of the probability of being a SNe~Ia for the \des\ sample (see Figure~\ref{fig:Pia}) is essentially bimodal, so there was not enough residual contamination for BEAMS to accurately model both populations. Therefore, we adopted a conservative cut of $P_{\rm Ia}>0.8$ and then treated DES as a pure sample. The classifiers used in Unite have been shown to achieve purity and efficiency levels of $\gtrsim98\%$ using a binary-classification threshold of $0.5$ and a more conservative threshold was chosen to minimise any residual contamination. 

Although we took this approach, future surveys such as LSST are expected to discover orders of magnitude more SNe. In this regime, it may be possible to retain a larger fraction of SNe in the intermediate probability range $0.2 < P_{\rm Ia} < 0.8$, where sufficient residual contamination exists for BEAMS to reliably model both populations. This would enable the use of a less conservative cut, improving statistical precision without including SNe with extremely large uncertainties, which are effectively uninformative.

\subsection{Cosmology with a low redshift cut}\label{sec:appendix_lowzcut}
At $z<0.01$, peculiar velocities contribute significantly to the observed redshift. In particular, at low redshift an Eddington bias occurs due to there being higher volume in equal $dz$ bins at higher redshift -- so peculiar velocities scatter more high-redshift supernovae to low redshift than vice versa.

To mitigate the bias in redshifts due to volumetric effects, \pplus\ cut SN below this redshift threshold when performing cosmology fits. Here, we have instead included the low-redshift supernovae but modelled the Eddington bias as part of our bias correction. To test the impact on our results, we reperform fits to Unite alone keeping only those SNe with $z>0.01$. Note that we impose the redshift cut in the same manner as \pplus\ by directly masking the SN data and covariance matrix, without redetermining bias corrections or nuisance parameters. We summarise our results in Table~\ref{tab:appendix_results} and compare posteriors in Figures~\ref{fig:flcdm_appendix},~\ref{fig:lcdm_appendix},~\ref{fig:fwcdm_appendix}, and~\ref{fig:fw0wacdm_appendix}. 

We find that including or excluding these very low-$z$ data points yields no discernible difference in our cosmological fits and consequently, we choose not to apply this cut. However, our public likelihood code includes the option to anchor Unite to Cepheid host distances. For this scenario, a default redshift cut of $z>0.0233$ is applied, and no peculiar velocity correction is applied, see \citet[][in prep.]{lee25h0} for more details.
\begin{figure}[ht!]
    \centering \includegraphics[width=\linewidth]{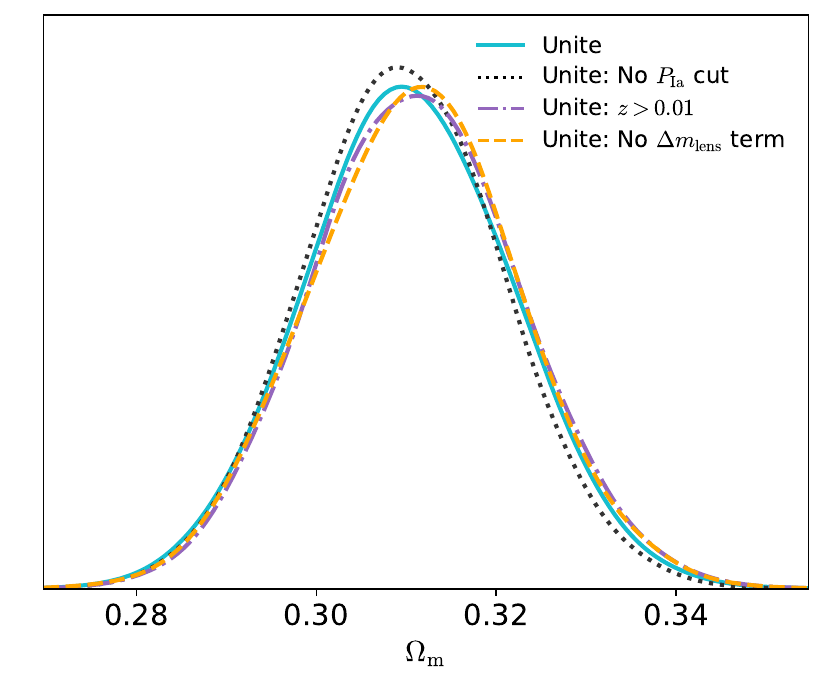}
    \caption{The posteriors for the constraints on the Flat-$\Lambda$CDM model from the nominal Unite analysis (blue), Unite with no type Ia probability cut and redetermined bias corrections (dotted black), Unite with a $z>0.01$ redshift cut applied to cosmology fits (dash-dotted purple; the same cut used by \pplus\ during cosmology fitting to mitigate against peculiar velocity biases) and Unite with no lensing correction term (yellow dashed). }
    \label{fig:flcdm_appendix}
\end{figure}

\subsection{Cosmology without the lensing correction term}\label{sec:appendix_nolens}
This is the first major supernova cosmology analysis to implement a lensing correction term, $\Delta m_{\rm lens}$ by default. In Table~\ref{tab:appendix_results}, we report the cosmological constraints had this term been omitted and compare posteriors in Figures~\ref{fig:flcdm_appendix},~\ref{fig:lcdm_appendix},~\ref{fig:fwcdm_appendix}, and~\ref{fig:fw0wacdm_appendix}.

We find consistent cosmological parameter constraints across all four models, with differences corresponding to shifts along parameter degeneracy directions. The lensing correction term produces a small displacement of the constraints away from $\Lambda$CDM.
\begin{figure}[ht!]
    \centering \includegraphics[width=\linewidth]{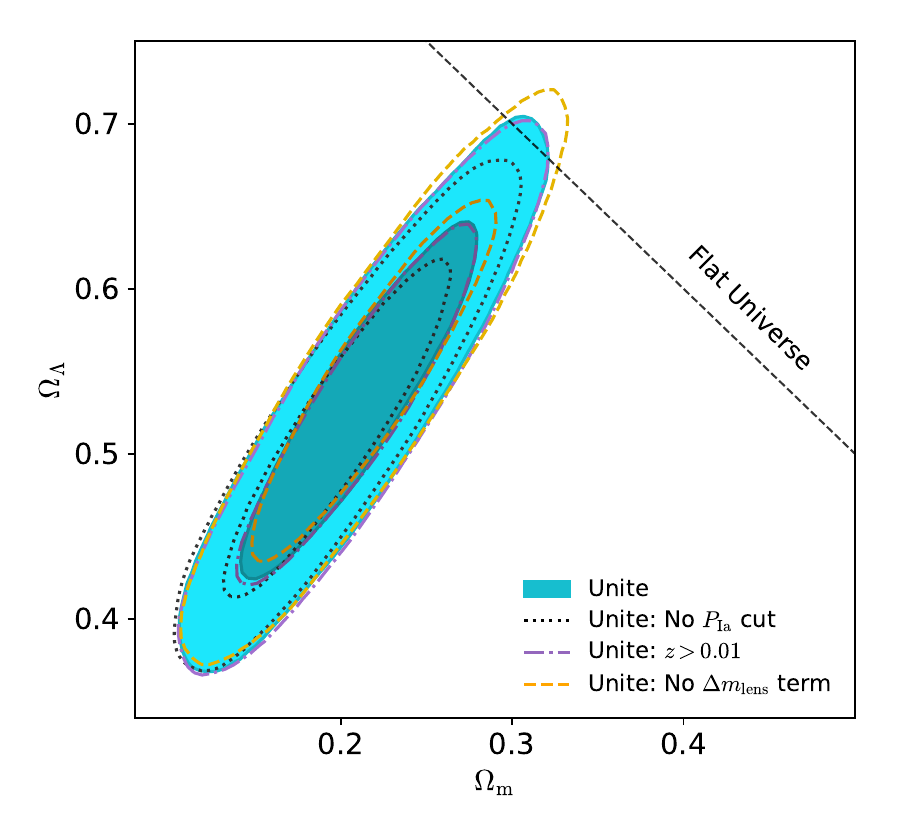}
\caption{Same as Figure~\ref{fig:flcdm_appendix}, but for the $\Lambda$CDM model.}
    \label{fig:lcdm_appendix}
\end{figure}
\begin{figure}
    \centering \includegraphics[width=\linewidth]{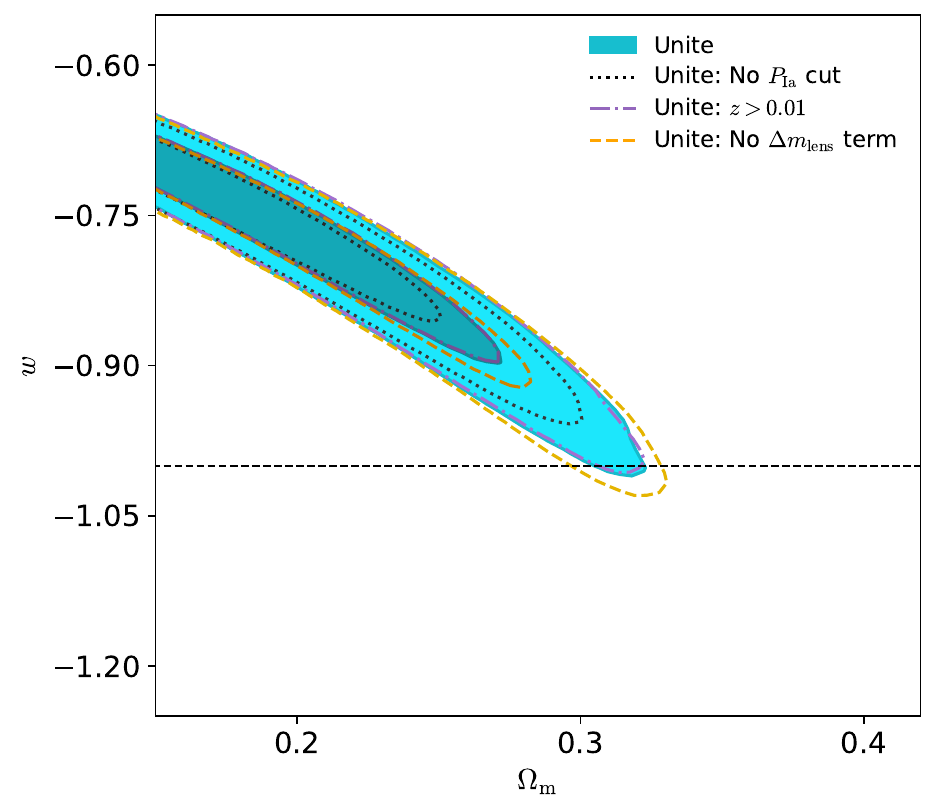}
\caption{Same as Figure~\ref{fig:flcdm_appendix}, but for the Flat-$w$CDM model.}
    \label{fig:fwcdm_appendix}
\end{figure}
\begin{figure}
    \centering \includegraphics[width=\linewidth]{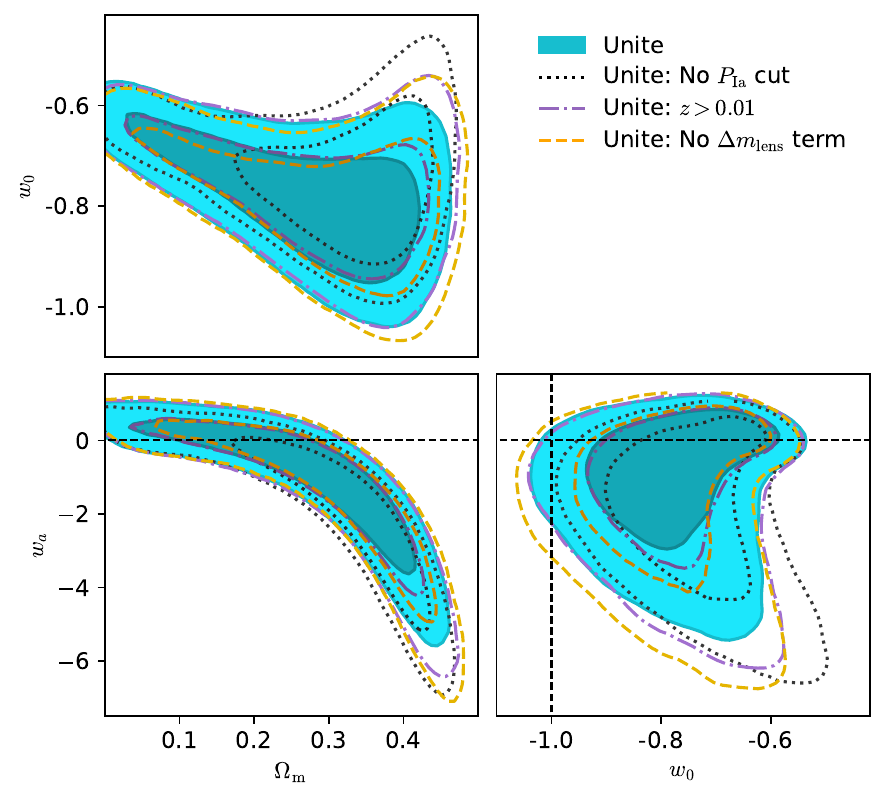}
\caption{Same as Figure~\ref{fig:flcdm_appendix}, but for the Flat-$w_0 w_a$CDM model.}
    \label{fig:fw0wacdm_appendix}
\end{figure}

\end{document}